\documentclass[aps,preprint,superscriptaddress,nofootinbib]{revtex4-2}
\usepackage{amsmath}
\allowdisplaybreaks[1]
\usepackage{mathtools}
\usepackage{xcolor}
\usepackage{physics}
\usepackage{graphicx}
\usepackage{amssymb}
\usepackage[colorlinks]{hyperref}
\usepackage{slashed}
\usepackage{bm}
\usepackage{multirow}
\usepackage{adjustbox}
\usepackage{subfigure}

\newcommand{\beq}{\begin{equation}}
	\newcommand{\eeq}{\end{equation}}

\long\def\comment#1{}

\usepackage{color}
\definecolor{darkblue}{rgb}{0.0,0,0.5}
\usepackage{ulem}

\begin{document}
	
\preprint{
{\vbox {
				\hbox{\bf MSUHEP-22-029}
	}}}
	\vspace*{0.2cm}
	
\title{Impact of Lattice Strangeness Asymmetry  Data in the CTEQ-TEA Global Analysis}

\author{Tie-Jiun Hou}
\email{tjhou@msu.edu}
\affiliation{School of Nuclear Science and Technology, University of South China, Hengyang, Hunan 421001, China}

\author{Huey-Wen Lin}
\email{hueywen@msu.edu}
\affiliation{Department of Physics and Astronomy, Michigan State University, East Lansing, MI 48824}
\affiliation{Department of Computational Mathematics, Science \& Engineering, Michigan State University, East Lansing, MI 48824}

\author{Mengshi Yan}
\email{msyan@pku.edu.cn}
\affiliation{Department of Physics and State Key Laboratory of Nuclear Physics and Technology, Peking University, Beijing 100871, China}

\author{C.-P. Yuan}
\email{yuanch@msu.edu}
\affiliation{Department of Physics and Astronomy, Michigan State University, East Lansing, MI 48824}

\begin{abstract}
We study the impact of lattice data on the determination of the strangeness asymmetry distribution $s_-(x) \equiv s(x) - {\bar s}(x)$ in the general CTEQ-TEA global analysis of parton distribution functions (PDFs) of the proton. Firstly, we find that allowing a nonvanishing $s_-(x)$, at the initial $Q_0=1.3$~GeV scale, in a global PDF analysis leads to a CT18As fit with similar quality to CT18A. Secondly, including the lattice data in the CT18As\_Lat fit greatly reduces the $s_-$-PDF error band size in the large-$x$ region.
To further reduce its error would require more precise lattice data, extended to smaller $x$ values.
We take ATLAS 7 TeV $W$ and $Z$ production data, SIDIS di-muon production data, $F_3$ structure function data, E866 NuSea data, and E906 SeaQuest data as examples to illustrate the implication of CT18As and CT18As\_Lat fits.
The parametrization dependence for PDF ratio $(s+\bar{s})/(\bar{u}+\bar{d})(x)$ is analyzed with CT18As2 and CT18As2\_Lat fits as results.
\end{abstract}
\date{\today}

\maketitle
\clearpage

\section{Introduction} \label{intro}

The Large Hadron Collider (LHC) has entered an era of precision physics. To match the experimental precision, it is necessary to have precise predictions in QCD theory, which require correspondingly precise parton distribution functions (PDFs), such as the recent CT18~\cite{Hou:2019efy}, MSHT20~\cite{Bailey:2020ooq} and NNPDF4.0~\cite{Ball:2021leu} PDFs obtained at the
next-to-next-to-leading order (NNLO) accuracy in QCD.

In the CT18 analysis, the strange quark and antiquark PDFs of the proton are assumed to be the same, $\bar{s}(x)=s(x)$, at $Q_0=1.3$~GeV, where the nonperturbative PDFs are specified, and a nonvanishing $(s-\bar{s})(x)$
is generated at higher energy scales by DGLAP evolution~\cite{Moch:2004pa,Catani:2004nc}.
In Ref.~\cite{Hou:2019efy}, noticeable tensions between the original NuTeV~\cite{Mason:2006qa} and CCFR~\cite{NuTeV:2001dfo} semi-inclusive deep-inelastic scattering (SIDIS) di-muon data and the precision ATLAS $\sqrt{s} = 7$~TeV $W$, $Z$ data~\cite{ATLAS:2016nqi} were found. In MSHT20~\cite{Bailey:2020ooq}, it was concluded that allowing $s(x) \neq \bar{s}(x)$ at the $Q_0$ scale can release some of these tensions.
In this work, we extend the CT18 analysis to allow a nonvanishing strangeness asymmetry,
$s_{-}(x) \equiv s(x) - \bar{s}(x)$, at the $Q_0$ scale, and the resulting PDF set is hereafter referred to as CT18As.

Besides the phenomenology approach of performing  global PDF analysis,
a nonperturbative approach from first principles, such as lattice QCD (LQCD), provides hope to resolve many of the outstanding theoretical disagreements and provides information in regions that are unknown or difficult to observe in experiments.
Recent breakthroughs, such as large-momentum effective theory (LaMET)~\cite{Ji:2013dva,Ji:2014gla,Ji:2020ect} (also called the quasi-PDF method), have made it possible for lattice calculations to provide information on the $x$-dependent PDFs. There have been many pioneering works showing great promise in obtaining quantitative results for the unpolarized, helicity and transversity quark and antiquark distributions~\cite{Lin:2014gaa,Lin:2014yra,Lin:2014zya,Chen:2016utp,Alexandrou:2014pna,Alexandrou:2015rja}
using the quasi-PDFs approach~\cite{Ji:2013dva}.
Increasingly many lattice works are being performed at physical pion mass since the first study in Ref.~\cite{Lin:2017ani}.
(A recent review of the theory and lattice calculations can be found in Refs.~\cite{Ji:2020ect,Lin:2017snn,Constantinou:2020hdm}.)
The first Bjorken-$x$--dependent strange (anti)quark PDF using LQCD calculations was reported in Ref.~\cite{Zhang:2020dkn}.
This calculation was done using a single lattice spacing, 0.12~fm, with extrapolation to physical pion masses.
In this work, we use the extrapolated lattice matrix elements to calculate the strangeness asymmetry distribution $s_{-}(x) \equiv s(x) - {\bar s}(x)$, which is then taken as an input to further constrain $s(x)$ and $\bar{s}(x)$ at the $Q_0$ scale in the CT18-like global analysis. The resulting PDF set is denoted as CT18As\_Lat.

The strangeness asymmetry $s_-(x)$ in the Lattice calculation and the CTEQ-TEA PDF analysis is reviewed in Sec.~\ref{sec:fitting}. The Sec.~\ref{results} describes the updated strangeness asymmetry results obtained in CT18As and CT18As\_Lat fits, by allowing a non-vanishing strangeness asymmetry $s_-(x)$ at the initial $Q_0=1.3$ GeV scale.
The implication of CT18As and CT18As\_Lat fits is studied in Sec.~\ref{pheno} by comparing numerical predictions to experimental data for some observables. Sec.~\ref{conclusion} contains our conclusion.
In Appendix~\ref{sec_app:CT18As2_Rs}, we study the parametrization dependence of the PDF ratio $(s+\bar{s})/(\bar{u}+\bar{d})(x)$ in the large-$x$ region and obtain alternative CT18As2 and CT18As2\_Lat PDFs.
In Appendix~\ref{sec_app:para}, we summarize the specific parametrization functional forms for $s(x)$ and $\bar{s}(x)$ PDFs.

\section{Strangeness Asymmetry from Lattice and CT18}
\label{sec:fitting}

\subsection{Lattice calculation of the strangeness asymmetry distribution $s_{-}(x)$ }
\label{fitting:svl_lat}

\begin{figure}[tb!]
\centering
\includegraphics[width=0.43\linewidth]{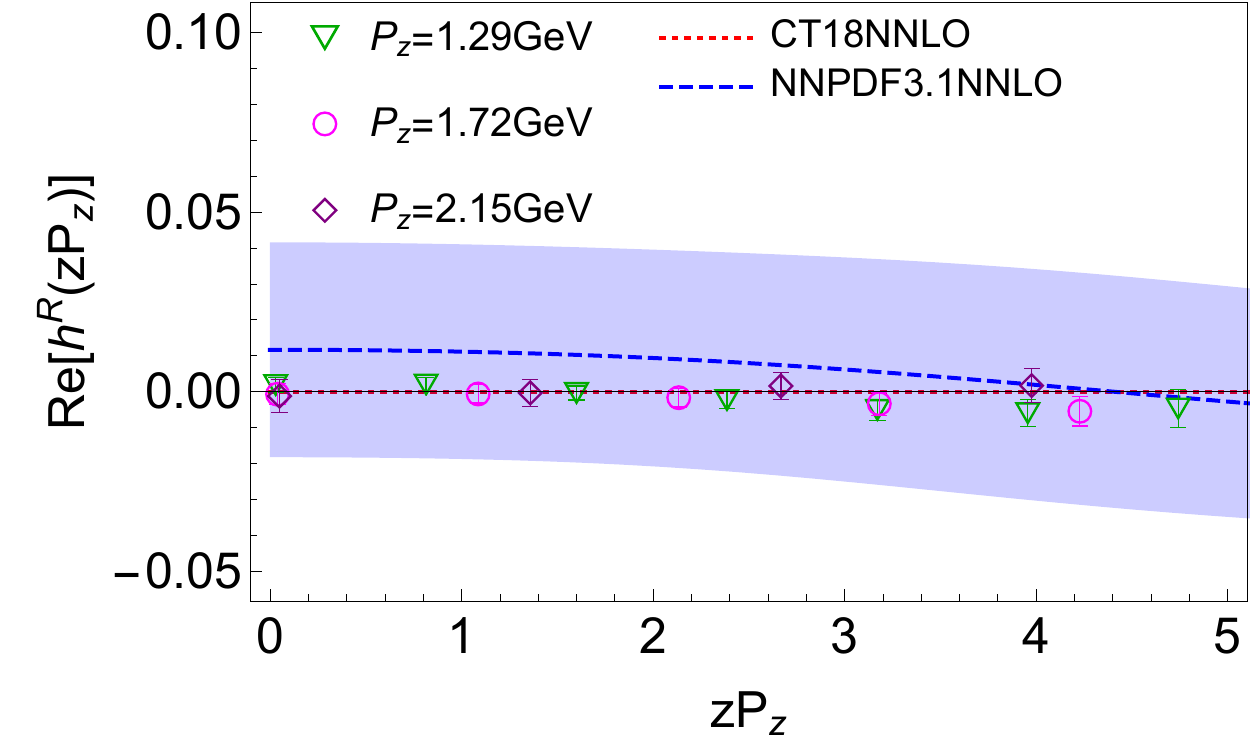}
\includegraphics[width=0.42\linewidth]{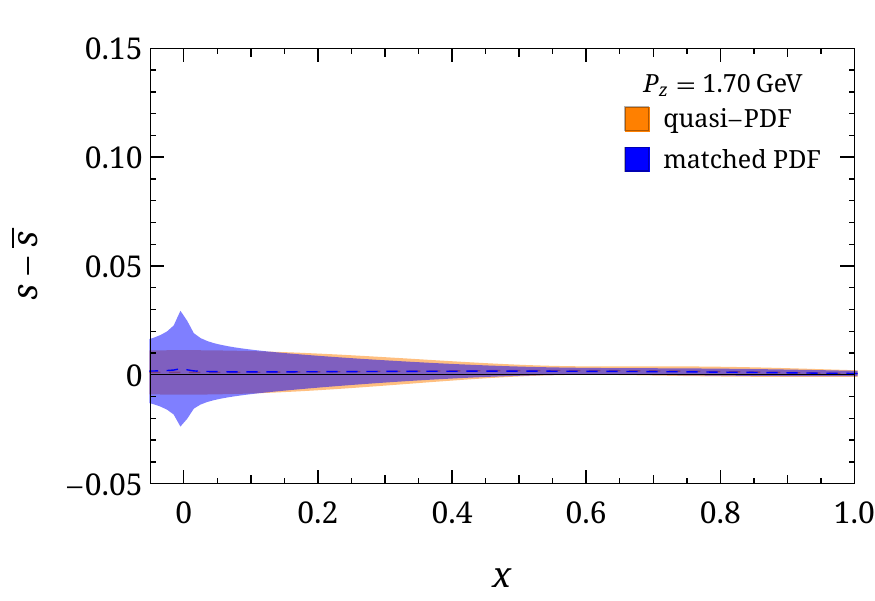}
\caption{\label{fig:LatStrange}
(left) The real parts of the strange quasi-PDF matrix elements in coordinate space from our calculations at physical pion mass with $P_z \in [1.3, 2.15]$~GeV~\cite{Zhang:2020dkn}, along with those from CT18 and NNPDF NNLO.
(right) The quasi (orange) and matched (blue) valence strange distribution from LQCD calculation.
}
\end{figure}

Since the strangeness asymmetry $s_{-}(x)\equiv s(x) - {\bar s}(x)$ is flavor-singlet, we can confidently calculate it using LaMET coordinate-space matrix elements on the lattice. In this work, we use the matrix elements from Ref.~\cite{Zhang:2020dkn} computed on a single 0.12-fm lattice ensemble with a $2+1+1$-flavor HISQ sea with 310-MeV pions generated by MILC Collaboration~\cite{MILC:2010pul,Bazavov:2012xda}.
The calculation uses two valence masses for the nucleon: light ($M_\pi\approx310$~MeV) and strange ($M_\pi\approx 690$~MeV). The two-point correlators include 344,064 (57,344) measurements in total and are extrapolated to physical pion mass.
The matrix elements are renormalized using the nonperturbative renormalization (NPR) in RI/MOM scheme, the same strategy as in previous works~\cite{Stewart:2017tvs,Chen:2017mzz}.
The left-hand side of Fig.~\ref{fig:LatStrange} shows the lattice real matrix elements at $M_\pi=135$~MeV (extrapolated linearly in $M_\pi^2$) with $P_z \in [1.3,2.2]$  compared with the CT18 NNLO (red band with dot-dashed line) and NNPDF3.1 NNLO (orange band with dotted line) gluon PDFs.
The real matrix elements are proportional to the integral of the difference between strange and antistrange PDFs ($\int dx \left(s(x)-\bar{s}(x)\right) \cos(xzP_z)$).
The lattice results of the real quasi-PDF matrix elements, as shown in Fig.~\ref{fig:LatStrange}, are consistent with zero at 95\% confidence level for most $zP_z$ points, indicating that the strange quark-antiquark asymmetry is likely very small.

To take advantage of existing lattice data to reach a wider region of $x$, we choose to focus on the result of $P_z \approx 1.7$~GeV.
We Fourier transform the renormalized matrix elements into quasi-PDFs by using the extrapolation formulation suggested in Ref.~\cite{Ji:2020brr} to fit the large-$|z|$ data to the formula $c_1(-izP_z)^{-d_1}+c_2 e^{izP_z}(izP_z)^{-d_2}$, inspired by the Regge behavior. Extrapolating the matrix elements into the region beyond the lattice calculation then suppresses Fourier-transformation artifacts.
The quasi-PDF can be related to the $P_z$-independent lightcone PDF
at scale $\mu$ in $\overline{\text{MS}}$ scheme through a factorization theorem~\cite{Ji:2014gla}
\begin{align}
\label{eq:matching}
	\tilde{q}_\psi(x,&P_z,\mu^{\overline{\text{MS}}},\mu^\text{RI},p^\text{RI}_z)=\int_0^1 \frac{dy}{|y|} \times \nonumber\\ &C\left(\frac{x}{y},\left(\frac{\mu^\text{RI}}{p_z^\text{RI}}\right)^2,\frac{yP_z}{\mu^{\overline{\text{MS}}}},\frac{yP_z}{p_z^\text{RI}}\right)
	q_\psi(y,\mu^{\overline{\text{MS}}}) +...
\end{align}
where $p_z^\text{RI}$ and $\mu^\text{RI}$ are the momentum of the off-shell strange quark, and the renormalization scale in the RI/MOM-scheme nonperturbative renormalization,
$C$ is a perturbative matching kernel used in our previous works~\cite{Chen:2018xof,Lin:2018qky,Chen:2018fwa,Chen:2019lcm}.
The quasi- and matched strangeness asymmetry distributions as functions of $x$ can be found on the right-hand side of Fig.~\ref{fig:LatStrange}; both are consistent with zero.
Note that the matching from quasi-PDF to PDF has residual systematics at
$O\left(\frac{\Lambda_\text{QCD}^2}{(xP_z)^2}\right)$ and $O\left(\frac{\Lambda_\text{QCD}^2}{(1-x)^2P_z^2}\right)$ at very small $x$ and $x$ near 1, respectively.
From the isovector nucleon PDF study, at this $P_z$ boost momentum, we can reasonably rely on lattice inputs for $x\in[0.3,0.8]$.
Beyond this region, the lattice errors could increase significantly due to the systematics at finite momentum.

\subsection{Strangeness asymmetry $s_{-}(x)$ in CTEQ-TEA PDF analysis}
\label{fitting:svl-CT18}

In the nominal CTEQ-TEA PDF fit~\cite{Hou:2019efy, Dulat:2015mca, Lai:2010vv, Nadolsky:2008zw, Lai:2007dq}, the active parton flavors to be parametrized at $Q_0 = 1.3$~GeV are $u$, $\bar{u}$, $d$, $\bar{d}$, $s$, $\bar{s}$, and $g$. In the parametrization of sea-quark distributions, $s(x, Q_0) = \bar{s}(x, Q_0)$ is imposed in the nominal CT PDFs.
In contrast, the CTEQ6.5S0 PDF~\cite{Lai:2007dq} and its earlier version~\cite{Olness:2003wz}, done at the NLO, focus on the strangeness sector, where the strangeness asymmetry $s_{-}(x, Q_0)$ is explicitly parametrized at $Q_0$.

In this work, we follow the strategy presented in Ref.~\cite{Lai:2007dq} to perform a global PDF analysis with nonzero $s_{-}(x, Q_0)$, but with updated experimental data and nonperturbative parametrization forms of active partons at the $Q_0$ scale, together with NNLO theory predictions. The resulting PDF set is denoted ``CT18As''.  More specifically, in the CT18As analysis,
we start from the alternative PDF set, CT18A NNLO~\cite{Hou:2019efy}, rather than the nominal CT18 NNLO fit. This is because the ATLAS $\sqrt{s} = 7$~TeV $W$, $Z$ combined cross-section measurement~\cite{ATLAS:2016nqi} (ID=248) data set is included in the CT18A fit, while it is absent in the nominal CT18 fit.
In CT18 analysis, this data is found to prefer larger total strangeness, $s_{+} \equiv s(x) + {\bar s} (x)$ in the small-$x$ region, and to have tensions with other di-muon data~\cite{Mason:2006qa, Mason:2006qa}, which is sensitive to the strangeness distribution.

The CT18As fit adopts the same nonperturbative PDF forms as the CT18A fit at the $Q_0$ scale, except for the strange quark and antiquark PDFs, which are determined by $s_{+}(x)$ and $s_{-}(x)$.
The parametrization of the strangeness asymmetry distribution $s_{-}(x)$ should respect the number sum rule for strangeness,
\begin{equation}
 \int_0^1 \ dx\,s_{-}(x) =  \int_0^1 \ dx\,s(x) - {\bar s}(x) = 0.
 \label{eq:sv_nsr}
\end{equation}
In principle, a parametrization with any number of crossings, with $s_{-}(x) = 0$, is possible, as long as Eq.~\eqref{eq:sv_nsr} is satisfied. Here, we focus on parametrization forms with only one crossing in the range $x \in [10^{-6}, 1]$.

To obtain CT18As\_Lat PDFs, we take
the lattice data for the strangeness asymmetry presented in Sec.~\ref{fitting:svl_lat} as a constraint to the global PDF fit. We use the Lagrange-multiplier method, since we regard the lattice $s_{-}(x)$ results as additional data on top of the CT18A data set. Hence, CT18As\_Lat is an update to CT18As with the inclusion of the lattice $s_{-}(x)$ data evaluated at the $Q_0$ scale.
\comment{
	The uncertainty of lattice calculation is treated as the uncorrelated error during the fitting. Therefore, the quality-of-fit receives the contribution of lattice calculation,
\begin{eqnarray}
 \chi^2 &=& \chi^2_{\text{Exp.}} + \chi^2_{\text{Lat.}} \nonumber \\
 &=& \chi^2_{\text{Exp.}} + \sum_i \Big{(} \frac{s_v^{\text{para.}}(x_i) - s_v^{\text{Lat.}}(x_i)}{ \Delta s_v^{\text{Lat.}}(x_i) } \Big{)}^2,
 \label{eq:chi2_LM}
\end{eqnarray}
where $\chi^2_{\text{Exp.}}$ is the total $\chi^2$ for fitting experimental data.
}

\section{Updated Strangeness Asymmetry Results}
\label{results}

\begin{table}[htbp]
\begin{center}
\begin{tabular}{lccc}
PDF         & $s_{-}(x,Q_0)${  }  & Lat. data  & $\chi^2_\text{tot}$ \\
\hline
CT18A        & 0  & No & 4376 \\
CT18As       & $\neq 0$ & No  & 4344 \\
CT18As\_Lat  & $\neq 0$  &Yes  & 4361 \\
\hline
\end{tabular}
\end{center}
\caption{
\label{tab:quality}
The total goodness-of-fit  $\chi^2_\text{tot}$ of the CT18A, CT18As, and CT18As\_Lat fits, respectively, at $Q_0=1.3$~GeV. The total number of data points (without including the lattice data) of each fit is 3674.
}
\end{table}

\begin{figure}[htbp]
\centering
\includegraphics[width=\textwidth]{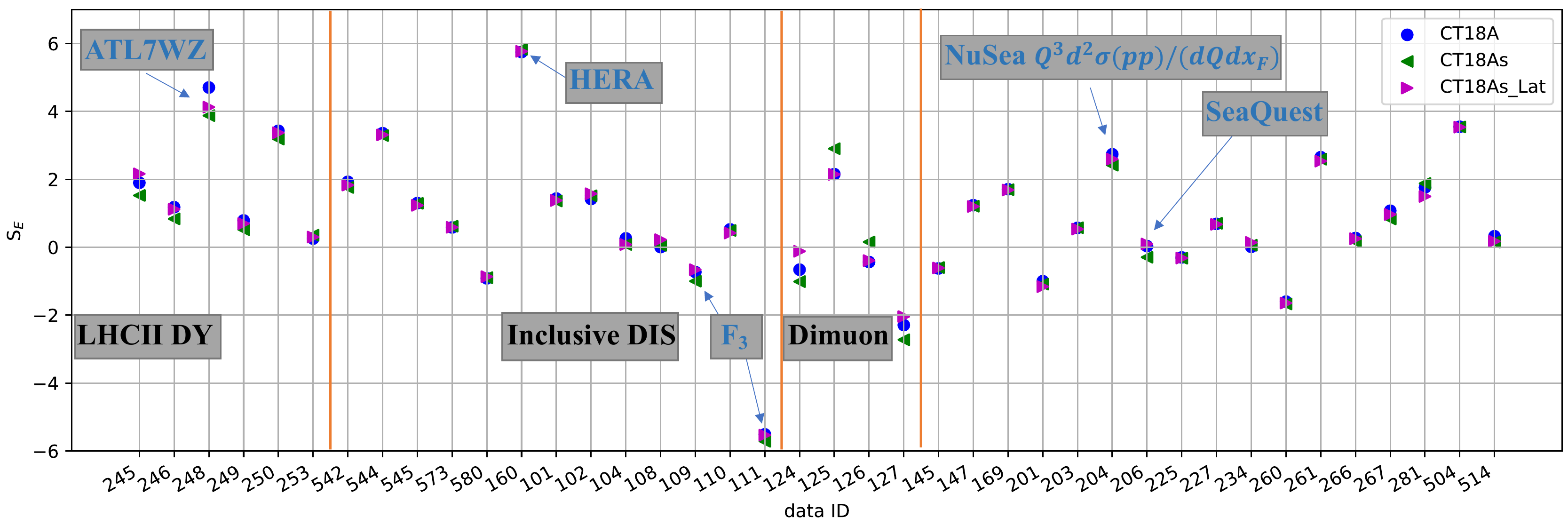}
\caption{
The effective Gaussian variables $S_E$ for individual data sets included in these fits. The data indices ``ID'' are as defined in Tables~I and II from the prior CT18 study~\cite{Hou:2019efy}. Note that $S_E$ for the E906 SeaQuest data~\cite{SeaQuest:2021zxb} (ID=206) is predicted in the above plot but not included in the PDF fits for this study, nor for the nominal CT18 PDF fit.
}
\label{fig:spartyness}
\end{figure}

\begin{table}[htbp]
\begin{center}
\begin{tabular}{ll|c|ccc}
ID & Experimental data set & $N_{\text{pt},E}$ & CT18A & CT18As & CT18As\_Lat \\
\hline
 245 & LHCb 7~TeV $W$/$Z$ rap.~\cite{LHCb:2015okr} & 33 & 1.52 & 1.40 & 1.60 \\
 246 & LHCb 8~TeV $Z\to ee$ rap.~\cite{LHCb:2015kwa} & 17 & 1.41 & 1.27 & 1.39 \\
 248 & ATLAS 7~TeV 4.6~fb$^{-1}$ $W$/$Z$ combined cross sec.~\cite{ATLAS:2016nqi} & 34 & 2.58 & 2.22 & 2.33 \\
 249 & CMS 8~TeV $W A_\text{ch}$~\cite{CMS:2016qqr} & 11 & 1.30 & 1.17 & 1.26 \\
 250 & LHCb 8~TeV $W$/$Z$ rap.~\cite{LHCb:2015mad} & 34 & 2.05 & 1.95 & 2.02 \\
 109 & CDHSW $F_3^p$~\cite{Berge:1989hr} & 96 & 0.89 & 0.86 & 0.90 \\
 111 & CCFR $xF_3^p$~\cite{Seligman:1997mc} & 86 & 0.37 & 0.36 & 0.37 \\
 124 & NuTeV $\nu\mu\mu$ SIDIS~\cite{Mason:2006qa} & 38 & 0.84 & 0.77 & 0.96 \\
 125 & NuTeV $\bar{\nu}\mu\mu$ SIDIS~\cite{Mason:2006qa} & 33 & 1.60 & 1.86 & 1.60 \\
 126 & CCFR $\nu\mu\mu$ SIDIS~\cite{NuTeV:2001dfo} & 40 & 0.89 & 1.02 & 0.90 \\
 127 & CCFR $\bar{\nu}\mu\mu$ SIDIS~\cite{NuTeV:2001dfo} & 38 & 0.55 & 0.49 & 0.59 \\
 204 & E866 Drell-Yan process $Q^3d^2\sigma_{pp}/(dQ\,dx_F)$~\cite{NuSea:2003qoe} & 184 & 1.31 & 1.27 & 1.29 \\
 206 & E906 Drell-Yan process $\sigma_{pd}/(2\sigma_{pp})$~\cite{SeaQuest:2021zxb} & 6 & 0.91 & 0.74 & 0.95 \\
 234 & D0 Run-2 muon $A_\text{ch}$, $p_{Tl}>20$~GeV~\cite{D0:2007pcy} & 9 & 0.93 & 0.95 & 0.99 \\
 266 & CMS 7~TeV 4.7~fb$^{-1}$, muon $A_\text{ch}$ $p_{Tl}>35$~GeV~\cite{CMS:2013pzl} & 11 & 1.06 & 1.01 & 1.05 \\
 267 & CMS 7~TeV 840~pb$^{-1}$, electron $A_\text{ch}$ $p_{Tl}>35$~GeV~\cite{CMS:2012ivw} & 11 & 1.45 & 1.33 & 1.40 \\
 281 & D0 Run-2 9.7~fb$^{-1}$ electron $A_\text{ch}$, $p_{Tl}>25$~GeV~\cite{D0:2014kma} & 13 & 1.79 & 1.86 & 1.64 \\
\hline
\end{tabular}
\end{center}
\caption{
\label{tab:chi2_npt}
The $\chi^2/N_\text{pt}$ of selected data sets included in CT18A, CT18As, and CT18As\_Lat fits with non-negligible changes in $\chi^2/N_\text{pt}$. The $N_{\text{pt},E}$ is the number of data points for the individual data set $E$. Note that the E906 SeaQuest data~\cite{SeaQuest:2021zxb} (ID=206) is not included in PDF fits for this study, as well as the nominal CT18 PDF fit; its $\chi^2/N_\text{pt}$ values are just predicted by the corresponding PDFs.
}
\end{table}

In this section, we discuss the quality of various fits and compare the resulting PDFs.
The qualities of the CT18A, CT18As, and CT18As\_Lat fits are compared in Table~\ref{tab:quality}, which shows that they all have the similar $\chi^2$, meaning that these three PDFs are comparable in describing the experimental data.
The difference in their $\chi^2_\text{tot}$ is much smaller than the tolerance (with a difference of 100 units) used in the CT18 analysis to define the PDF uncertainty at the 90\% confidence level (CL).

We present the qualities of fit of each individual data set $E$ by comparing the effective Gaussian variable $S_E = \sqrt{2\chi^2_E} - \sqrt{2N_{\text{pt},E}-1}$ in Fig.~\ref{fig:spartyness}; see Ref.~\cite{Hou:2019efy} and references therein.
Alternative to the usual $\chi^2$, the effective Gaussian variable provides an estimation of quality of fit. $S_E > 1$ means that data set $E$ is not fitted well, while $S_E < 1$ represents overfitting. Moreover, if all deviations of theory from data are purely of random fluctuation, the distribution of $S_E$ is expected to recover the standard normal distribution $\mathcal{N}(0,1)$.
In Fig.~\ref{fig:spartyness}, variation of the effective Gaussian variable $S_E$ for individual data sets $E$ suggests the potential sensitivity of the strangeness asymmetry to the data set $E$. There are three groups of data that show variations in $S_E$:
1) NuTeV~\cite{Mason:2006qa} and CCFR~\cite{NuTeV:2001dfo} SIDIS di-muon production measurements (ID=124--127), which directly probe (anti)strange PDFs; 2) CDHSW~\cite{Berge:1989hr} (ID=109) and CCFR~\cite{Seligman:1997mc} (ID=111) measurements of the $F_3$ structure function, which are directly related to the valence-sector PDFs, so to the strangeness asymmetry $s_-(x)$; 3) data sets, which are sensitive to sea-quark PDFs, such as LHC DY data sets, E866 NuSea $Q^3 d^2\sigma_{pp}/(dQ\,dx_F)$ data~\cite{NuSea:2003qoe} (ID=204) and E906 SeaQuest data~\cite{SeaQuest:2021zxb} (ID=206).
A detailed comparison of theory prediction and data for the above-mentioned groups of data will be deferred to Sec.~\ref{pheno}.
For the last group, we pick the ATLAS 7 TeV $W$ and $Z$ differential cross-section measurement~\cite{ATLAS:2016nqi} (ID=248) and E906 SeaQuest data~\cite{SeaQuest:2021zxb} (ID=206) as the representatives.
In addition to these groups, the HERA I+II reduced cross-section data~\cite{H1:2015ubc} (ID=160), as marked in Fig.~\ref{fig:spartyness}, receives negligible impacts from varying prescriptions of the strangeness asymmetry.
Similar to Fig.~\ref{fig:spartyness}, the quality of fit, $\chi^2/N_\text{pt}$, for the selected data sets in CT18A, CT18As, and CT18As\_Lat fits with non-negligible variations in $\chi^2/N_\text{pt}$ is presented in Table~\ref{tab:chi2_npt}.

Before proceeding, we note that, in the CT18As and CT18As\_Lat PDF fits, the charged-current NNLO QCD correction for SIDIS processes~\cite{Berger:2016inr} is taken into account, in contrast to the original CT18 and CT18A fits~\cite{Hou:2019efy}.
As noted in the section V.4 of the CT18 paper~\cite{Hou:2019efy}, the NNLO prediction provides a
marginally better agreement with the data.

\begin{figure}[htbp]
	\includegraphics[width=0.49\textwidth]{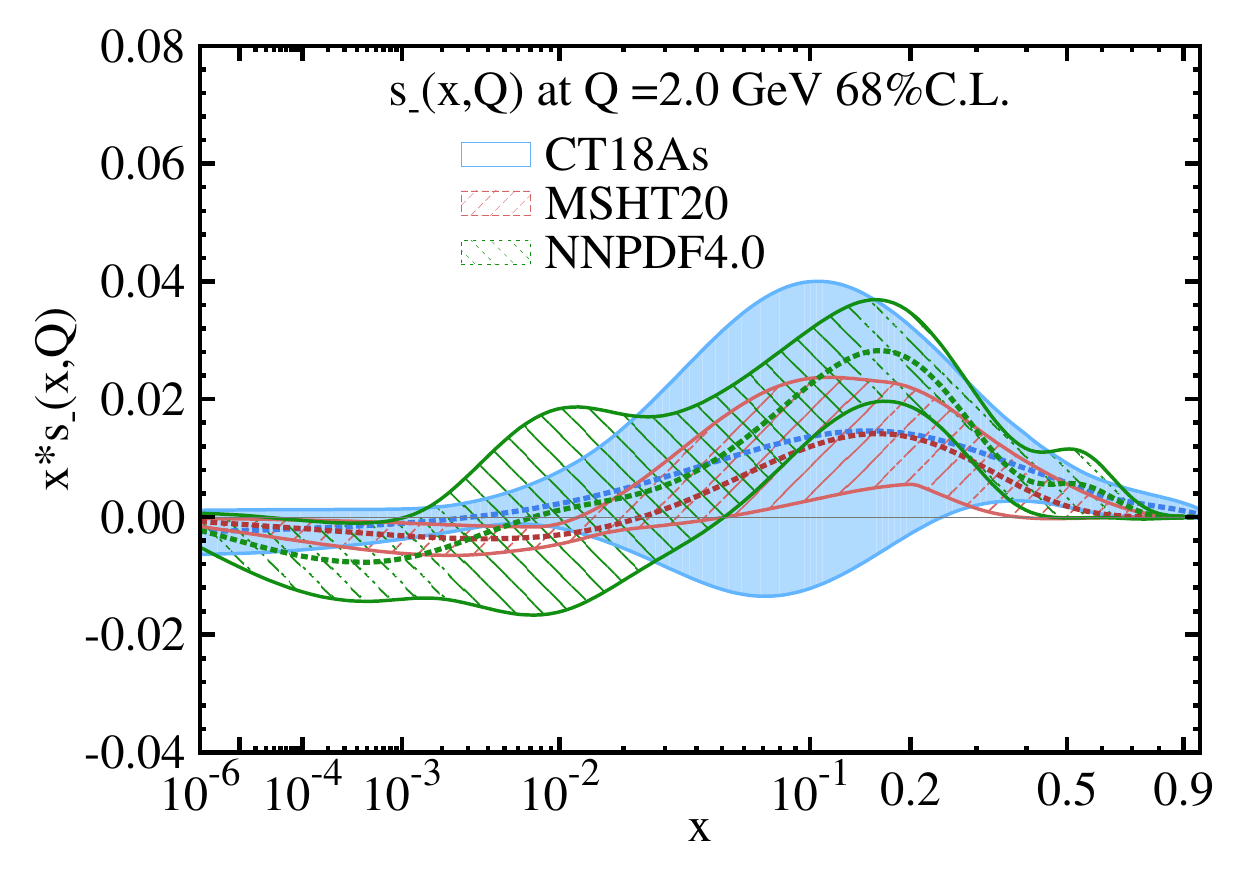}
	\includegraphics[width=0.49\textwidth]{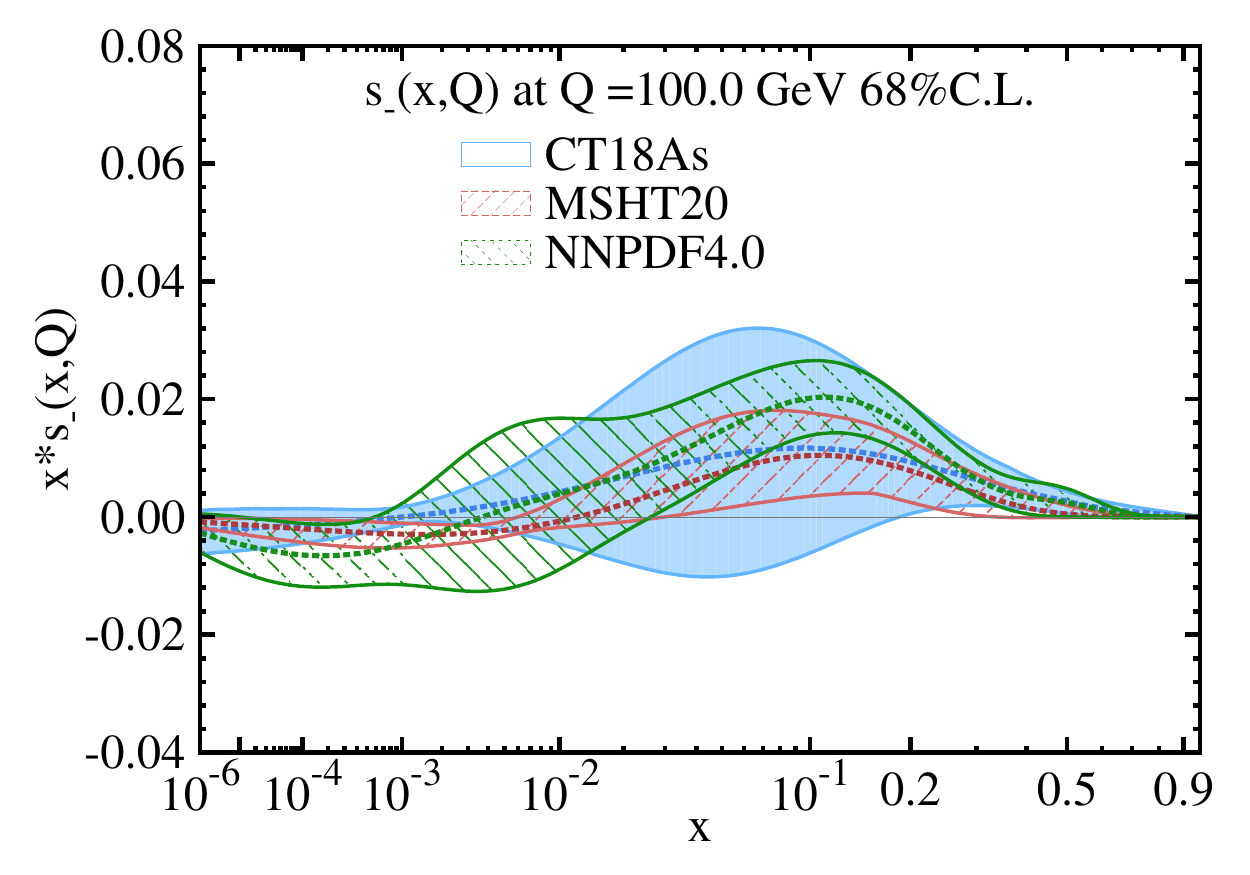}
	\includegraphics[width=0.49\textwidth]{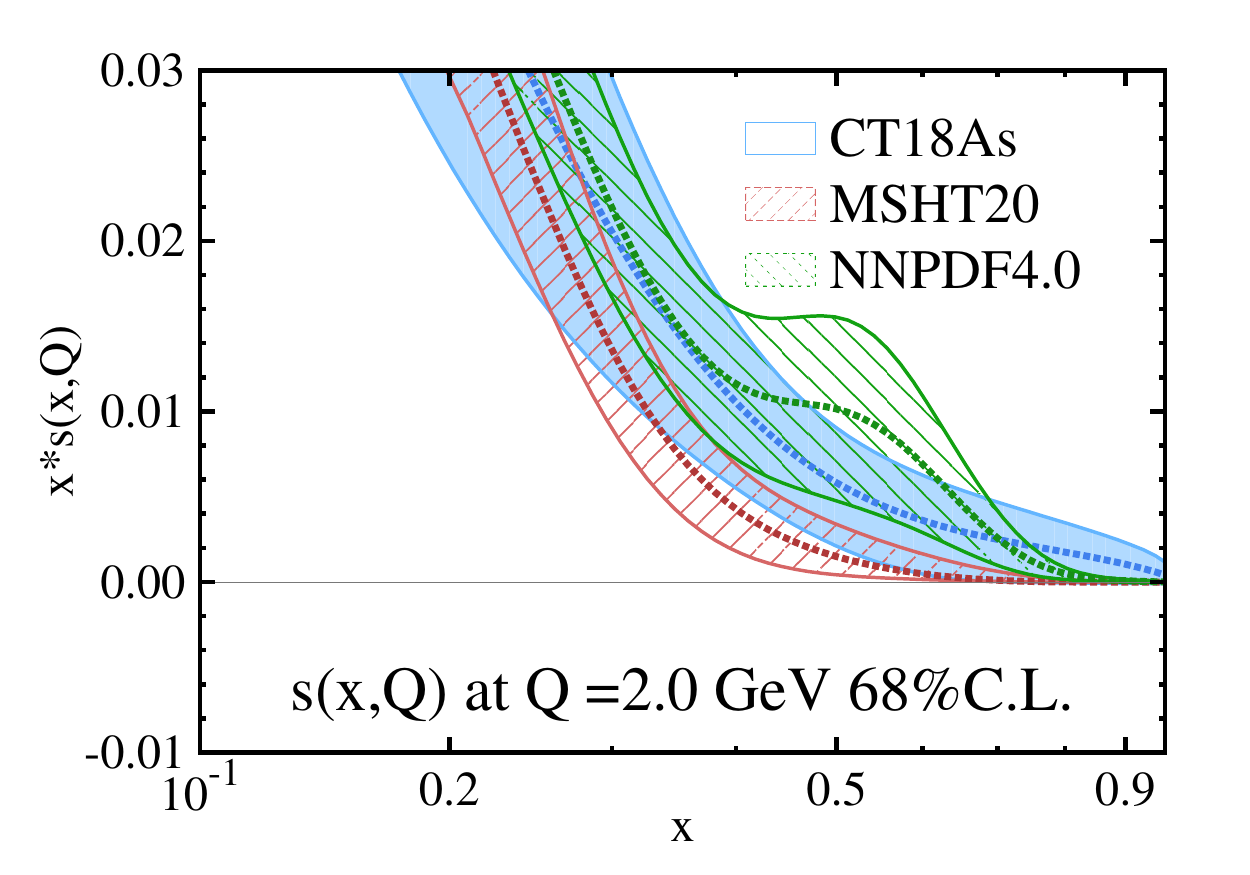}
    \includegraphics[width=0.49\textwidth]{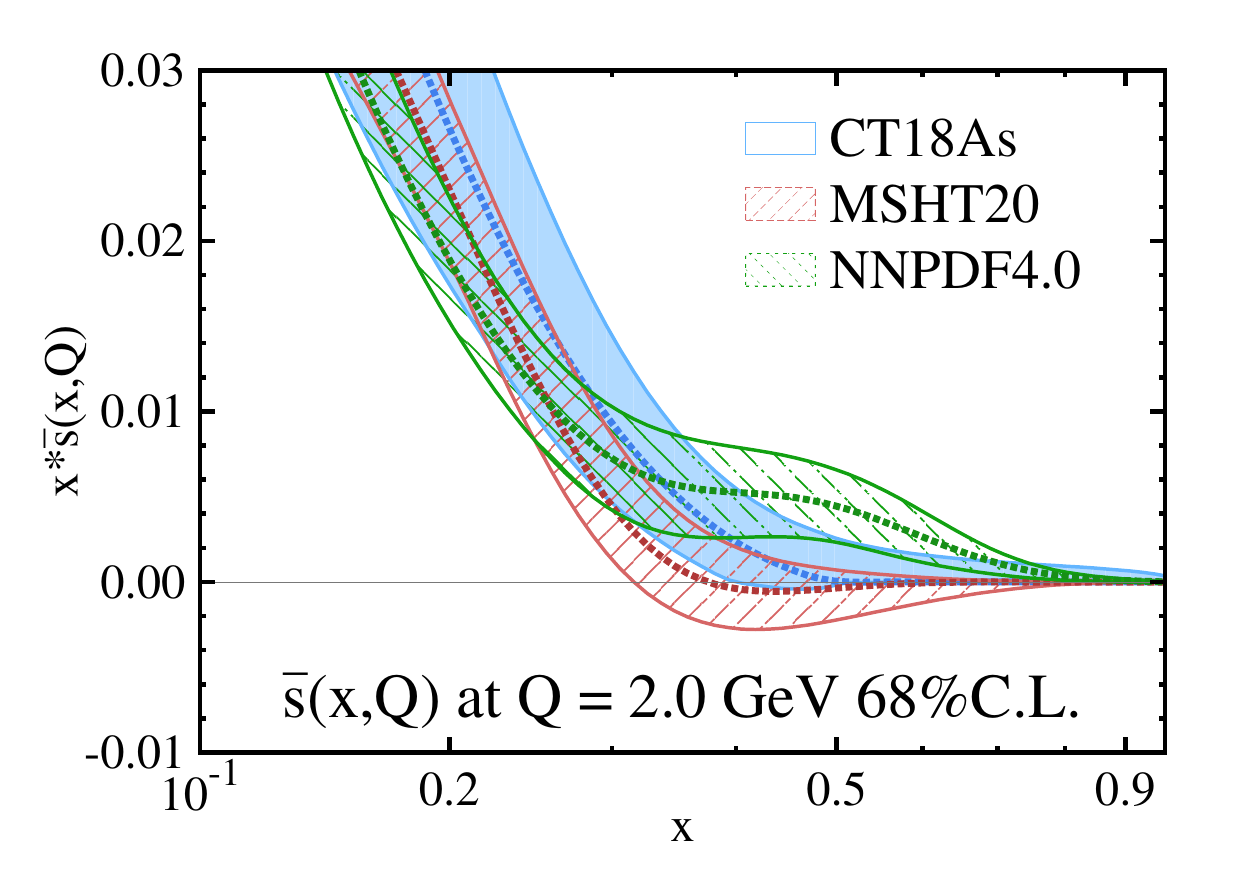}
\caption{
\label{fig:CT18Aas_Lat_other_PDF}
The CT18As $s_-(x)$ distributions at 2~GeV (top-left), 100~GeV (top-right), $s(x)$ at 2~GeV (bottom-left) and $\bar{s}(x)$ at 2~GeV (bottom-right) are compared to those of MSHT20~\cite{Bailey:2020ooq} and NNPDF4.0~\cite{Ball:2021leu}.
}
\end{figure}

The $s_-(x)$ distributions at 2.0~GeV and 100~GeV, and $s(x)$ and $\bar{s}(x)$ at 2.0~GeV of CT18As are compared to PDF fitting results by other groups, as shown in Fig.~\ref{fig:CT18Aas_Lat_other_PDF}.
As shown in the top panels, CT18As agrees with MSHT20~\cite{Bailey:2020ooq} in terms of the $s_-$ central values. For $x \sim 0.1$, NNPDF4.0~\cite{Ball:2021leu} presents the largest $s_-(x)$ central value. In the range of $0.05 < x < 0.4$, CT18As shows a wide error band, so that CT18As is consistent with $s_-(x)$ PDF obtained by other groups.
For $s(x)$ and $\bar{s}(x)$ PDFs at 2.0~GeV on the bottom panels, three PDFs are in agreement for $x<0.3$. For $x>0.3$, three PDFs present different shapes and the CT18As lies in the middle of the MSHT20~\cite{Bailey:2020ooq} and NNPDF4.0~\cite{Ball:2021leu} NNLO PDFs.

\begin{figure}[htbp]
\includegraphics[width=0.49\textwidth]{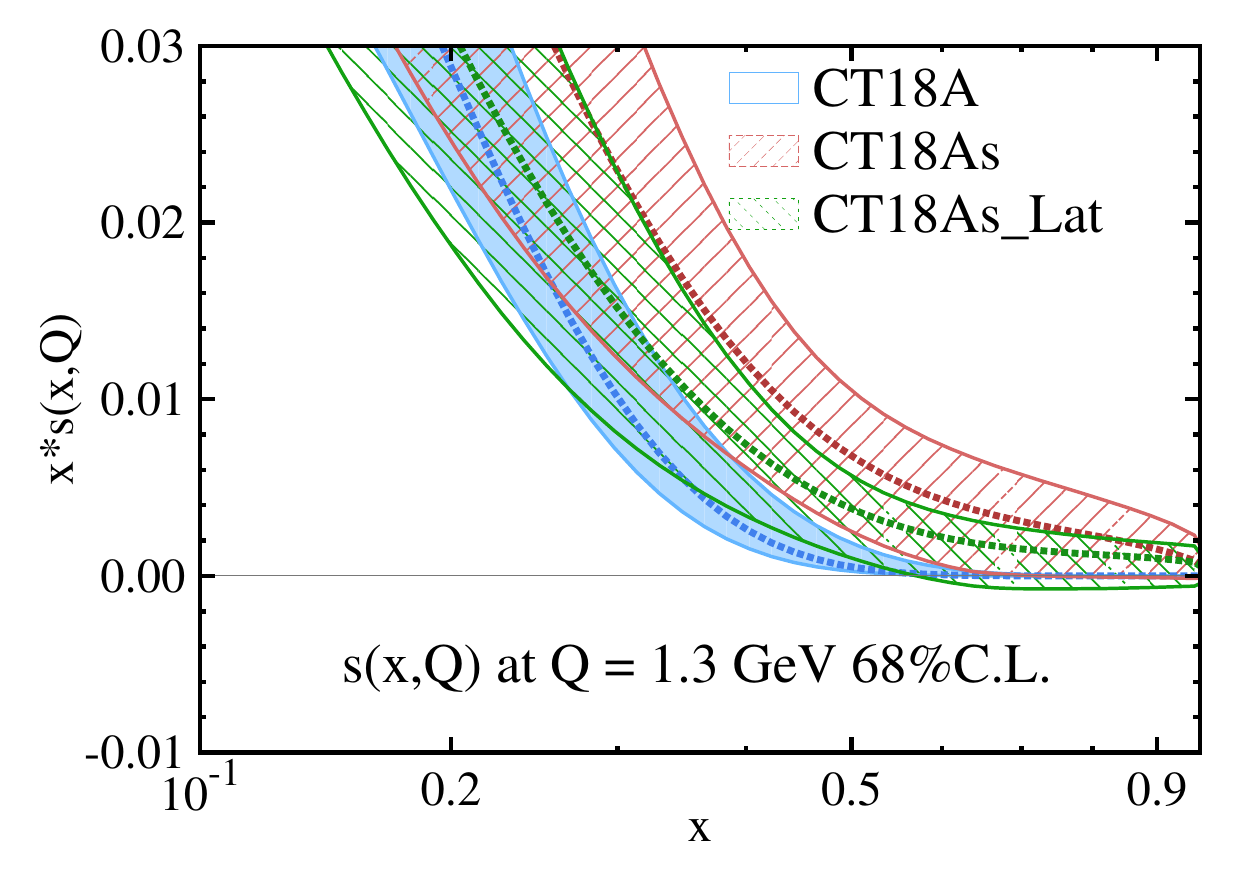}
\includegraphics[width=0.49\textwidth]{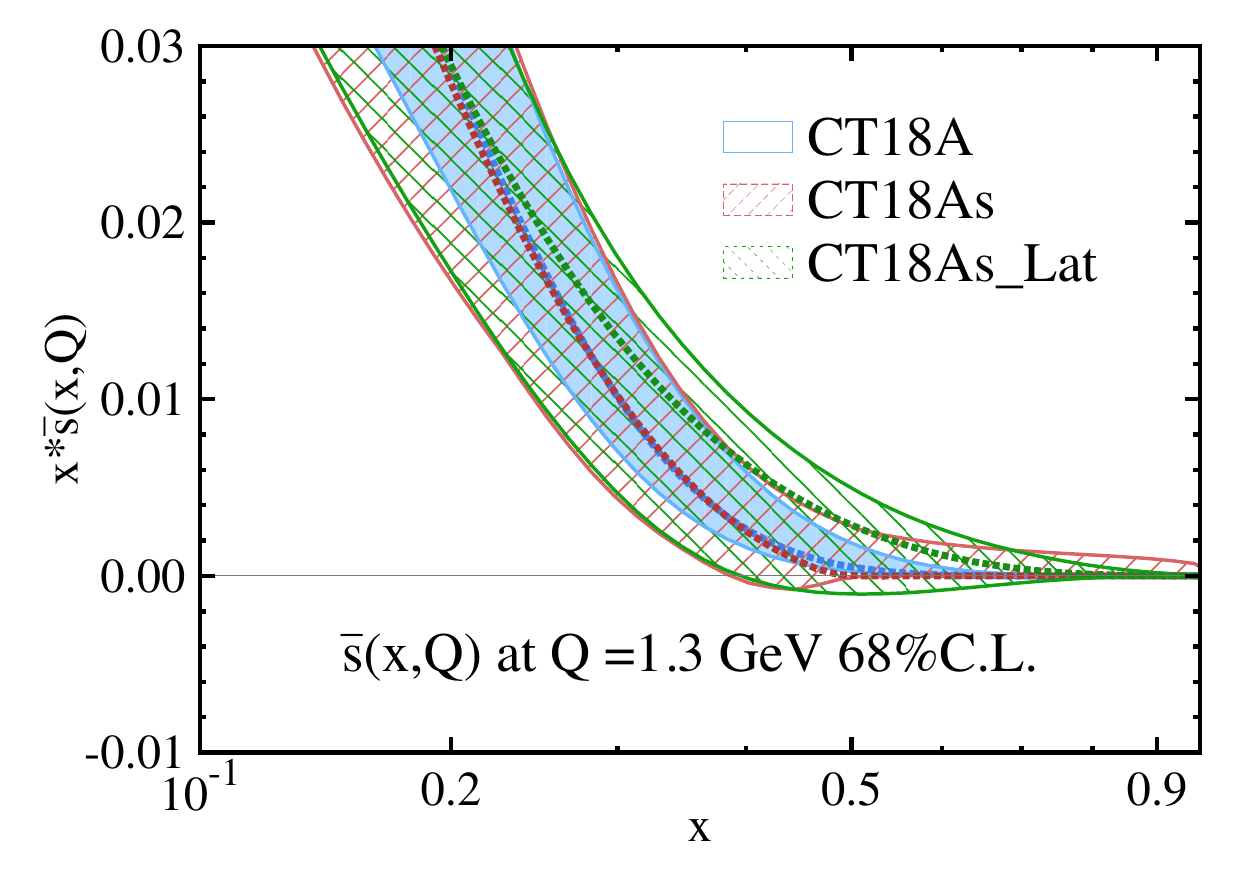}
\includegraphics[width=0.49\textwidth]{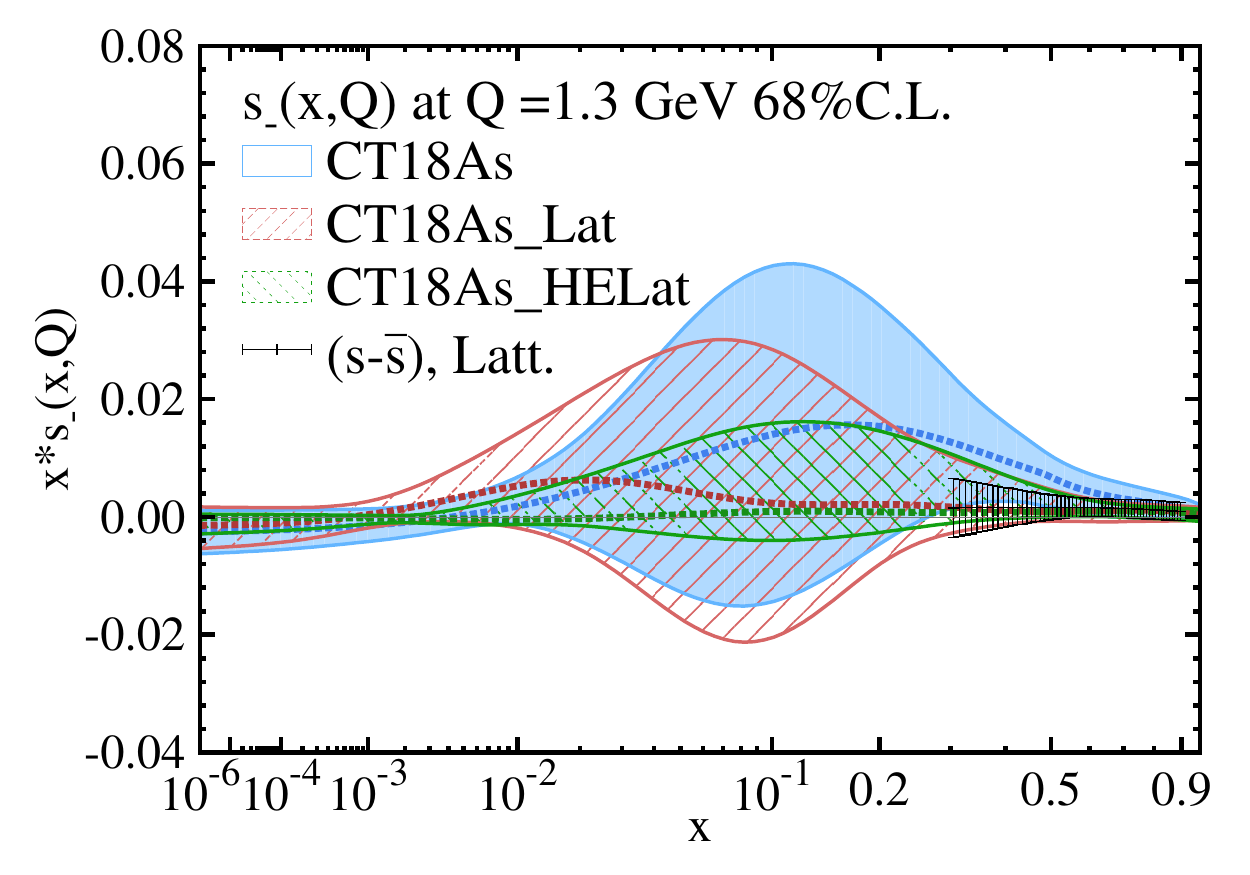}
\includegraphics[width=0.49\textwidth]{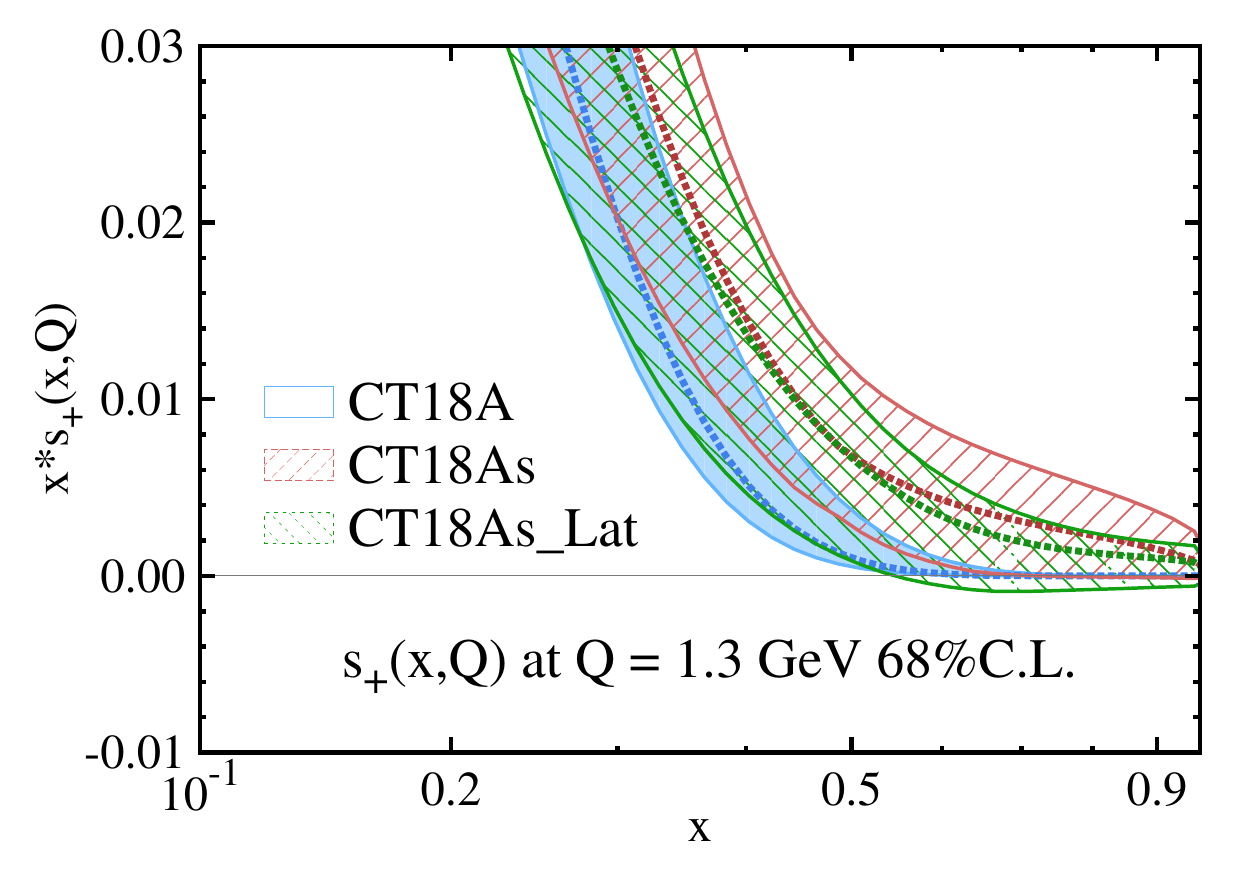}
\includegraphics[width=0.49\textwidth]{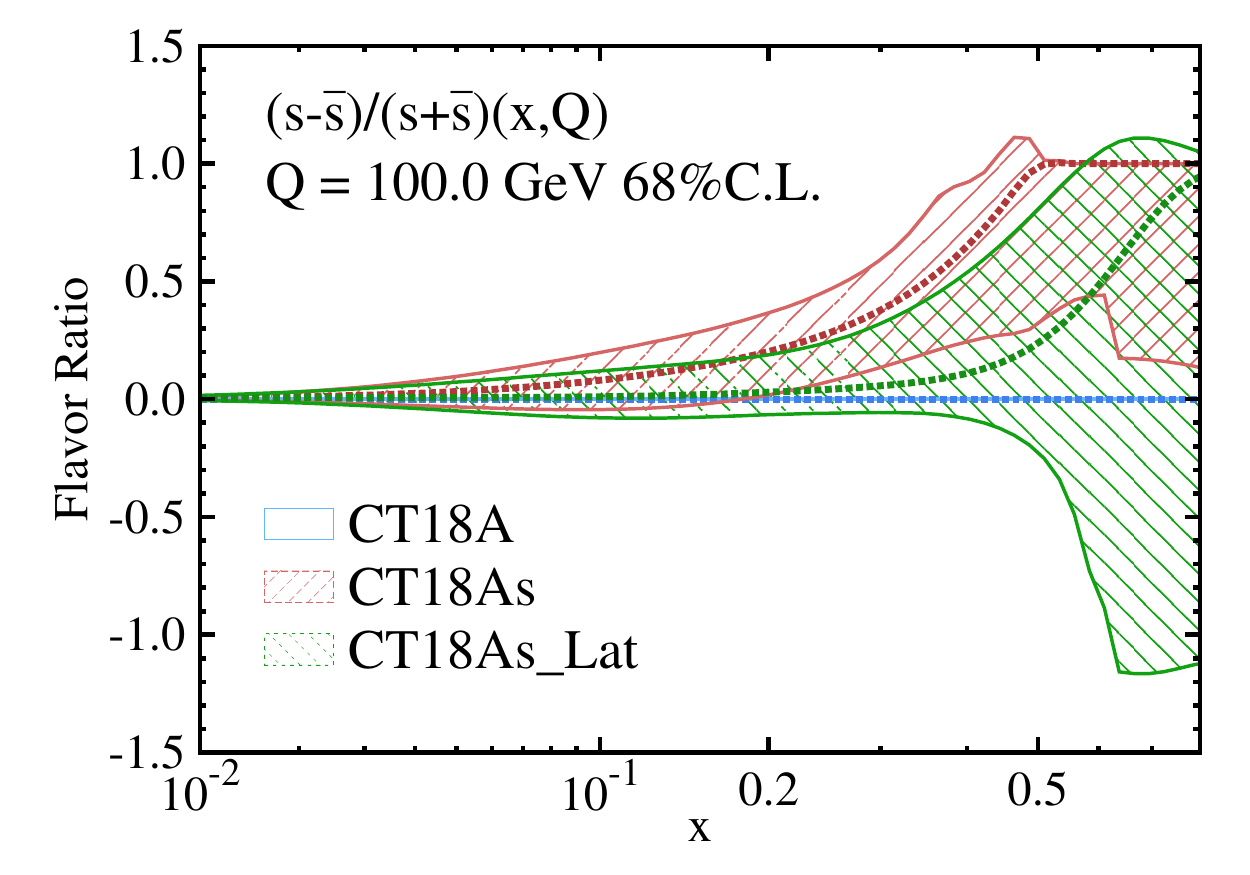}
\includegraphics[width=0.49\textwidth]{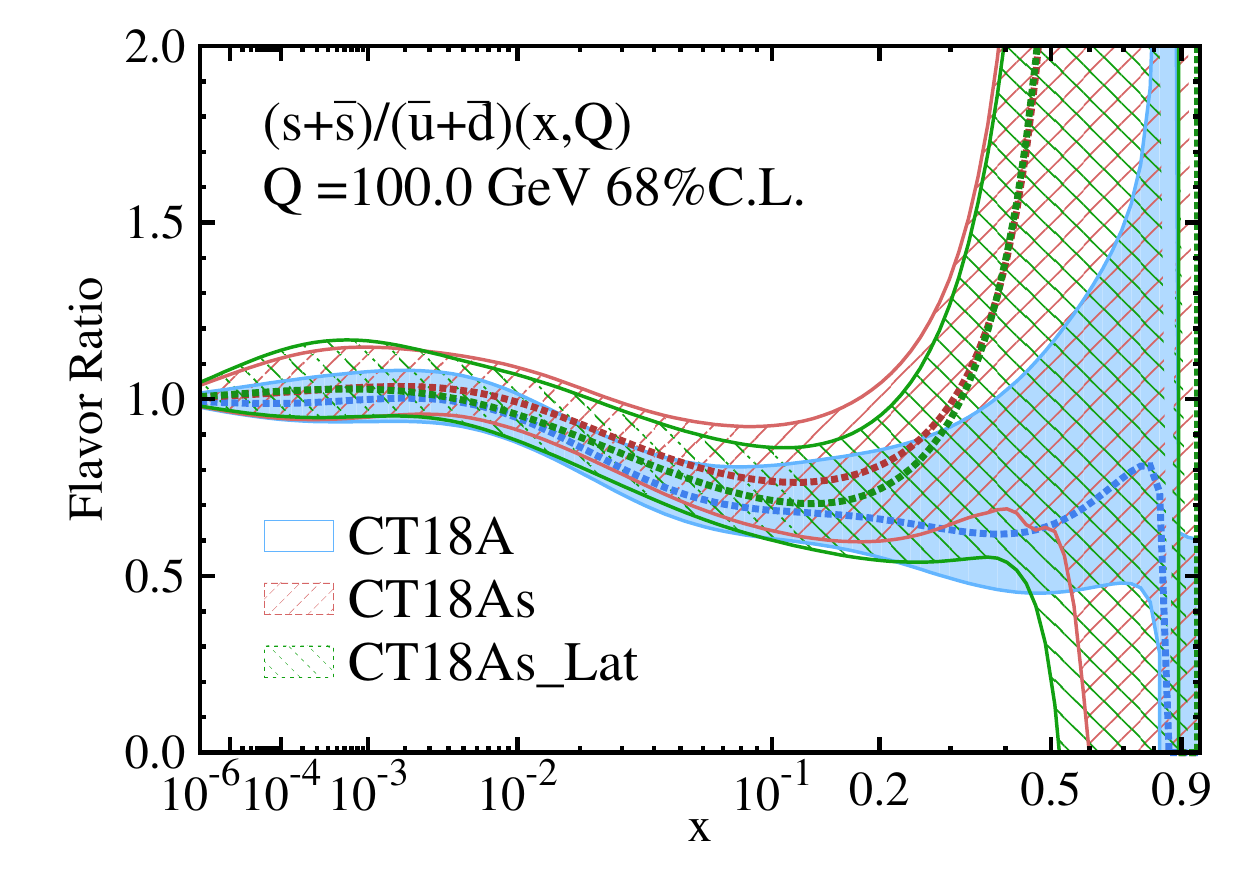}
\caption{
\label{fig:strange_PDFs}
The comparison of $s(x)$ (top-left), $\bar{s}(x)$ (top-right), and $s_-(x)$ (middle-left), $s_+(x)$ (middle-right) PDFs at the initial $Q_0$ scale, as well as PDF ratios $(s+\bar{s})/(\bar{u}+\bar{d})(x)$ (bottom-left) and $(s+\bar{s})/(s-\bar{s})(x)$ (bottom-right) at $Q = 100$~GeV, for CT18A, CT18As, and CT18As\_Lat.
Note that in the middle-left panel, predictions of the strangeness asymmetry of CT18A and CT18As\_Lat are compared to the current lattice data and expected improvement if current lattice data errors are reduced by a half (green backslashed area, i.e., CT18As\_HELat).
For CT18A, no strangeness asymmetry $s_-(x)$ is allowed at the initial $Q_0$ scale in the nonperturbative parametrization, so CT18A is absent in the comparison plot of $s_-(x)$.
}
\end{figure}

We compare CT18A, CT18As, and CT18As\_Lat at the initial $Q_0$ scale for $s(x)$, $\bar{s}(x)$, and $s_-(x)$, as well as  $s_+(x)$ and PDF ratio $(s+\bar{s})/(\bar{u}+\bar{d})(x)$ at $Q = 100$~GeV in Fig.~\ref{fig:strange_PDFs}.
In the top panels of Fig.~\ref{fig:strange_PDFs}, we compare $s(x)$ and $\bar{s}(x)$ at the initial scale, among CT18A, CT18As, and CT18As\_Lat PDFs. Moving from CT18A to CT18As, allowing a nonvanishing strangeness asymmetry $s_-(x)$ at the initial scale enhances the $s(x)$ PDF but affects $\bar{s}(x)$ less significantly. As the constraint of the lattice data tightens, the CT18As\_Lat $s(x)$ PDF becomes closer to that of CT18A.
It is appears that the error bands of $s(x)$ and $\bar{s}(x)$ PDFs for CT18As and CT18As\_Lat allow a negative value for these PDFs.
However, we note that in the parametrization of $s(x)$ and $\bar{s}(x)$ PDFs, cf. Appendix~\ref{sec_app:para}, we force $s(x)$ and $\bar{s}(x)$ to be non-negative across the whole $x$ range.
We checked that these error bands allowing a negative PDF is due to the numerical construction of the Hessian uncertainty, and that all eigenvector PDFs are non-negative.
In the middle-left panel of Fig.~\ref{fig:strange_PDFs}, the impact of lattice data on the strangeness asymmetry $s_-(x)$ is exhibited. The lattice data points are distributed in the region of $0.8>x>0.3$, and they are consistent with a very small strangeness asymmetry with high precision.
Compared to the error band of CT18As, the uncertainty in lattice data points is quite small, so that including the lattice data in the CT18As\_Lat fit greatly reduces the size of the $s_-$-PDF error band in the large-$x$ region. The amount of reduction of the CT18As\_Lat error band in the much smaller $x$ region is likely to depend on the chosen nonperturbative parametrization form of $s_-(x)$ at $Q_0=1.3$~GeV. Hence, it is important to have more precise lattice data, extended to smaller $x$ values.
Based on the CT18As\_Lat PDF, we further investigate how much lattice data with higher precision would be able to constrain the $s_-$ distribution. We again fit the lattice data, but reduce the uncertainty of lattice data points by half, resulting another PDF labelled ``CT18As\_HELat". The half-error lattice data shows strong power in further constraining $s_-$, reducing the error band of $s_-$ by nearly a factor of two in the large-$x$ region.
In the middle-right panel, the comparison of the total strangeness $s_+(x)$ at $Q_0=1.3$ GeV is shown. In CT18As, the central value of the total strangeness $s_+(x)$ is enhanced across a wide range of $x$ relative to CT18A. The uncertainty of $s_+$ in CT18As is also enlarged.
The similar behaviour can also be observed in the ratios of strange asymmetry to total strangeness $s_-$/ $s_+$  and  total strangeness to light quarks $(s+\bar{s})/(\bar{u}+\bar{d})(x)$ at $Q=100$ GeV, as shown in the bottom panel of  Fig.~\ref{fig:strange_PDFs}.
Despite of the large uncertainty of the PDF ratio  $(s+\bar{s})/(\bar{u}+\bar{d})(x)$
in the large-$x$ region, the enhancement of $(s+\bar{s})/(\bar{u}+\bar{d})(x)$ in CT18As suggests a greater total strangeness than light-quark content.
This feature is caused by the choice of the more flexible non-perturbative parametrization form of the  (anti)strange PDF adopted in the CT18As fit, as compared to that in CT18. In Appendix~\ref{sec_app:CT18As2_Rs}, we present the result of an alternative fit (termed CT18As2 fit) with additional theory prior to constrain the ratio of $(s+\bar{s})/(\bar{u}+\bar{d})(x)$ in the limit of $x$ approaching to 1. After including the lattice data, the resulting CT18As2\_Lat fit leads to similar conclusions as the CT18As\_Lat fit.

\begin{figure}[htbp]
\includegraphics[width=0.49\textwidth]{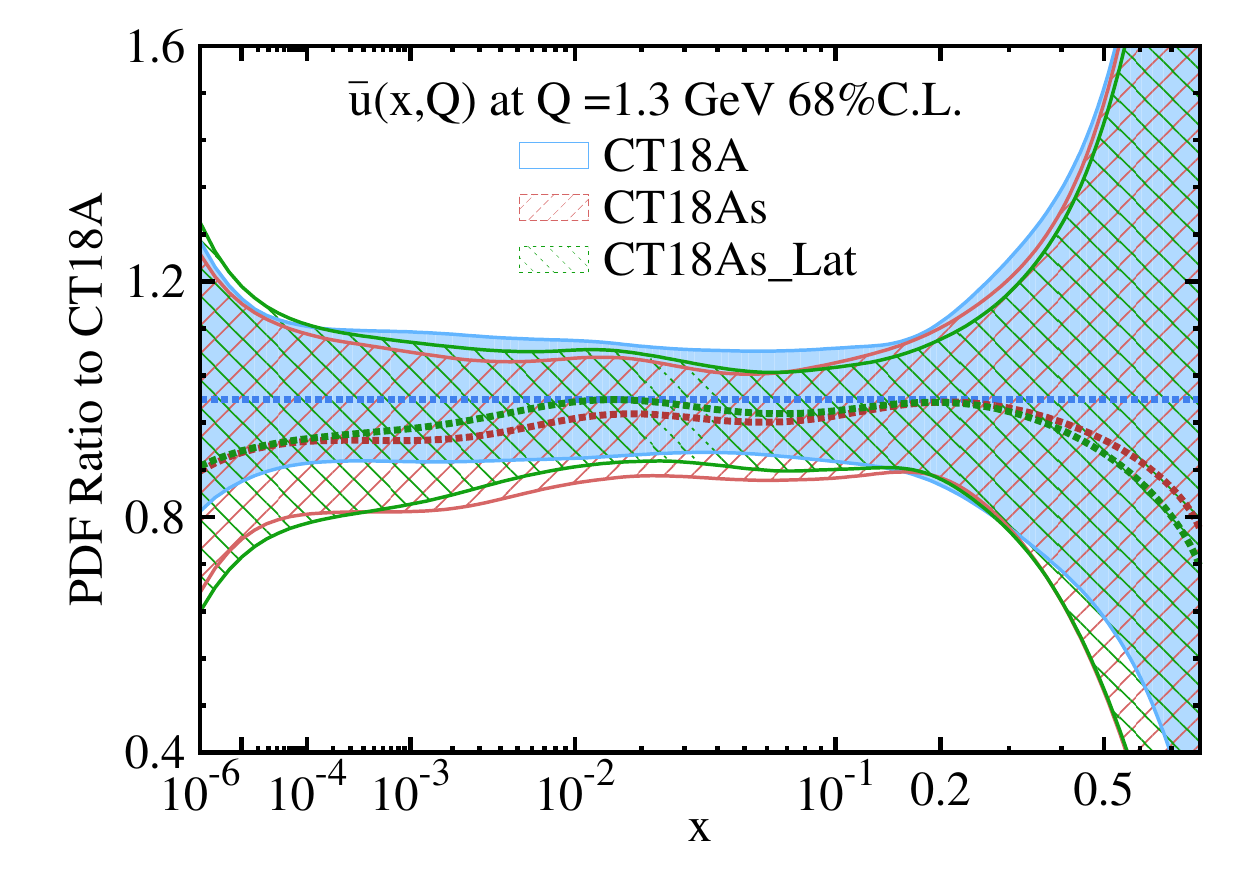}
\includegraphics[width=0.49\textwidth]{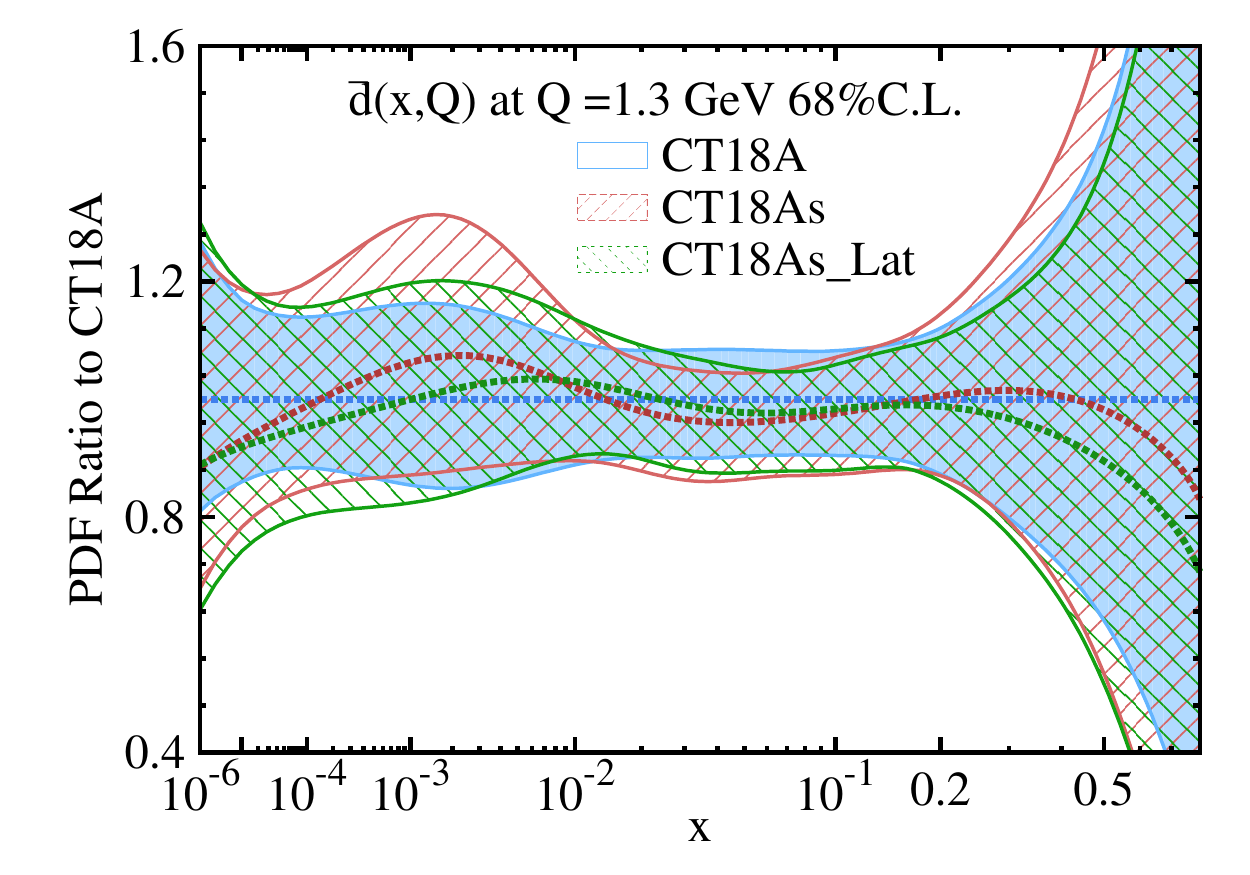}
\includegraphics[width=0.49\textwidth]{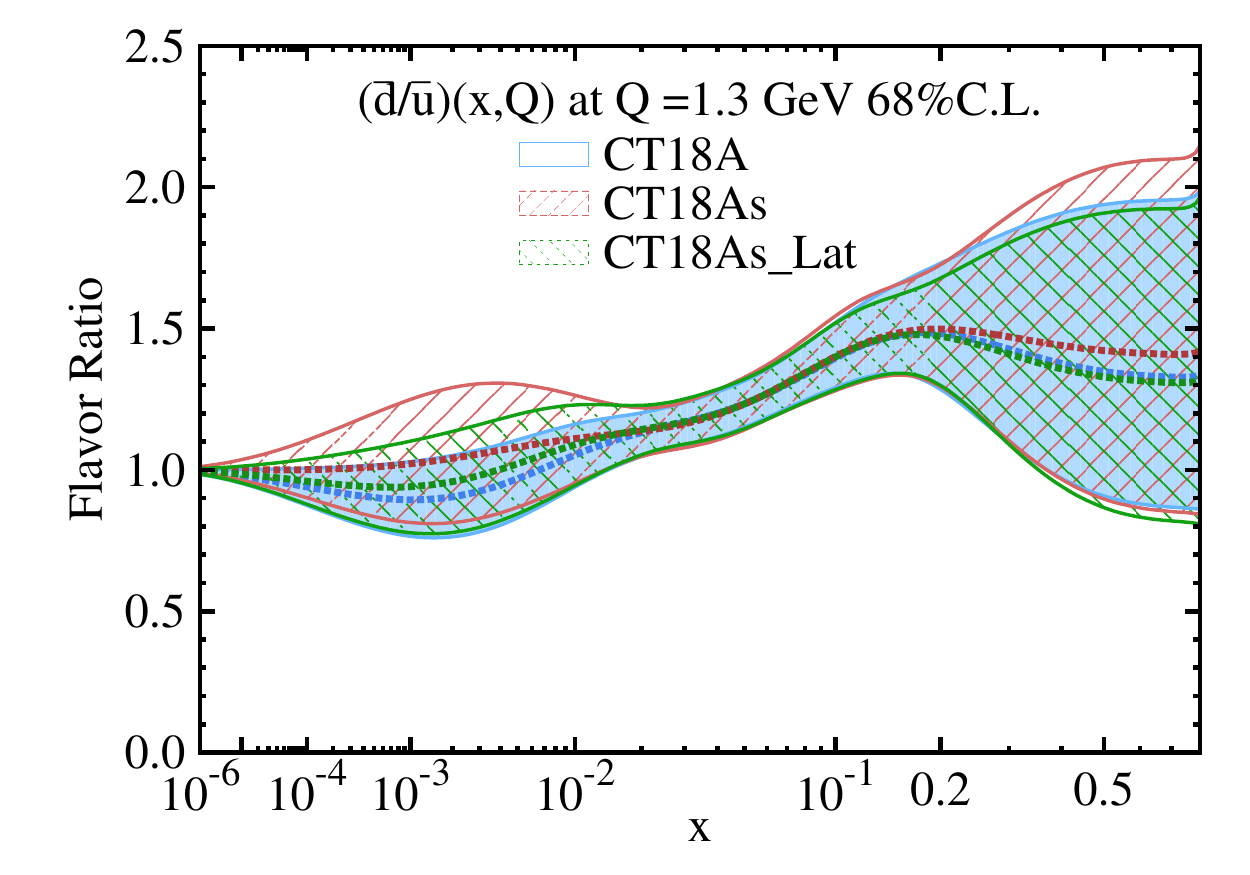}
\caption{
\label{fig:ubr_dbr}
The comparison of $\bar{u}(x)$ (upper-left), $\bar{d}(x)$ (upper-right), and $\bar{d}/\bar{u}(x)$ (bottom) parton distributions from the CT18A, CT18As, and CT18As\_Lat analyses at the initial $Q_0$ ($=1.3$~GeV) scale.
}
\end{figure}

In Fig.~\ref{fig:ubr_dbr}, we compare the $\bar{u}(x)$, $\bar{d}(x)$, and $\bar{d}/\bar{u}(x)$ PDFs at the initial $Q_0=1.3$~GeV.
The $\bar{u}(x)$ of CT18As has been decreased for $x<0.1$ and $x>0.3$ compared to the $\bar{u}(x)$ of CT18A, as shown in the top-left panel of Fig.~\ref{fig:ubr_dbr}. After adding in the lattice data, the $\bar{u}(x)$ of CT18As\_Lat moves closer to that of CT18A for $0.01<x<0.1$, while in the large-$x$ region, CT18As\_Lat still has a smaller magnitude in $\bar{u}(x)$.
We also observer in the top-right panel of Fig.~\ref{fig:ubr_dbr} that  $\bar{d}(x)$ of CT18As for $0.01 < x < 0.1$ and $\bar{d}(x)$ of CT18As\_Lat for $x > 0.2$ are suppressed in comparison to $\bar{d}(x)$ of CT18A, while they are enhanced at $x$ around a few $10^{-3}$.
Combining variations in $\bar{u}(x)$ and $\bar{d}(x)$, the PDF ratio $\bar{d}/\bar{u}(x)$ of CT18As floats up for $x > 0.2$  or $x$ around a few $10^{-3}$, as shown in the bottom panel.

Lastly in Table~\ref{tab:moments}, we summarize the second and third moments of the strangeness asymmetry $s_-$ and the total strangeness $s_+$ obtained in our phenomenological PDF fits and LQCD calculations. For $\langle x \rangle_{s_+}$ at 2.0 GeV, phenomenological calculations from PDF fits are consistent with the recent LQCD calculations by ETMC~\cite{Alexandrou:2020sml} and $\chi$QCD~\cite{Yang:2018nqn} using the traditional moment method.

\begin{table}[htbp]
\begin{center}
\begin{tabular}{c|cccc}
 $Q=1.3$ GeV & CT18A  & CT18As  & CT18As\_Lat & \\ \hline
 $\langle x \rangle_{s_-}$ & 0.0 & 0.0074(112) & 0.0016(70) & \\
 $\langle x^2 \rangle_{s_-}$ & 0.0 & 0.0024(27) &  0.00057(120) &  \\
 $\langle x \rangle_{s_+}$ & 0.038(12) & 0.048(19) & 0.044(18) & \\
 $\langle x^2 \rangle_{s_+}$ & 0.0035(17) & 0.0060(33) & 0.0051(30) & \\ \hline
 $Q=2.0$ GeV & CT18A  & CT18As  & CT18As\_Lat & LQCD \\ \hline
   &  &  & & $0.052(12)$~\cite{Alexandrou:2020sml}  \\
 $\langle x \rangle_{s_+}$ & 0.043(10) & 0.052(17) & 0.048(16) & $0.051(26)(5)$~\cite{Yang:2018nqn}  \\
\end{tabular}
\end{center}
\caption{
\label{tab:moments}
The second and the third moments of the strangeness asymmetry $s_-$ and the total strangeness $s_+$ from phenomenological PDF fits and LQCD calculations (from ETMC~\cite{Alexandrou:2020sml} and $\chi$QCD~\cite{Yang:2018nqn}) at 1.3 (top-panel) and 2.0 (bottom-panel) GeV. The uncertainty corresponds to 68\% confidence level.
Of these moments, $\langle x \rangle_\mathrm{s_+}$ and $\langle x^2 \rangle_\mathrm{s^-}$ are calculable in lattice
QCD.
}
\end{table}

\section{Phenomenology}
\label{pheno}

In this section, we illustrate the impact of allowing nonzero strangeness asymmetry at the initial $Q_0$ scale and of using the lattice data for the strangeness asymmetry by comparing the experimental data and theory predictions for ATLAS 7-TeV $W$ and $Z$ differential cross sections as functions of dilepton pseudo-rapidity~\cite{ATLAS:2016nqi} (ID=248), di-muon production from (anti)neutrino-DIS process by NuTeV~\cite{Mason:2006qa} (ID=124--125) and CCFR~\cite{NuTeV:2001dfo} (ID=126--127), $F_3$ structure function measurements by CDHSW~\cite{Berge:1989hr} (ID=109) and CCFR~\cite{Seligman:1997mc} (ID=111), E866 NuSea $Q^3 d^2\sigma_{pp}/(dQ\,dx_F)$~\cite{NuSea:2003qoe} (ID=204), and E906 SeaQuest~\cite{SeaQuest:2021zxb} (ID=206) experiments.

\subsection{ATLAS 7-TeV $W$ and $Z$ production at the LHC}
\label{subsec:ID248}

\begin{figure}[htbp]
	\includegraphics[width=0.49\textwidth]{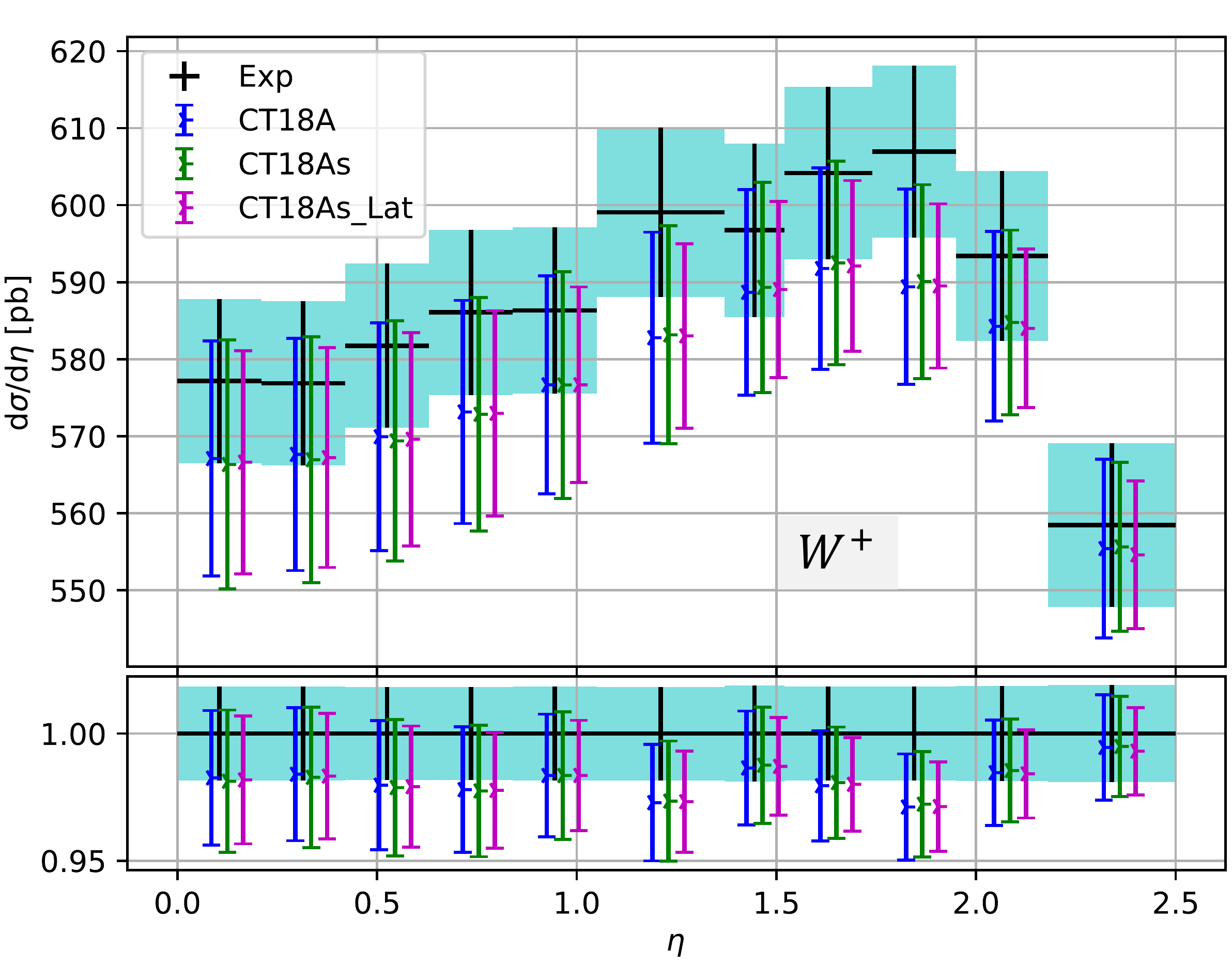}
	\includegraphics[width=0.49\textwidth]{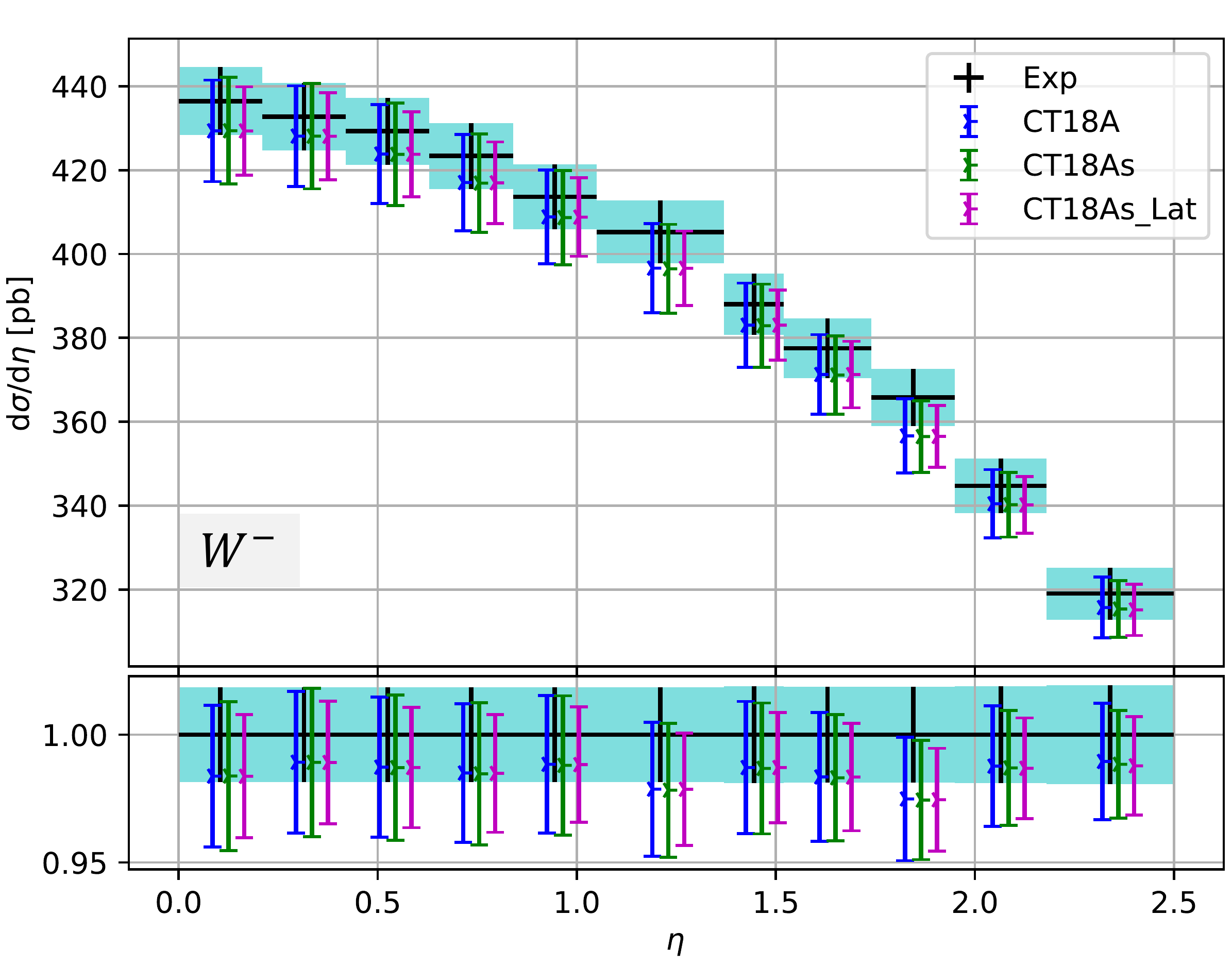}
	\includegraphics[width=0.49\textwidth]{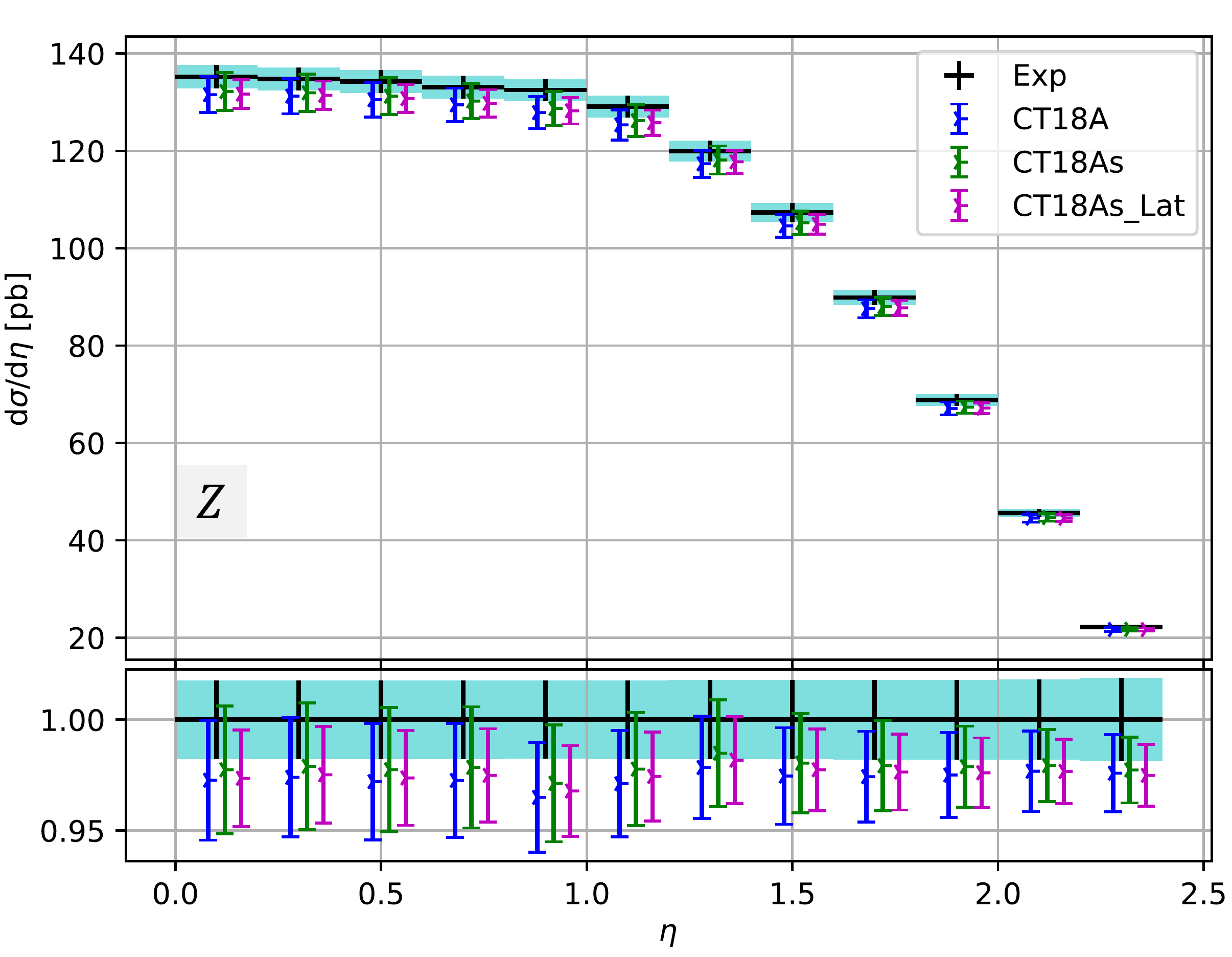}
	\caption{
	\label{fig:ID248}
	Comparison of CT18A, CT18As, and CT18As\_Lat predictions to the	experimental values of ID=248 ATLAS 7-TeV $W$ and $Z$ differential cross sections for $W^+$ (top-left), $W^-$ (top-right), $Z$ (bottom) as functions of dilepton pseudorapidity~\cite{ATLAS:2016nqi}.
}
\end{figure}

\begin{table}[htbp]
\begin{center}
\begin{tabular}{l|cccc|cc}
 & $Z$ & $W^+$ & $W^-$ & $R^2$ ($R^2$/131) & reduced $\chi^2$ & total $\chi^2$ \\ \hline
CT18A    & 17.48 & 15.29 & 13.82 & 40.99 (0.31) & 46.59 & 87.58 \\
CT18As   & 15.78 & 15.72 & 11.96 & 32.13 (0.25) & 43.46 & 75.59 \\
CT18As\_Lat   & 17.22 & 14.58 & 12.94 & 34.36 (0.26) & 44.74 & 79.10 \\  \hline
$N_{\text{pt}}$ & 12 & 11 & 11 &  & 34 & 34 \\
\end{tabular}
\end{center}
\caption{
\label{tab:ID248_reduced_chi2}
The reduced $\chi^2$ for $Z$ and $W^{\pm}$ production and the contribution of the nuisance parameter ($R^2$) to the total $\chi^2$ for the ATLAS 7-TeV $W$ and $Z$ data (ID=248).
The reduced $\chi^2$ quantifies the quality of fit to the shifted data.
ATLAS 7-TeV $W$ and $Z$ data~\cite{ATLAS:2016nqi} (ID=248) contains 131 correlated systematic errors.
}
\end{table}

The data for ATLAS 7-TeV $W$- and $Z$production differential cross sections~\cite{ATLAS:2016nqi} (ID=248) were found to be in tension with NuTeV~\cite{Mason:2006qa} and CCFR~\cite{NuTeV:2001dfo} DIS di-muon data; see Sec.~II.C of Ref.~\cite{Hou:2019efy}.
In Fig.~\ref{fig:ID248}, we compare the unshifted ATLAS 7-TeV $W$ and $Z$ data to the theoretical predictions of CT18A, CT18As, and CT18As\_Lat.
The central values of the predictions for all the $W$ and $Z$ data are below the experimental measurements and on the edge or even outside of the experimental error bands.
However, considering PDF-induced uncertainties, all predictions are consistent with the experimental measurements.
The differences among the predictions of CT18A, CT18As, and CT18As\_Lat for $W^{\pm}$ production (top panels of Fig.~\ref{fig:ID248}) are small, compared to the large uncertainty.
In Table~\ref{tab:ID248_reduced_chi2}, by allowing a nonvanishing strangeness asymmetry at $Q_0$ scale, the reduced $\chi^2$ for the $W^-$ production data is improved, while it is almost unchanged for the $W^+$ production data. The improvement relative to the $W^-$ production (via $s {\bar u}, s {\bar c} \to W^-$) data can be understood from Fig.~\ref{fig:strange_PDFs}, where $s(x)$ is enhanced with a nonvanishing strangeness asymmetry, while $\bar{s}(x)$ is less affected.

As for the Z-boson production (bottom panel of Fig.~\ref{fig:ID248}), the CT18As prediction is slightly larger than CT18A. Since the production of Z-bosons via the Drell-Yan process is dominated by quark-antiquark fusion, the enhancement in the $Z$ production rate reflects a higher magnitude in the combination of quark and antiquark PDFs.
This can be seen in Fig.~\ref{fig:strange_PDFs}, which shows the total strangeness $s_+(x)$ receiving a higher magnitude if nonzero strangeness asymmetry $s_-(x)$ is allowed.
Relative to CT18As, the CT18As\_Lat prediction is shifted such that it becomes closer to that of CT18A. Meanwhile, the predicted uncertainty of CT18As\_Lat shrinks compared to CT18As.

\subsection{SIDIS di-muon production data}

\begin{table}[htbp]
\begin{center}
\begin{tabular}{l|ccc|ccc}
	& \multicolumn{3}{c|}{NuTeV ($\nu \mu^+ \mu^-$) SIDIS} & \multicolumn{3}{c}{NuTeV (${\bar \nu} \mu^+ \mu^-$) SIDIS} \\
 & Reduced $\chi^2$ & $R^2$ & total $\chi^2$ & Reduced $\chi^2$ & $R^2$ & total $\chi^2$ \\ \hline
 CT18A & 30.43 & 1.47 & 31.90 & 42.20 & 10.59 & 52.79 \\
 CT18As & 21.83 & 7.44 & 29.27 & 48.23 & 13.27 & 61.50 \\
 CT18As\_Lat & 33.22 & 3.08 & 36.30 & 40.01 & 12.75 & 52.76 \\
\end{tabular}
\end{center}
\caption{
\label{tab:NuTeV_reduced_chi2}
The reduced $\chi^2$ and the contribution of the nuisance parameter ($R^2$) to the total $\chi^2$ for the NuTeV measurements of di-muon production in neutrino-ion (ID=124, $N_{\text{pt}}=38$) and antineutrino-ion (ID=125, $N_{\text{pt}}=33$) collisions~\cite{Mason:2006qa}.
The NuTeV dimuon production data sets have only one systematic error, which is the normalization error (10\%).
}
\end{table}

\begin{table}[htbp]
\begin{center}
\begin{tabular}{l|ccc|ccc}
	& \multicolumn{3}{c|}{ CCFR ($\nu \mu^+ \mu^-$) SIDIS } & \multicolumn{3}{c}{CCFR (${\bar \nu} \mu^+ \mu^-$) SIDIS} \\
 & Reduced $\chi^2$ & $R^2$ & total $\chi^2$ & Reduced $\chi^2$ & $R^2$ & total $\chi^2$ \\ \hline
 CT18A & 34.04 & 1.57 & 35.61 & 19.69 & 1.19 & 20.88 \\
 CT18As & 33.08 & 7.66 & 40.74 & 16.56 & 1.95 & 18.51 \\
 CT18As\_Lat & 32.65 & 3.34 & 35.99 & 20.11 & 2.27 & 22.38 \\
\end{tabular}
\end{center}
\caption{
\label{tab:CCFR_reduced_chi2}
Similar to Table~\ref{tab:NuTeV_reduced_chi2}.
The reduced $\chi^2$ and the contribution of the nuisance parameter ($R^2$) to the total $\chi^2$ for the CCFR measurements of di-muon production in neutrino-ion (ID=126, $N_{\text{pt}}=40$) and antineutrino-ion (ID=127, $N_{\text{pt}}=38$) collisions~\cite{NuTeV:2001dfo}.
The CCFR dimuon production data sets have only one systematic error, which is the normalization error (10\%).
}
\end{table}

\begin{figure}[htbp]
\centering
\includegraphics[width=\textwidth]{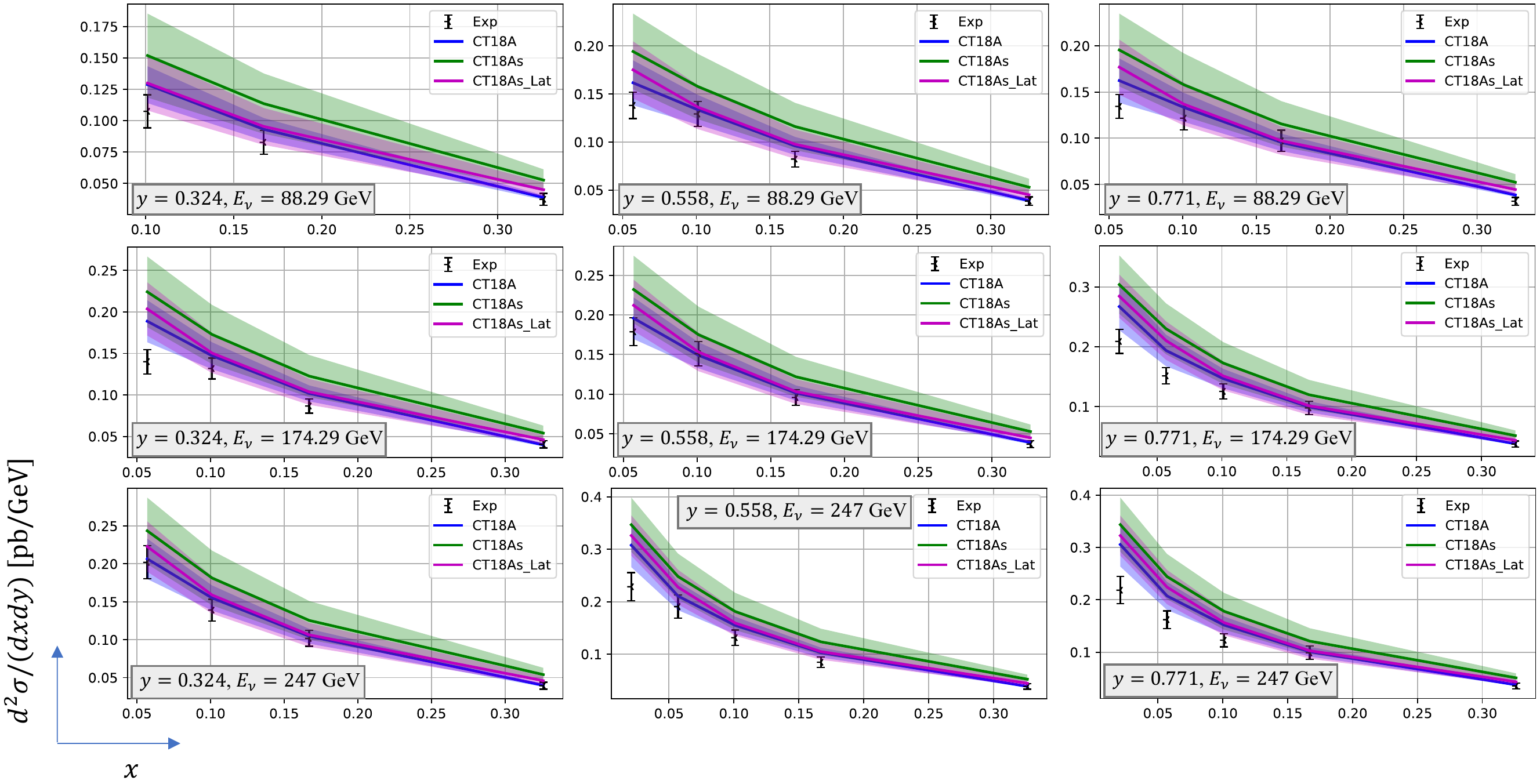}
\caption{
Comparison of data and theories for NuTeV measurements of di-muon production~\cite{Mason:2006qa} in neutrino-ion collisions (ID=124). The unshifted data is presented in the form of $d^2\sigma / dx dy$ [pb/GeV] as a function of $x$ for a certain values of $y$ and the neutrino energy $E_{\nu}$.
}
\label{fig:ID124}
\end{figure}

\begin{figure}[htbp]
\centering
\includegraphics[width=\textwidth]{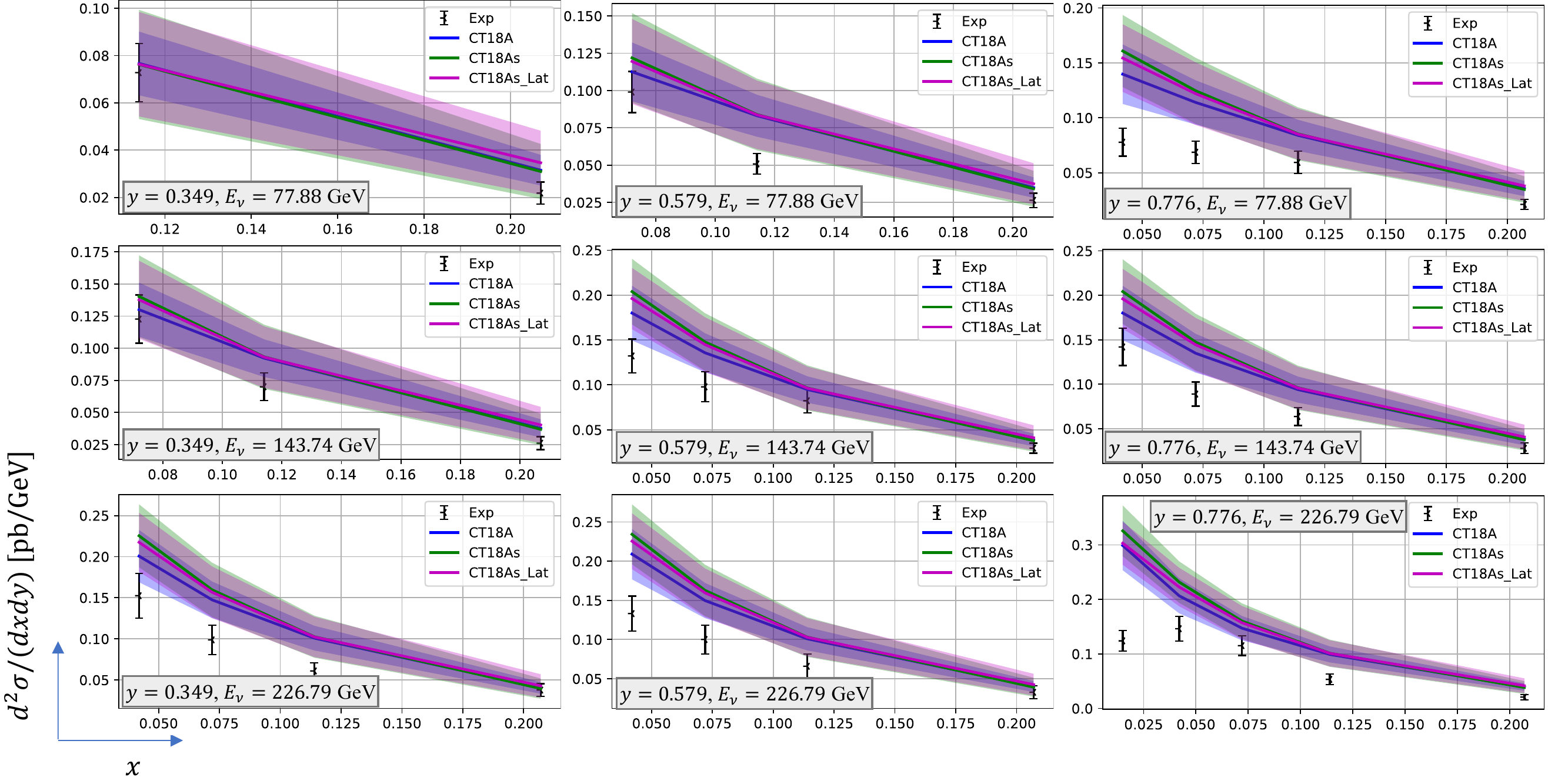}
\caption{
Similar to Fig.~\ref{fig:ID124}, but in antineutrino-ion collisions (ID=125).
}
\label{fig:ID125}
\end{figure}

\begin{figure}[htbp]
\centering
\includegraphics[width=\textwidth]{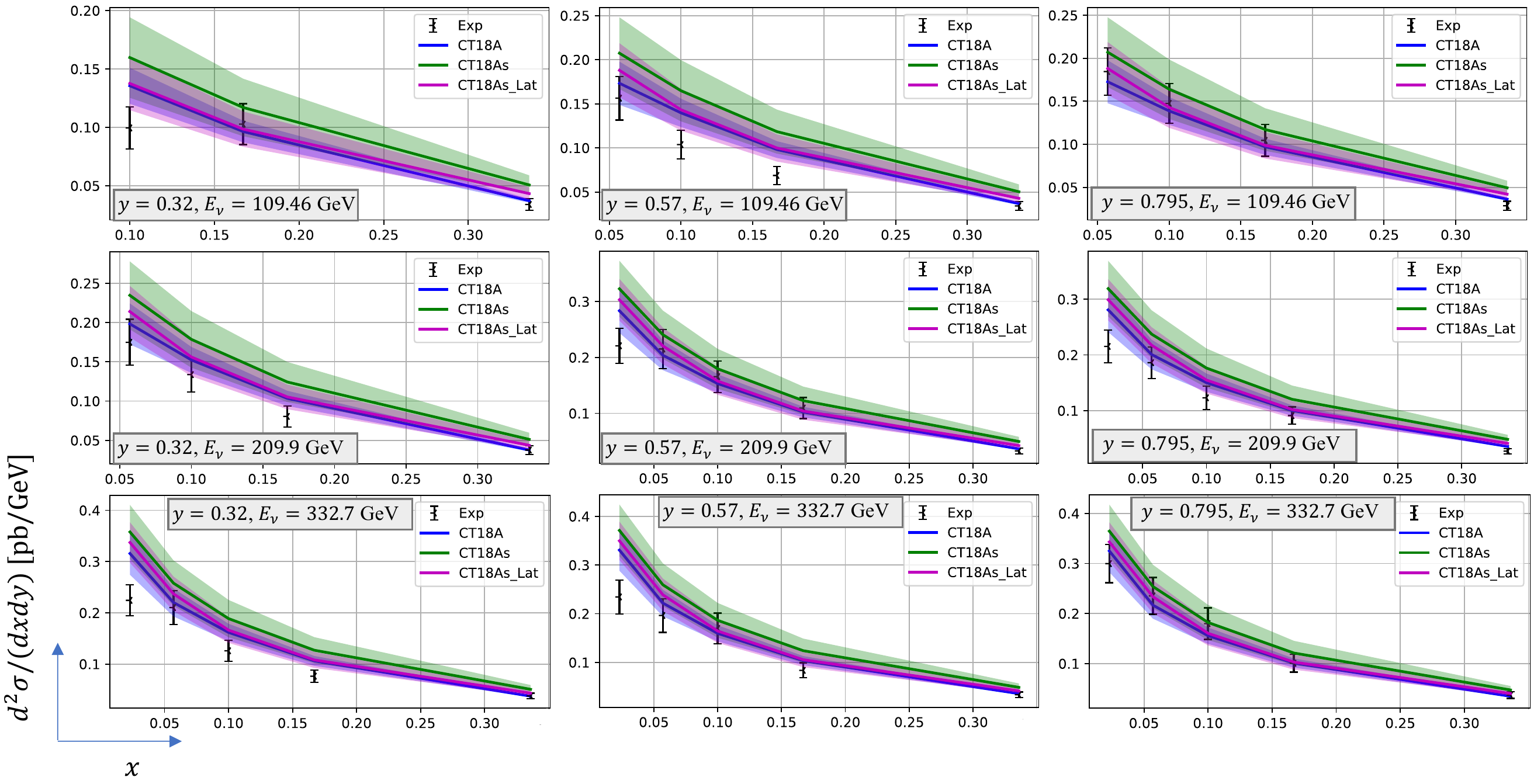}
\caption{
Comparison of unshifted data and theories for CCFR measurements of di-muon production~\cite{NuTeV:2001dfo} in neutrino-ion collisions (ID=126). The data is presented in the form of $d^2\sigma / dx\,dy$ [pb/GeV] as a function of $x$ for a certain values of $y$ and the neutrino energy $E_{\nu}$.
}
\label{fig:ID126}
\end{figure}

\begin{figure}[htbp]
\centering
\includegraphics[width=\textwidth]{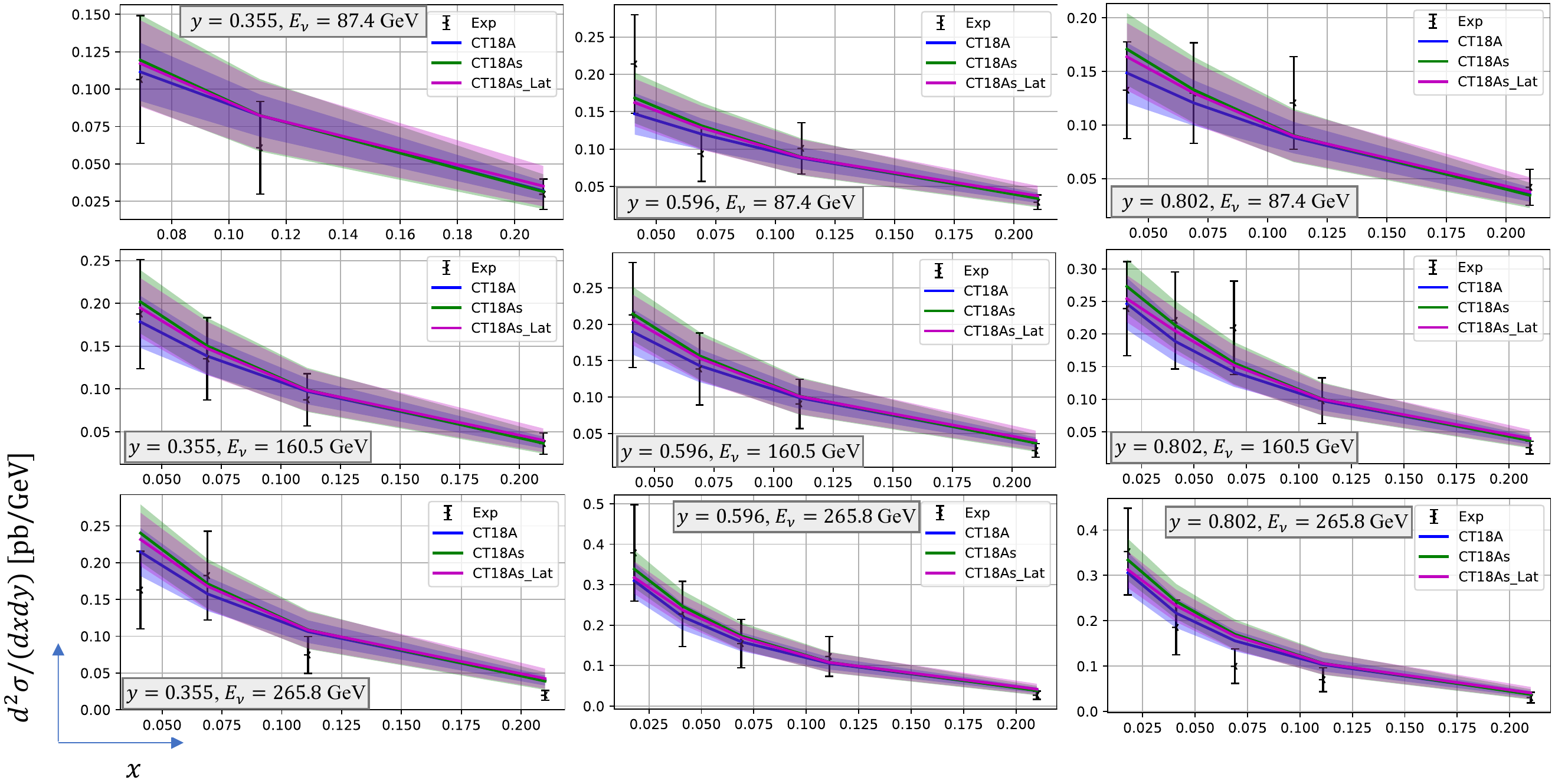}
\caption{
Similar to Fig.~\ref{fig:ID126}, but in antineutrino-ion collisions (ID=127).
}
\label{fig:ID127}
\end{figure}

The SIDIS di-muon production data selected for our PDF fits comprises NuTeV~\cite{Mason:2006qa} and CCFR~\cite{NuTeV:2001dfo} experiments (ID=124--127). At leading order, these data sets directly probe $s(x)$ and $\bar{s}(x)$, so they play an important role in determining strange quark and quark PDFs.
In Figs.~\ref{fig:ID124} and \ref{fig:ID125}, the comparison of the unshifted data of the NuTeV measurements of SIDIS di-muon production in neutrino-ion (ID=124) and antineutrino-ion (ID=125) collisions~\cite{Mason:2006qa} to the differential cross-sections, predicted with CT18A, CT18As, and CT18As\_Lat PDFs, is presented.
The reduced $\chi^2$ and the nuisance parameter contribution ($R^2$) to the total $\chi^2$ for the NuTeV dimuon data are summarized in Table~\ref{tab:NuTeV_reduced_chi2}.
The similar comparison for the CCFR measurements of di-muon production are shown in Figs.~\ref{fig:ID126} and~\ref{fig:ID127}, and Table~\ref{tab:CCFR_reduced_chi2}.

In Fig.~\ref{fig:ID124}, the central values of CT18As predictions for the NuTeV measurement with neutrino beam (ID=124), which directly probes $s(x)$ at leading order, are found to be so enhanced that they are outside of the experimental uncertainties and deviate from central values of CT18A.
Table~\ref{tab:NuTeV_reduced_chi2} shows that the CT18A and CT18As have a comparable total $\chi^2$ value, but the CT18As can fit the shifted data better with a larger $R^2$.
The tight constraint on strangeness asymmetry $s_-(x)$ of the LQCD calculation strongly impacts the strange PDF, as seen in Fig.~\ref{fig:strange_PDFs}. Consequently, the prediction of CT18As\_Lat in Fig.~\ref{fig:ID124} and the reduced $\chi^2$ shown in Table~\ref{tab:NuTeV_reduced_chi2} are closer to those of CT18A. The uncertainty of CT18As\_Lat prediction is then reduced from the wide error band of CT18As prediction, but is still larger than the uncertainty of CT18A prediction.
A similar comparison is presented for the NuTeV measurement with antineutrino beam (ID=125) in Fig.~\ref{fig:ID125}.
The central values for predictions of all three PDFs are close to each other, though with a much larger $\chi^2/N_\text{pt}$, about $1.6-1.9$, in comparison to that (less than 1) found in the NuTeV measurement with neutrino beam (ID=124), cf. Table~\ref{tab:NuTeV_reduced_chi2}.
In terms of PDF-induced uncertainty in predictions, CT18A has a smaller uncertainty band when compared to CT18As and CT18As\_Lat, in which a nonvanishing strangeness asymmetry $s_-(x)$ is allowed at the initial $Q_0$ scale.

We show the comparisons of data and theory for the CCFR di-muon production~\cite{NuTeV:2001dfo} in Figs.~\ref{fig:ID126} and \ref{fig:ID127}.
The reduced $\chi^2$ and the nuisance parameter contribution ($R^2$) for the CCFR dimuon data are summarized in Table~\ref{tab:CCFR_reduced_chi2}.
We observed the similar phenomena as in CCFR measurements as those seen in the NuTeV di-muon production~\cite{Mason:2006qa} with neutrino beam (ID=124).
The $\chi^2$ for the CCFR dimuon data with neutrino beam (ID=126) is slightly increased in CT18As by a large shift to the data. With the inclusion of the lattice data, the CT18As\_Lat obtain a comparable fit quality with the CT18A.
Unlike the fits to the NuTeV dimuon data with antineutrino beam (ID=125), all three PDFs can fit the CCFR dimuon data with antineutrino beam (ID=127) well.

As shown in Tables~\ref{tab:chi2_npt}, \ref{tab:NuTeV_reduced_chi2}, and ~\ref{tab:CCFR_reduced_chi2}, the descriptions to both NuTeV and CCFR dimuon data are in general good, except for the NuTeV dimuon data with antineutrino beam (ID=125).
For both NuTeV and CCFR dimuon data, a large shift to the raw data, hence large $R^2$ penalty, is required to fit the shifted data. In terms of the total $\chi^2$, no significant improvement is observed for the NuTeV and CCFR dimuon data altogether, if a non-vanishing strangeness asymmetry $s_-(x, Q_0)$ is allowed.
Further introducing the LQCD calculation of the strangeness asymmetry $s_-(x)$ in the CT18As\_Lat results in comparable descriptions of these data sets to the CT18A.
Though, as been seen in Sec.~\ref{subsec:ID248}, improvements for the $R^2$ penalty, reduced $\chi^2$, and hence the total $\chi^2$ for ATLAS 7 TeV $W$ and $Z$ data~\cite{ATLAS:2016nqi} (ID=248) are found in the CT18As, we did not find a clear evidence that allowing a nonvanishing strangeness asymmetry in the CT18As fit could release the above-mentioned tension, as compared to the CT18A fit.

Before closing this subsection, we would like to give a final comment about the large $R^2$ values observed in the above discussion. In the CT analyses, the CCFR and NuTeV dimuon cross sections are calculated by assuming the $c\to \mu$ branching ratio of 0.099, as in Section 5.2.1 of \cite{Mason:2006qa}.  The normalization uncertainty of 10\% is treated as fully correlated over the $\nu$ channel and similarly over the $\bar \nu$ channel.
In this work, we confirmed the finding in Ref.~\cite{Hou:2019efy} that reducing the $c\to \mu$ branching ratio from 0.099 to 0.092, as adopted by MMHT \cite{Harland-Lang:2015nxa}, only marginally increases $s(x,Q)$ in CT18 at $x>0.1$, while slightly reducing the CCFR and NuTeV $\chi^2$ values in the CT18A fit. Roughly, this will reduce the $R^2$ values of the NuTeV and CCFR SIDIS di-muon data listed in Tables~\ref{tab:NuTeV_reduced_chi2} and \ref{tab:CCFR_reduced_chi2}, by about one to two units, with a much smaller effect on their reduced $\chi^2$.
Similar reduction (by about one to two units) in the total $\chi^2$ value of ATLAS 7 TeV $W$ and $Z$ data is also observed.

\subsection{$F_3$ structure function}

\begin{figure}[htbp]
\centering
\includegraphics[width=\textwidth]{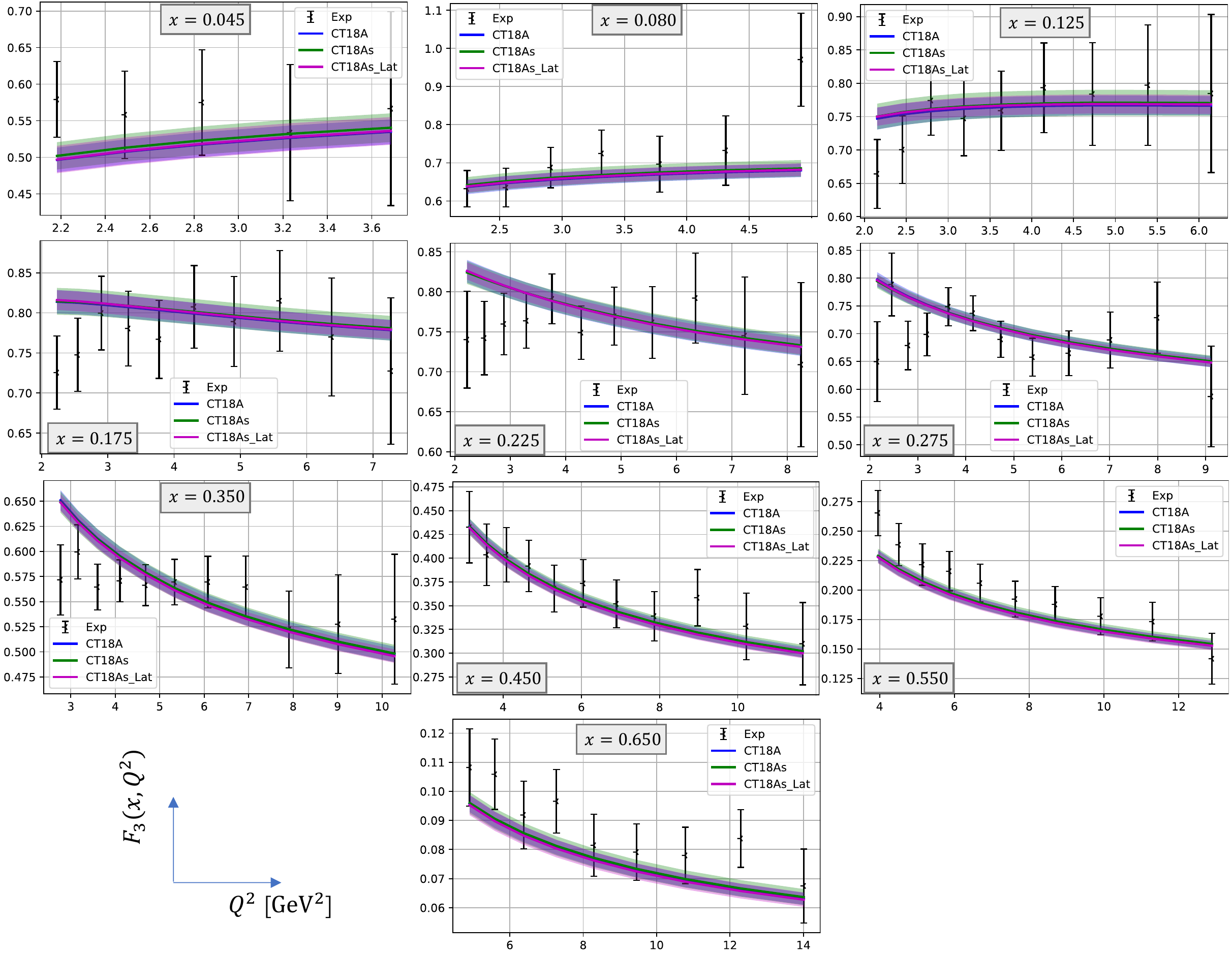}
\caption{
Comparison of data and theories for the CDHSW $F^p_3$ structure function measurements~\cite{Berge:1989hr} (ID=109). The unshifted data are presented in the form of $F_3$ as a function of $Q^2$ for a certain values of $x$.
}
\label{fig:ID109}
\end{figure}

\begin{figure}[htbp]
\centering
\includegraphics[width=\textwidth]{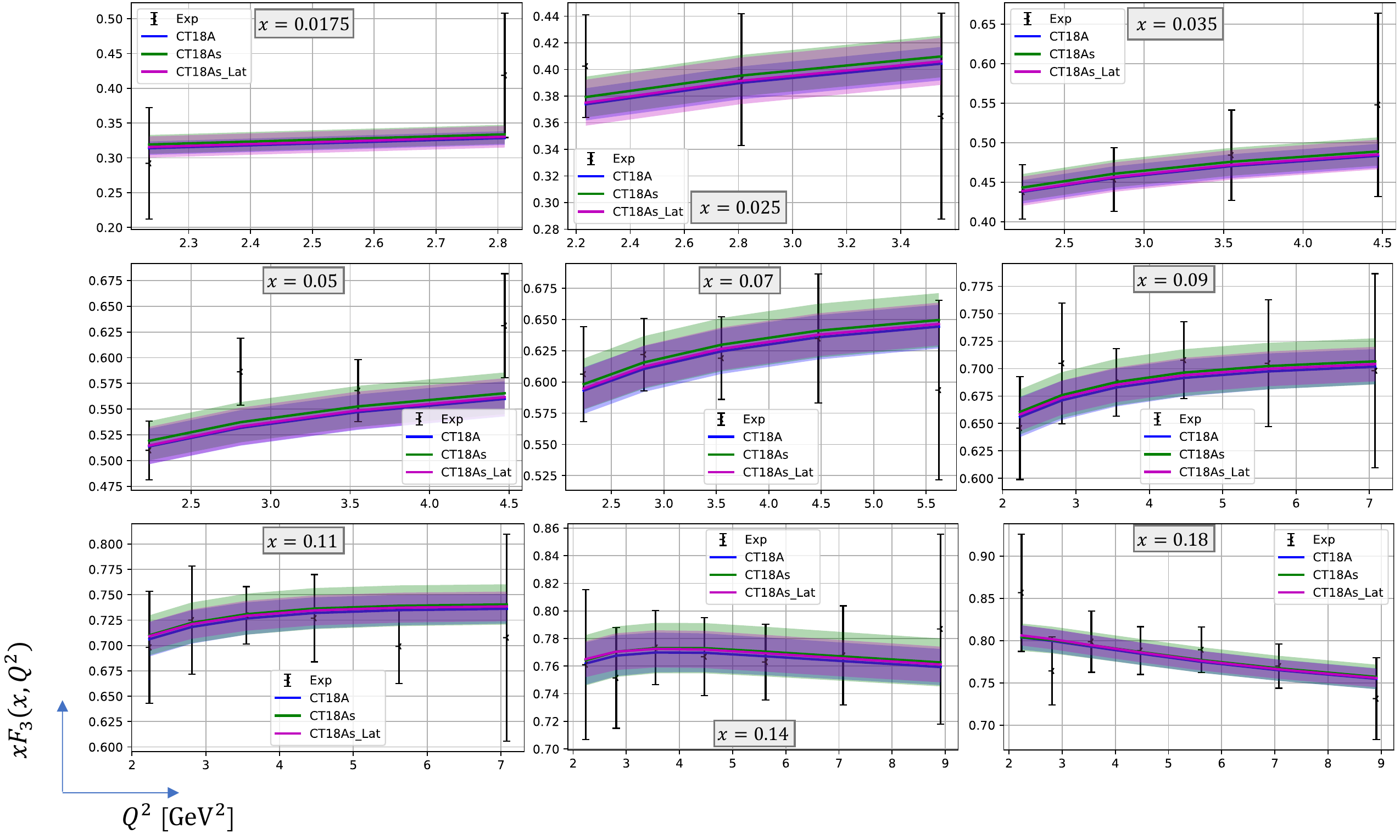}
\caption{
Comparison of data and theories for the CCFR $xF^p_3$ structure function measurements~\cite{Seligman:1997mc} (ID=111). The unshifted data are presented in the form of $xF_3$ as a function of $Q^2$ for a certain values of $x$.
Larger values of $x$ are shown in Fig.~\ref{fig:ID111_2} below.
}
\label{fig:ID111_1}
\end{figure}

\begin{figure}[htbp]
\centering
\includegraphics[width=\textwidth]{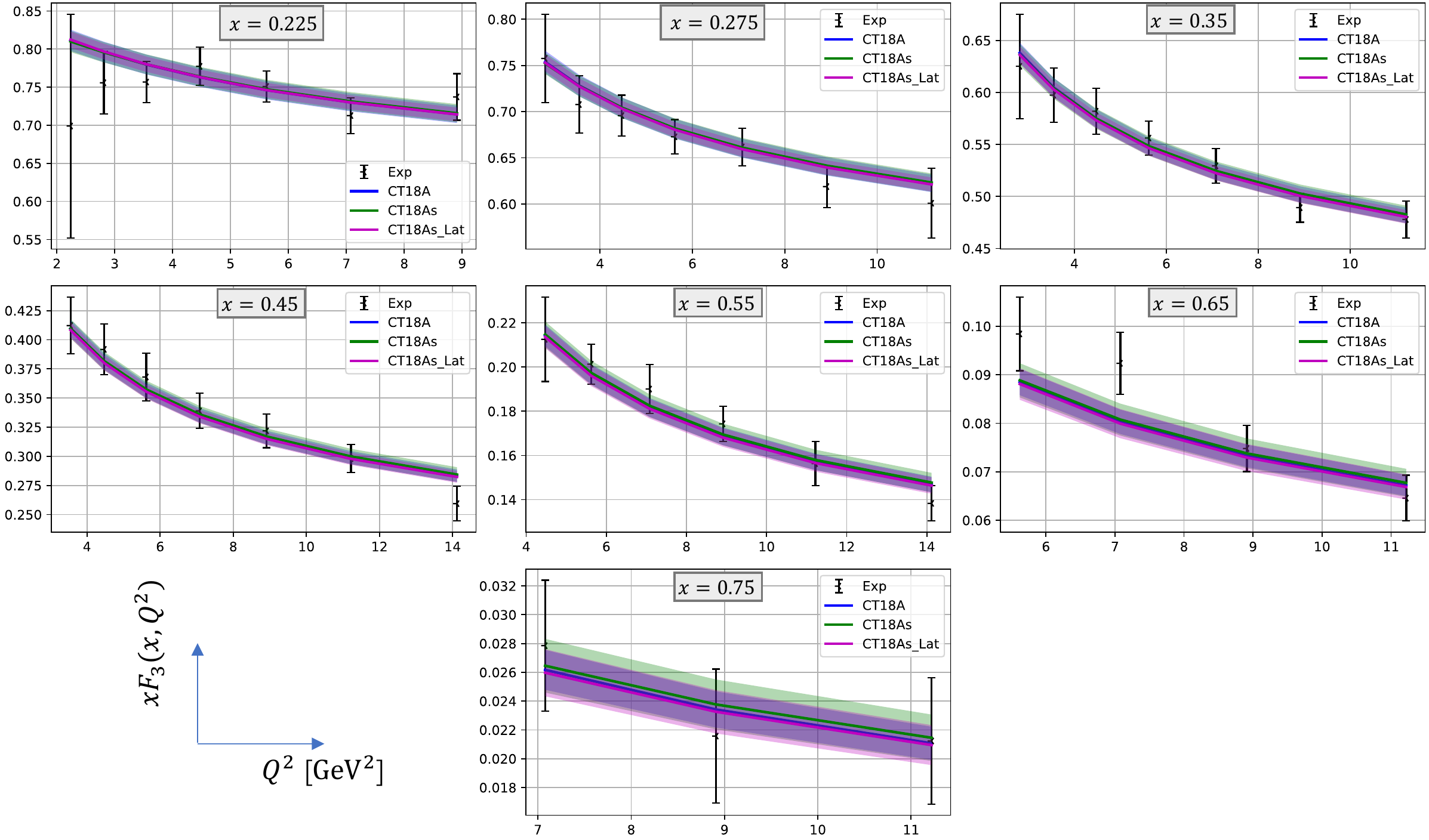}
\caption{
Following Fig.~\ref{fig:ID111_1}, the comparison of data and theories for CCFR $xF^p_3$ structure function measurements~\cite{Seligman:1997mc} (ID=111) for larger $x$.
}
\label{fig:ID111_2}
\end{figure}

The $F_3$ structure function at leading order is proportional to the valence-sector PDFs, so it is sensitive to the strangeness asymmetry $s_-(x)$. Since $u_v(x)$ and $d_v(x)$ PDFs already receive tight constraints from many other data sets included in the fit, it is expected that the variation in predictions of the $F_3$ structure function should reflect the information in the $s_-(x)$ distribution.
The CERN-Dortmund-Heidelberg-Saclay-Warsaw (CDHSW) $F^p_3$ structure function measurement~\cite{Berge:1989hr} (ID=109) and CCFR $xF^p_3$ structure function measurement~\cite{Seligman:1997mc} (ID=111) are two $F_3$-measurement data sets  included in CT18A, CT18As, and CT18As\_Lat. In Figs.~\ref{fig:ID109}-\ref{fig:ID111_2}, we study the implications of the $s_-(x)$ distribution in the comparison of data and theory for the CDHSW and CCFR $F_3$ measurements.

We show that the predictions of CT18A, CT18As, and CT18As\_Lat for the CDHSW $F^p_3$ structure function are consistent with experiment in Fig.~\ref{fig:ID109}.
For $x \leq 0.125$ and $x \geq 0.35$, the central values of the predictions of CT18As have a slightly higher magnitude than those of CT18A and CT18As\_Lat, in which less strangeness asymmetry is predicted. But for $x$ around $0.1 \sim 0.2$, where the predicted strangeness asymmetry in CT18As peaks, the difference among predictions of the three PDFs is not obvious. This is because the $F_3$ prediction also receives contributions from $u_v(x)$ and $d_v(x)$, whose magnitudes are much greater than the strangeness asymmetry $s_-(x)$.
For $x$ around $0.1 \sim 0.2$, the predicted uncertainty of CT18As is larger than those of CT18A and CT18As\_Lat.
In this range of $x$, CT18As has the largest uncertainty for the strangeness asymmetry $s_-(x)$, cf. Fig.~\ref{fig:strange_PDFs}.
In Figs.~\ref{fig:ID111_1} and \ref{fig:ID111_2}, a similar comparison of data and theory is done for the CCFR $xF^p_3$ structure function measurement.
An upward shift of the central value and an enlarged uncertainty in the CT18As prediction are also observed for this case.

\subsection{E866 NuSea data and E906 SeaQuest data}

\begin{figure}[htbp]
\centering
\includegraphics[width=\textwidth]{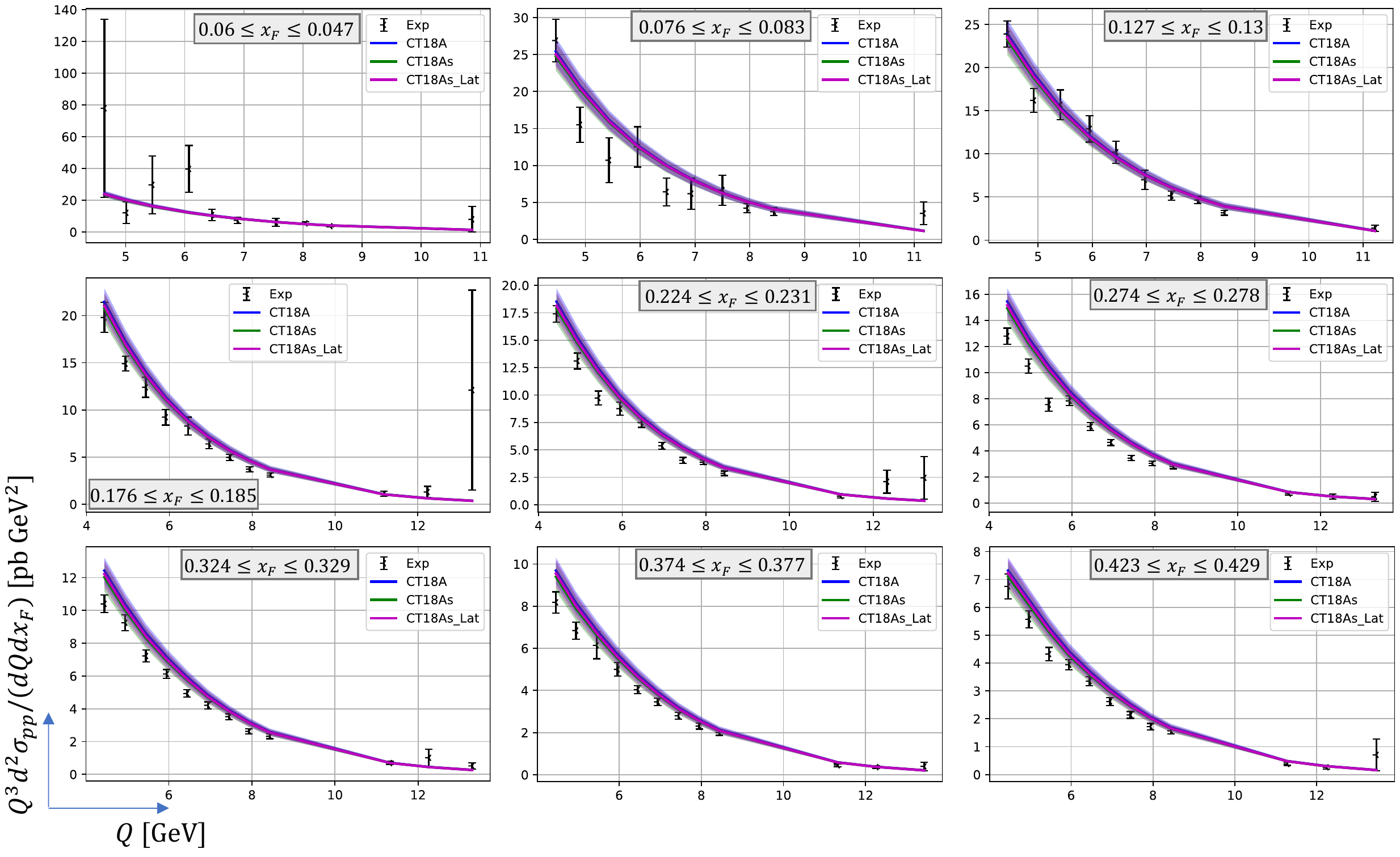}
\caption{
Comparison of data and theories for the E866 NuSea measurement of $Q^3 d^2\sigma_{pp}/(dQ\,dx_F)$~\cite{NuSea:2003qoe} (ID=204).
The unshifted data is presented in the form of $Q^3 d^2\sigma_{pp}/(dQ\,dx_F)$ as a function of invariant mass $Q$ for ranges of $x_F$.
Larger values of $x_F$ are shown in Fig.~\ref{fig:ID204_2}.
}
\label{fig:ID204_1}
\end{figure}

\begin{figure}[htbp]
\centering
\includegraphics[width=\textwidth]{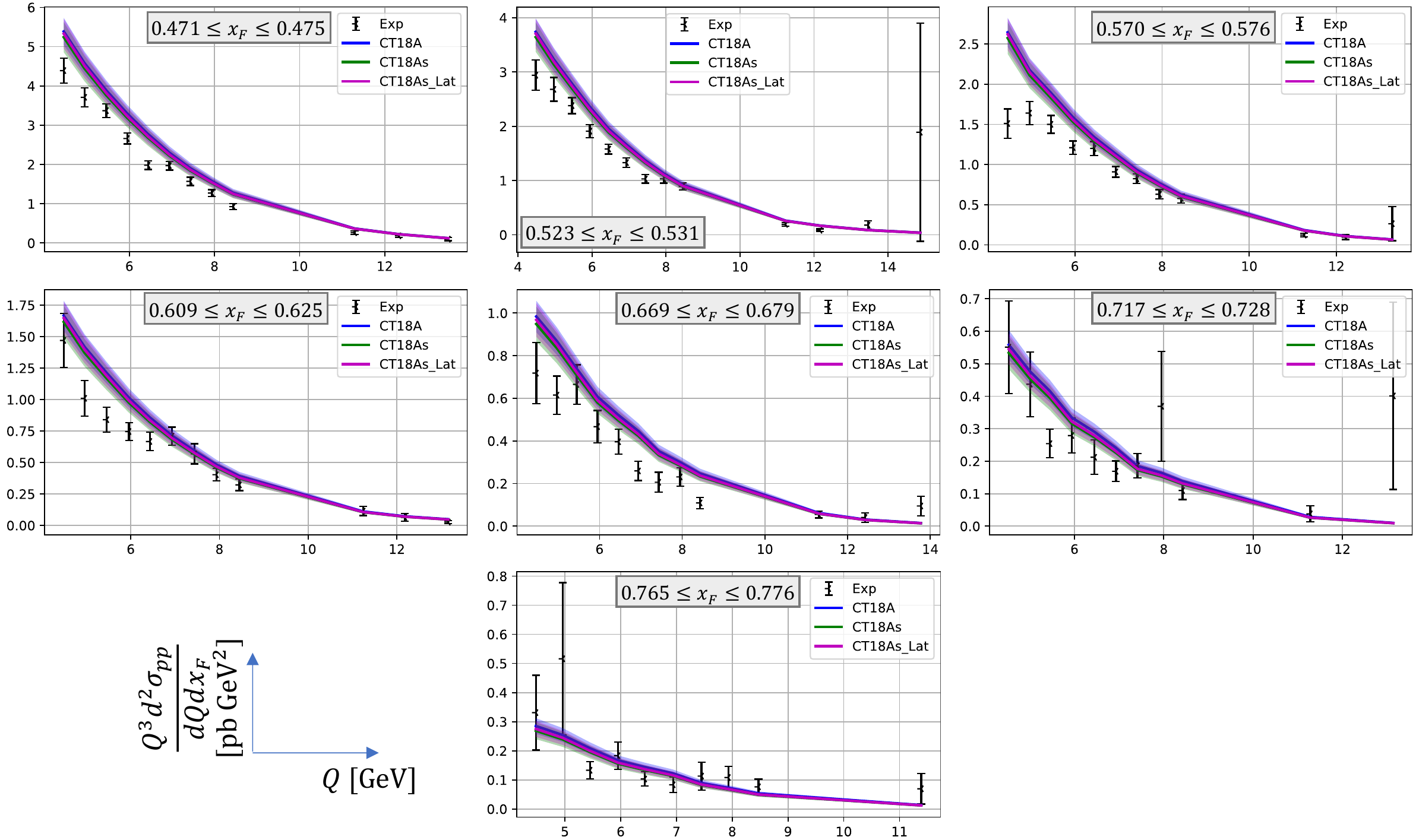}
\caption{
Following Fig.~\ref{fig:ID204_1}, the comparison of the unshifted data and theories for E866 NuSea $Q^3 d^2\sigma_{pp}/(dQ\,dx_F)$ measurement~\cite{NuSea:2003qoe} (ID=204) for larger values of $x_F$.
}
\label{fig:ID204_2}
\end{figure}

In Fig.~\ref{fig:spartyness}, we find that the fixed-target E866 NuSea $Q^3 d^2\sigma_{pp}/(dQ\,dx_F)$ data~\cite{NuSea:2003qoe} (ID=204) has effective Gaussian variables $S_E>2$, suggesting that this data cannot be well fitted.
In Figs.~\ref{fig:ID204_1} and \ref{fig:ID204_2}, the E866 NuSea data is compared to theoretical predictions of CT18A, CT18As, and CT18As\_Lat.
We observe that, for $0.17 < x_F < 0.73$, there is a trend that central values of numerical predictions are above the experimental data points.
The difference among theoretical predictions is only noticeable for bins with $x_F > 0.6$. Allowing a nonzero strangeness asymmetry at the initial $Q_0$ scale pulls theory predictions downward, and hence produces a slightly better fit to E866 NuSea $Q^3 d^2\sigma_{pp}/(dQ\,dx_F)$ data.
Fig.~\ref{fig:ubr_dbr} shows that allowing a nonvanishing strangeness asymmetry at $Q_0$ scale would enhance the total strangeness, whose cost is the suppression on $\bar{u}$ and $\bar{d}$ for $0.01 < x <0.1$, due to the conservation of the total momentum sum rule. Considering that $u(x)$ and $d(x)$ in large-$x$ region would be less affected by variation in the strangeness asymmetry $s_-(x)$, the decreases in $\bar{u}(x)$ and $\bar{d}$ leads to less numerical value of production cross-section $Q^3 d^2\sigma_{pp}/(dQ\,dx_F)$ prediction.

We also study the implication of CT18A, CT18As, and CT18As\_Lat by comparing theory and data for the E906 SeaQuest experiment~\cite{SeaQuest:2021zxb} (ID=206). In Fig.~\ref{fig:spartyness}, effective Gaussian variables for all three PDFs are close to zero, indicating that all three PDFs describe data well. We note that the E906 SeaQuest data is not included in the CT18A data set, neither in this study. Here, we only assess the impact of changing in strangeness asymmetry via comparing data to theoretical calculations.
The E906 SeaQuest data measures the ratio of cross-section $\sigma(pd)/2\sigma(pp)$ as a function of the momentum fraction $x_2$ of the target. The ratio of cross-section $\sigma(pd)/2\sigma(pp)$ approximates the ratio of antiquark PDFs $\sigma(pd)/2\sigma(pp) \approx \big{(} 1 + \bar{d}_p(x_2) / \bar{u}_p(x_2) \big{)}/2$, so that the E906 SeaQuest data, as well as its predecessor E866 NuSea data~\cite{NuSea:2001idv} (ID=203), provides useful information of antiquark asymmetry in the large-$x$ region.

\begin{figure}[htbp]
\centering
\includegraphics[width=0.6\textwidth]{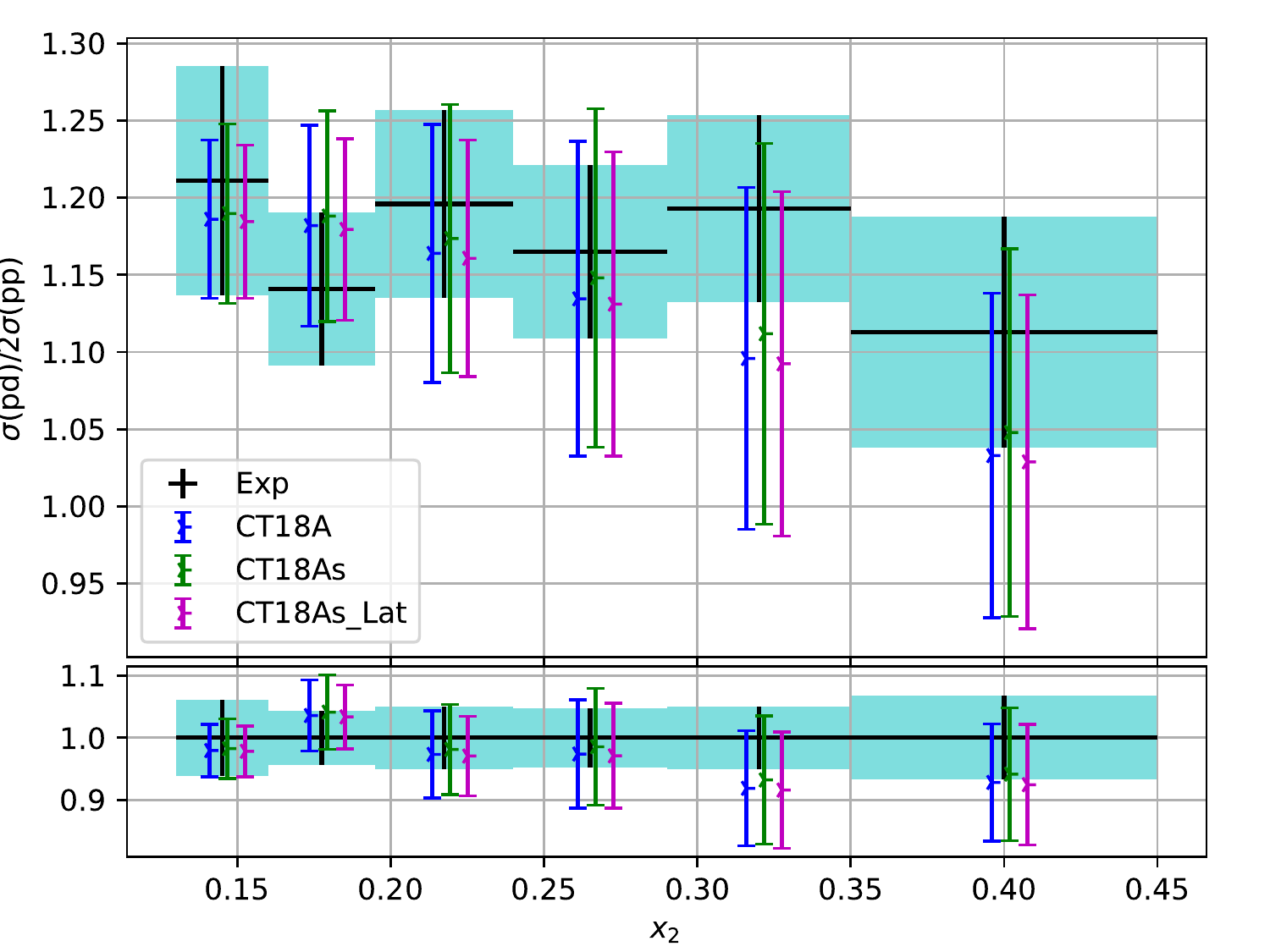}
\caption{
Comparison of the unshifted data and theories for E906 SeaQuest data~\cite{NuSea:2001idv} (ID=206).
}
\label{fig:SeaQuest}
\end{figure}

In Fig.~\ref{fig:SeaQuest}, we compare theory predictions of CT18A, CT18As, and CT18As\_Lat to the SeaQuest data, and find that they are all consistent with experimental values.
From prediction of CT18A to CT18As, introducing the strangeness asymmetry at the initial $Q_0$ scale would raise the central value and enlarge the uncertainty of theoretical prediction.
Adding in the lattice data, which is consistent with a negligible strangeness asymmetry, the prediction of CT18As\_Lat becomes close to the original prediction of CT18A.
As shown in Fig.~\ref{fig:ubr_dbr}, allowing a nonvanishing strangeness asymmetry at $Q_0$ scale in CT18As would raise the PDF ratio $\bar{d}/\bar{u}(x)$ for $x > 0.2$ so as to induce a different theory prediction for CT18As, as shown in Fig.~\ref{fig:SeaQuest}.

\begin{figure}[htbp]
\includegraphics[width=0.49\textwidth]{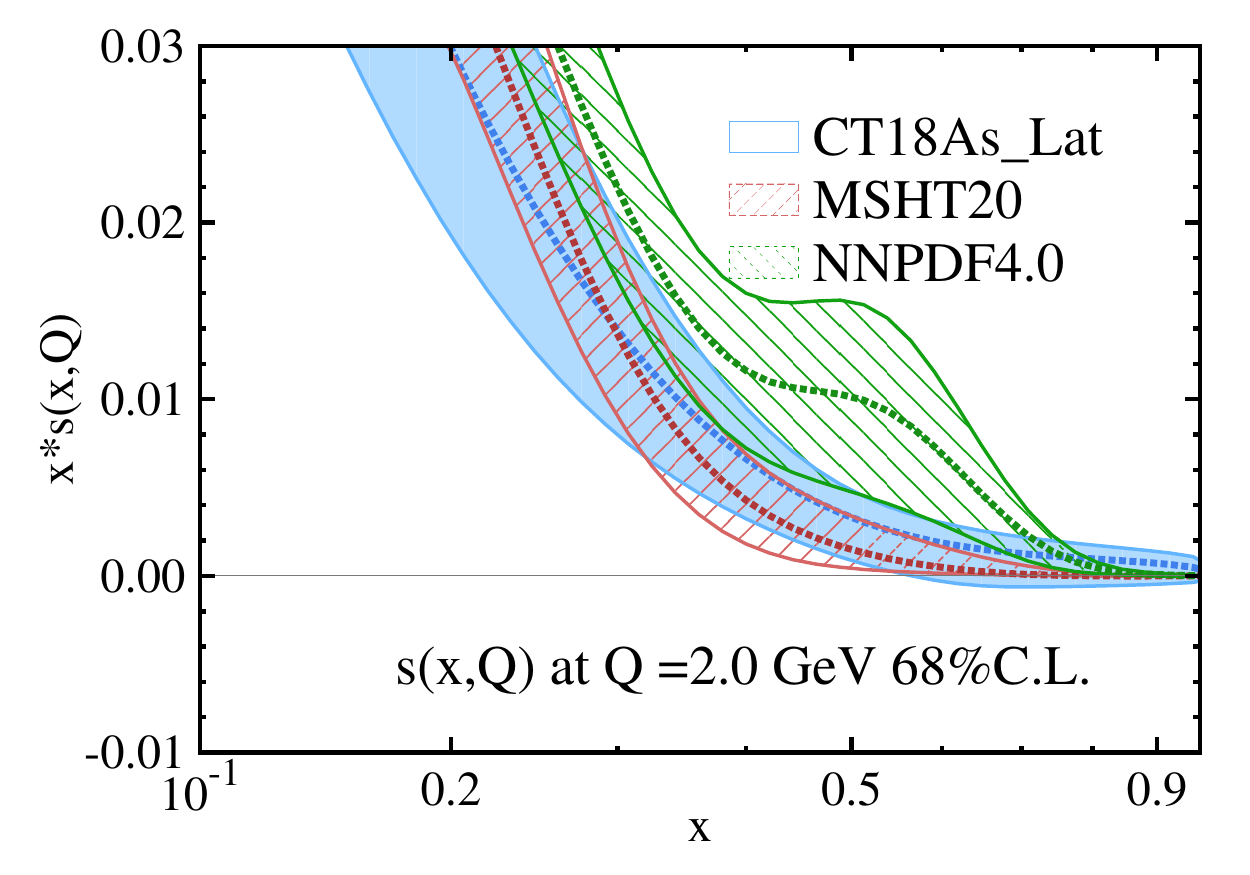}
\includegraphics[width=0.49\textwidth]{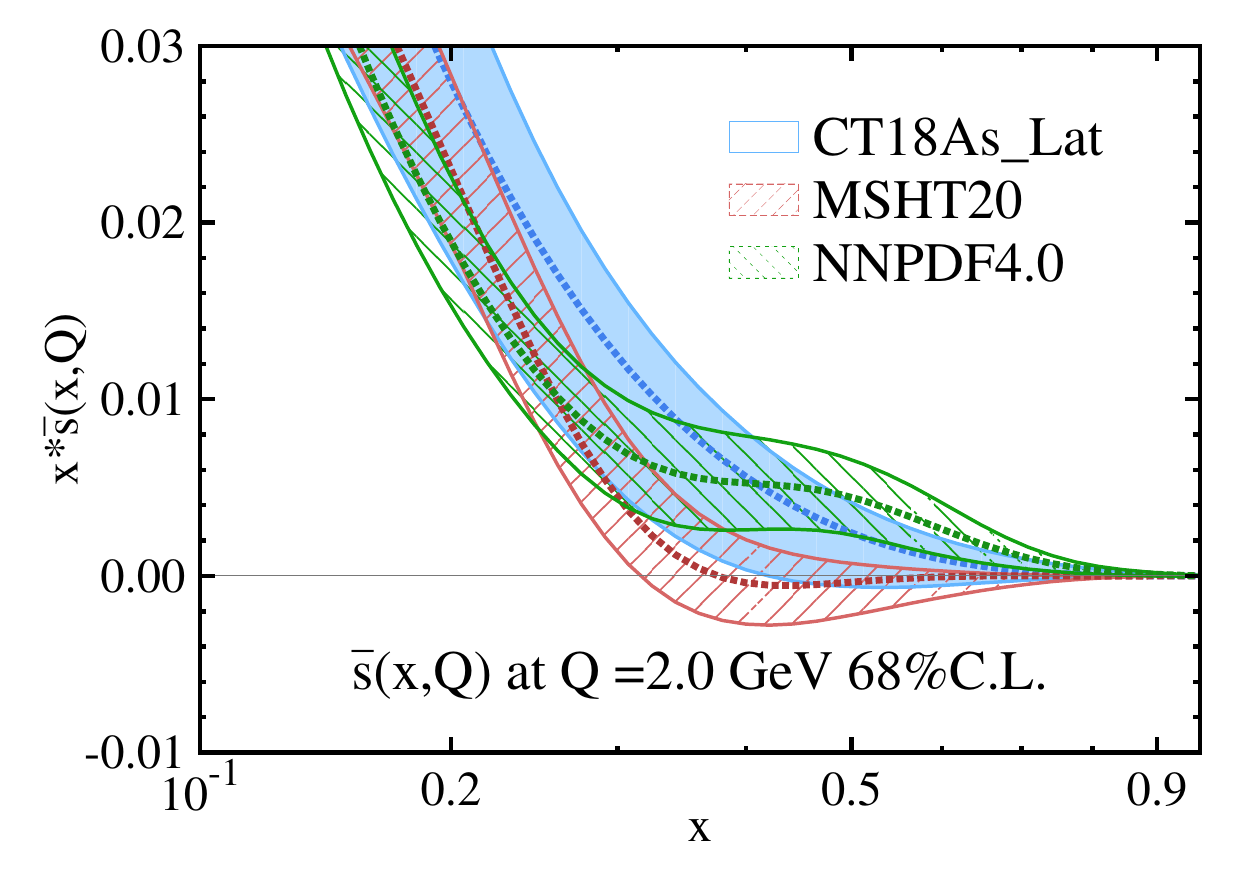}
\includegraphics[width=0.49\textwidth]{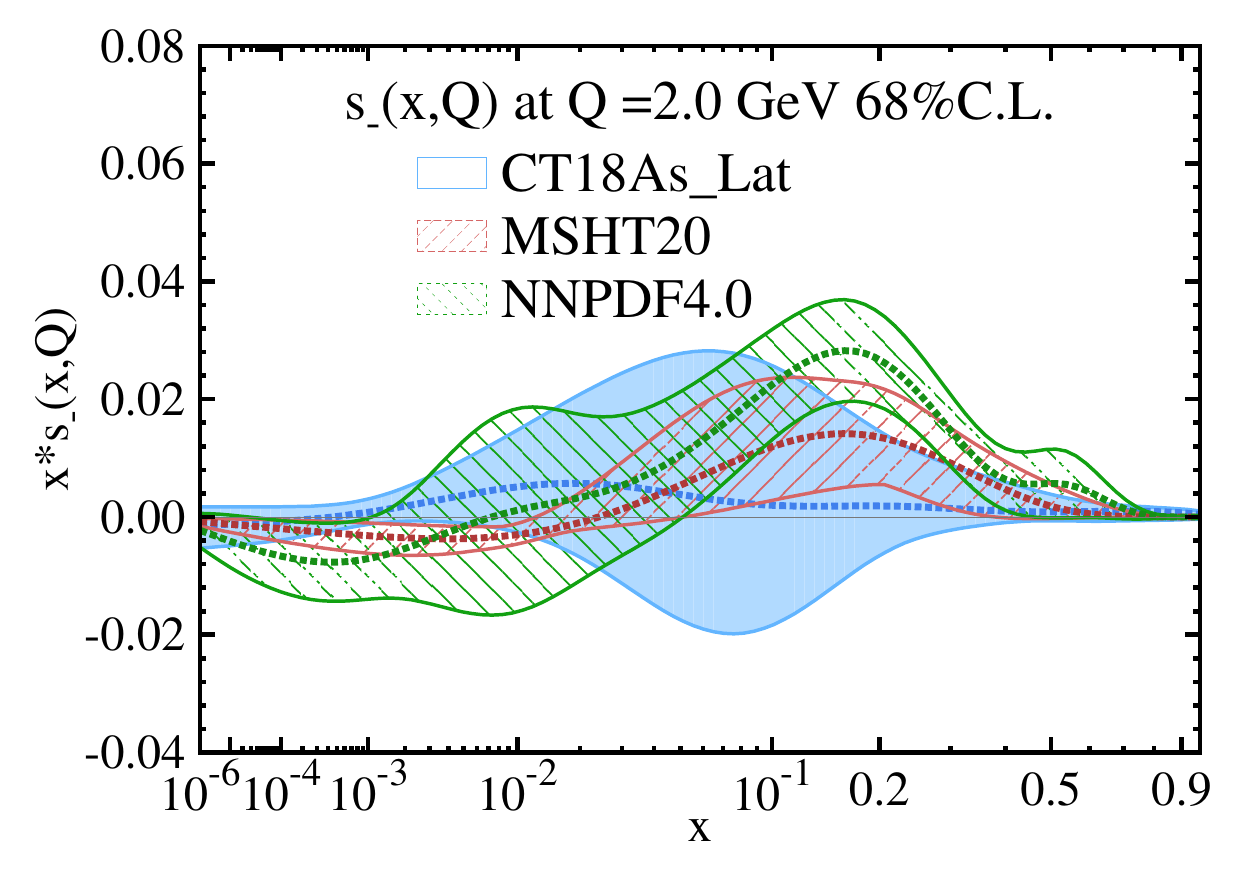}
\includegraphics[width=0.49\textwidth]{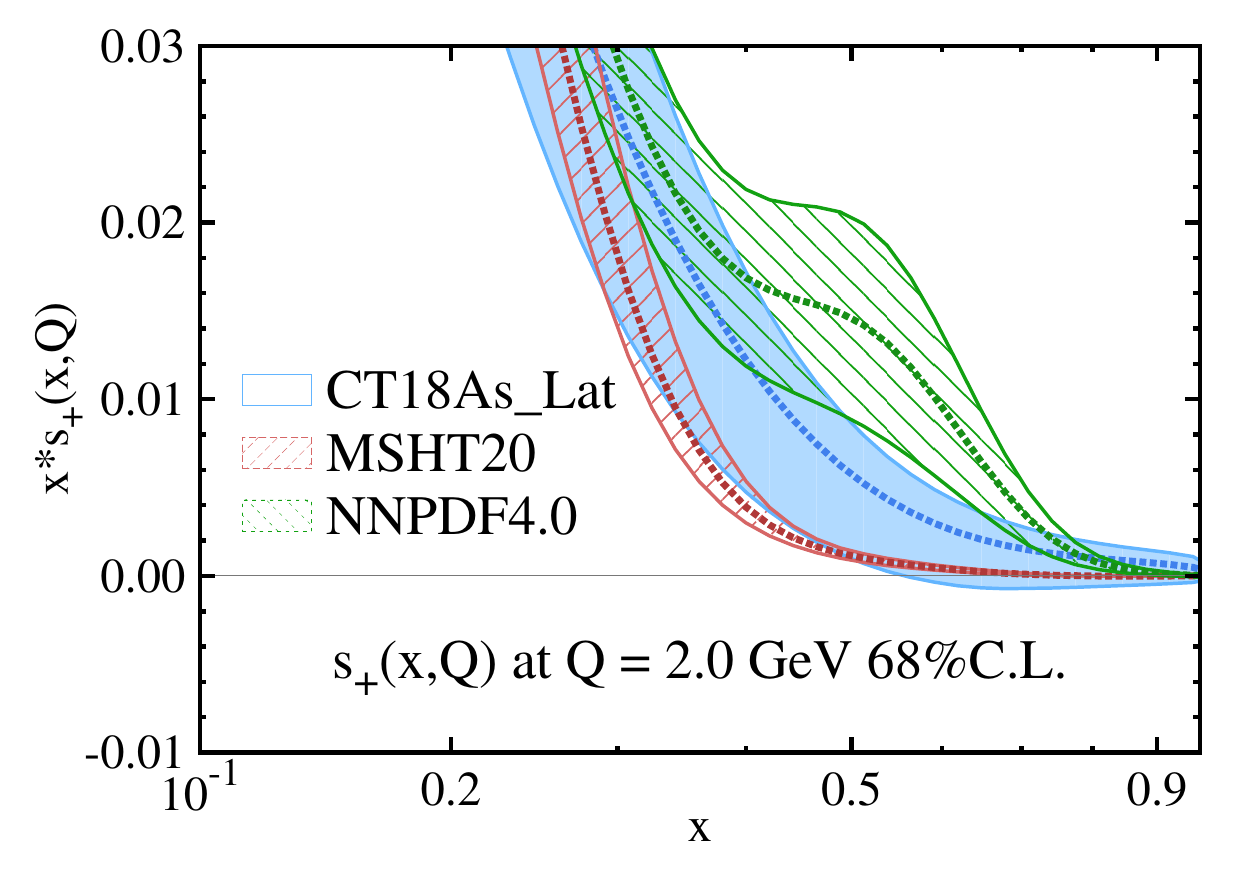}
\includegraphics[width=0.49\textwidth]{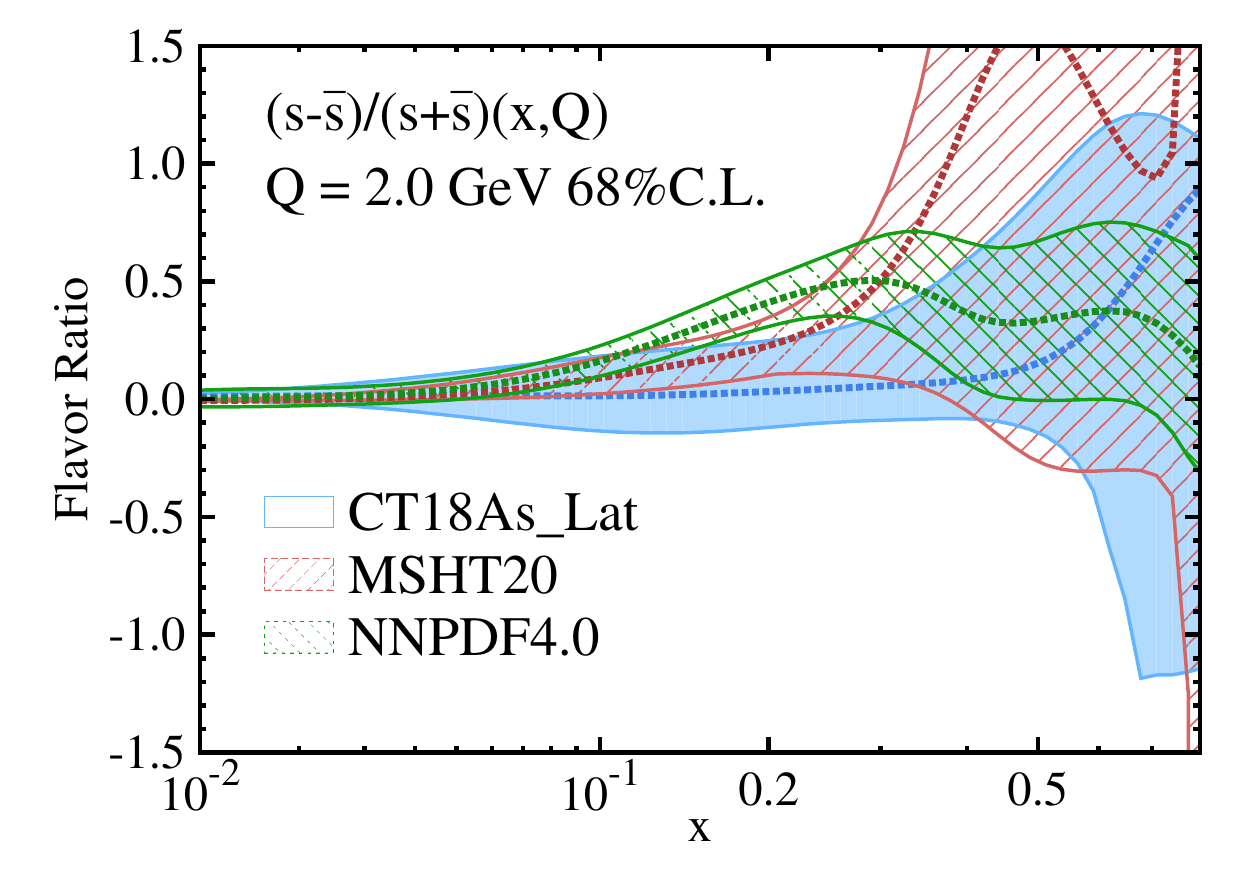}
\includegraphics[width=0.49\textwidth]{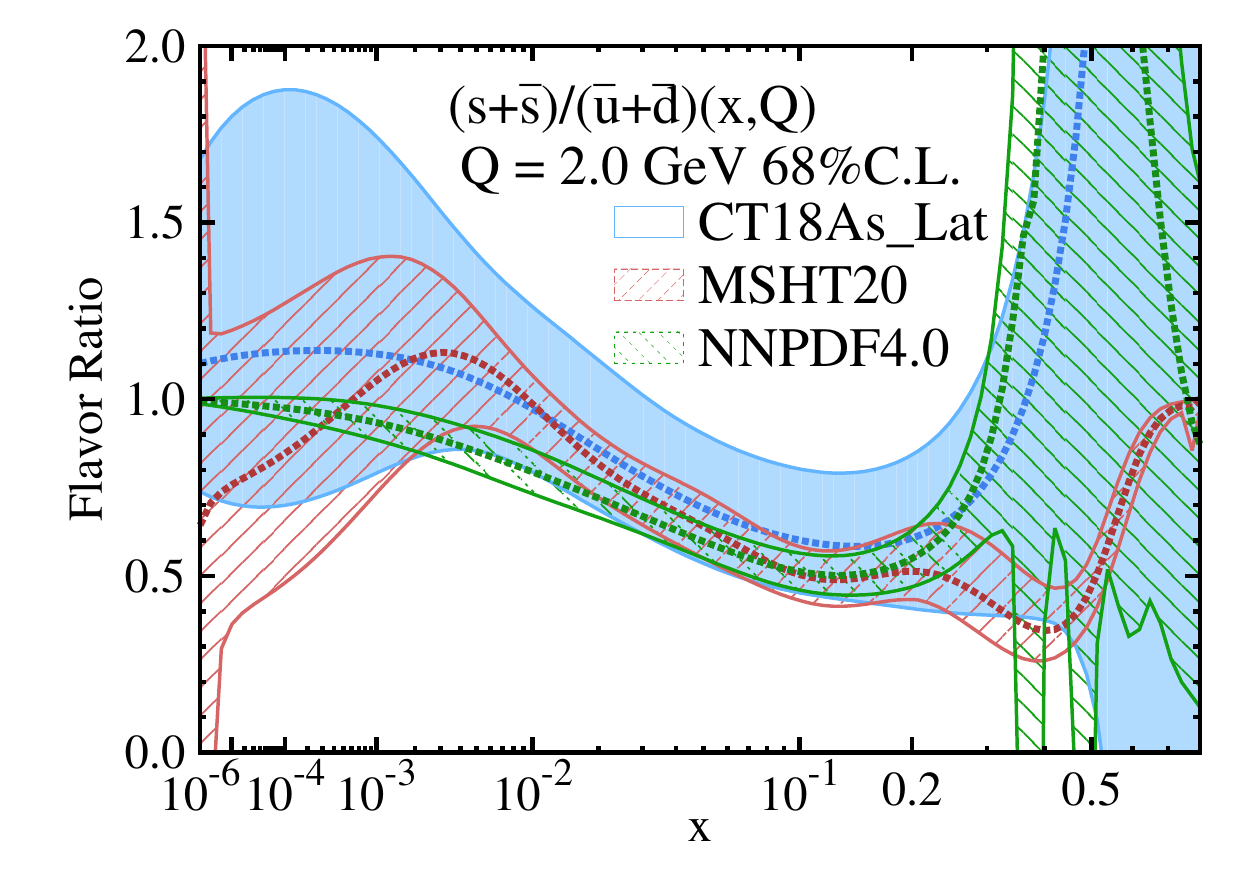}
\caption{
\label{fig:compare_other_PDFs}
The comparisons of $s(x)$, $\bar{s}(x)$, $s_-(x)$, $s_+(x)$, $(s+\bar{s})/(\bar{u}+\bar{d})(x)$, and $(s-\bar{s})/(\bar{s}+\bar{s})(x)$ for CT18As\_Lat, CT18As2\_Lat, MSHT20~\cite{Bailey:2020ooq}, and NNPDF4.0~\cite{Ball:2021leu} at $Q = 2.0$ GeV.
}
\end{figure}

\section{Conclusion and Outlook}
\label{conclusion}

In this work, we study the impact of the lattice data on the determination of the strangeness asymmetry distribution $s_-(x) \equiv s(x) - {\bar s}(x)$ at the initial $Q_0$ scale in the general CTEQ-TEA global analysis of parton distribution functions (PDFs) of the proton.
Following the recommendation made in Ref.~\cite{Hou:2019efy}, we start with the CT18A NNLO fit, rather than the nominal CT18 NNLO fit,
because we are interested in the (anti)strange quark distributions of the proton.
The ATLAS $\sqrt{s} = 7$~TeV $W$, $Z$ combined cross-section data~\cite{ATLAS:2016nqi} (ID=248) is included in the CT18A fit, while it is absent in the nominal CT18 fit.

We extend the nonperturbative parametrization in the CT18A analysis by allowing a strangeness asymmetry distribution $s_-(x) \equiv s(x) - \bar{s}(x)$ at the initial $Q_0$ ($=1.3$~GeV) scale. The resulting PDF set from the CT18A data set is labelled as CT18As, whose quality of fit is similar to that of the CT18A fit.
The constraint from the lattice data into the PDF global fit is added by using the Lagrange Multiplier method.
We found that the resulting PDF, named as CT18As\_Lat, present a different strangeness asymmetry distribution and a smaller uncertainty band than those of CT18As. We also investigate the possible constraint of the lattice data with higher precision by performing a PDF fit with errors in the original lattice data points reduced by half.
Our results conclude that the current lattice data is able to help further constraining the strangeness asymmetry $s_-(x)$ in PDF global analysis. Future precision improvement in the lattice calculation of this quantity could further improve the $s_-(x)$ for $x \in [10^{-2},0.6]$.

We also assess the impact of introducing a nonvanishing strangeness asymmetry $s_-(x)$ at the initial $Q_0$ scale and lattice data by comparing predictions of CT18A, CT18As, and CT18As\_Lat on a few selected experimental data. The predictions of different PDFs are in general consistent with each other.
As noted in Ref.~\cite{Hou:2019efy}, the CT18A fit reveals tensions between
the precision ATLAS $\sqrt{s} = 7$~TeV $W$, $Z$ data~\cite{ATLAS:2016nqi} and the  NuTeV~\cite{Mason:2006qa} and CCFR~\cite{NuTeV:2001dfo} SIDIS di-muon data.
In this study, we find that in the CT18As fit, the ATLAS 7 TeV $W$ and $Z$ production data~\cite{ATLAS:2016nqi} can be better described by an enhanced strange quark distribution $s(x)$, and $s(x) + {\bar s}(x)$, while the improvement in the quality of the fit to the NuTeV and CCFR SIDIS di-muon data is not evident.

We make available grids for the CT18As\_Lat and CT18As2\_Lat NNLO PDFs described above as a part of the LHAPDF library~\cite{Buckley:2014ana} and at the CTEQ-TEA website~\cite{CTEQ:hepforge}
The CT18As2\_Lat, defined in App.~\ref{sec_app:CT18As2_Rs}, is a variant fit similar to the CT18As\_Lat, but with an alternative parametrization for strange quark and antiquark PDFs.
As to be discussed in Appendix~\ref{sec_app:CT18As2_Rs}, the CT18As2 has a moderate PDF ratio $(s+\bar{s})/(\bar{u}+\bar{d})(x)$ in the large-$x$ region, so does the CT18As2\_Lat, while CT18As\_Lat and CT18As2\_Lat PDFs present comparable descriptions to experimental data.
A few comparisons of PDF combinations related to strange quark and antiquark PDFs for CT18As\_Lat, MSHT20~\cite{Bailey:2020ooq}, and NNPDF4.0~\cite{Ball:2021leu} are shown in Fig.~\ref{fig:compare_other_PDFs}, for reference.

With the release of CT18As\_Lat, we also update the CT18 PDF series, in the LHAPDF library, with a better numerical precision in providing PDFs for $Q$ crossing the heavy (charm and bottom) parton mass threshold, and for $x$ approaching to 1. There is no significant difference from the version of CT18 already on LHAPDF.

\section*{Acknowledgment}
We thank Rui Zhang for early-state participation in this work and our CTEQ-TEA colleagues for useful discussions.
This work is partially supported by the U.S. National Science Foundation
under Grant No.PHY-1653405, PHY-2013791 and  PHY-2209424.
HL thank MILC Collaboration for sharing the lattice configurations.
This research used Chroma software suite~\cite{Edwards:2004sx} and  resources of the National Energy Research Scientific Computing Center, a DOE Office of Science User Facility supported by the Office of Science of the U.S. Department of Energy under Contract No. DE-AC02-05CH11231 through ERCAP,
facilities of the USQCD Collaboration are funded by the Office of Science of the U.S. Department of Energy,
and  the Extreme Science and Engineering Discovery Environment (XSEDE)  supported by National Science Foundation Grant No. 1548562.
The work of HL is partially supported by the  Research  Corporation  for  Science  Advancement through the Cottrell Scholar Award.
CPY is also grateful for the support from
the Wu-Ki Tung endowed chair in particle physics.

\clearpage
\appendix

\section{Parametrization dependence for PDF ratio $(s+\bar{s})/(\bar{u}+\bar{d})(x)$ in large-$x$}
\label{sec_app:CT18As2_Rs}

In Sec.~\ref{results}, we found that the central prediction for the PDF ratio $(s+\bar{s})/(\bar{u}+\bar{d})(x)$ of CT18As and CT18As\_Lat for $x>0.2$ is enhanced comparing to CT18A, as shown in the bottom panel of Fig.~\ref{fig:strange_PDFs}.
This feature is caused by the choice of the more flexible nonperturbative parametrization form of the (anti)strange PDF adopted in the CT18As fit, in comparing to that in CT18.
Since the (anti)strange PDF is less constrained by data at such large $x$ values, we examine
in this Appendix an alternative fit with an additional theory prior to constrain the ratio of $(s+\bar{s})/(\bar{u}+\bar{d})(x)$ as $x$ approaching to 1.	
This alternative parametrization enforces strange and antistrange distributions having the same behavior in the large-$x$ limit as up and down quark distributions.
The resulting PDF is denoted as ``CT18As2''.
Specifically, it is done as follows.
The general functional form, in terms of the free parameters $a_k$ at the initial scale $Q_0$, is given in Eq.~(\ref{eq:PDF_para_general}) in the Appendix~\ref{sec_app:para} and summarized in the Appendix~C of the CT18 distributions paper~\cite{Hou:2019efy}.
The coefficients $a_1$ and $a_2$, cf. Eq.~(\ref{eq:PDF_para_general}), control the asymptotic behavior of $f_{(i)}(x,Q_0)$ in the limits $x\rightarrow 0$ and $1$ respectively.
In CT18As, we bind the high-$x$ exponents of the $\bar{u}$, and $\bar{d}$ distributions, $a^{\bar{u}}_2\! =\! a^{\bar{d}}_2$, to stabilize $\bar{d}/\bar{u}$ for $x\!\to\!1$, while allowing $a^{s}_2$ and $a^{\bar{s}}_2$ to be fit independently.
However, in CT18As2, we impose a stronger theory prior to bind the high-$x$ exponents of the $\bar{u}$, $\bar{d}$, ${s}$ and $\bar{s}$ distributions, $a^{\bar{u}}_2\! =\! a^{\bar{d}}_2 =\!
a^{s}_2\! =\! a^{\bar{s}}_2$, to stabilize both  $\bar{d}/\bar{u}$ and $(s+\bar{s})/(\bar{u}+\bar{d})$ for $x\!\to\!1$.

\begin{figure}[htbp]
\includegraphics[width=0.49\textwidth]{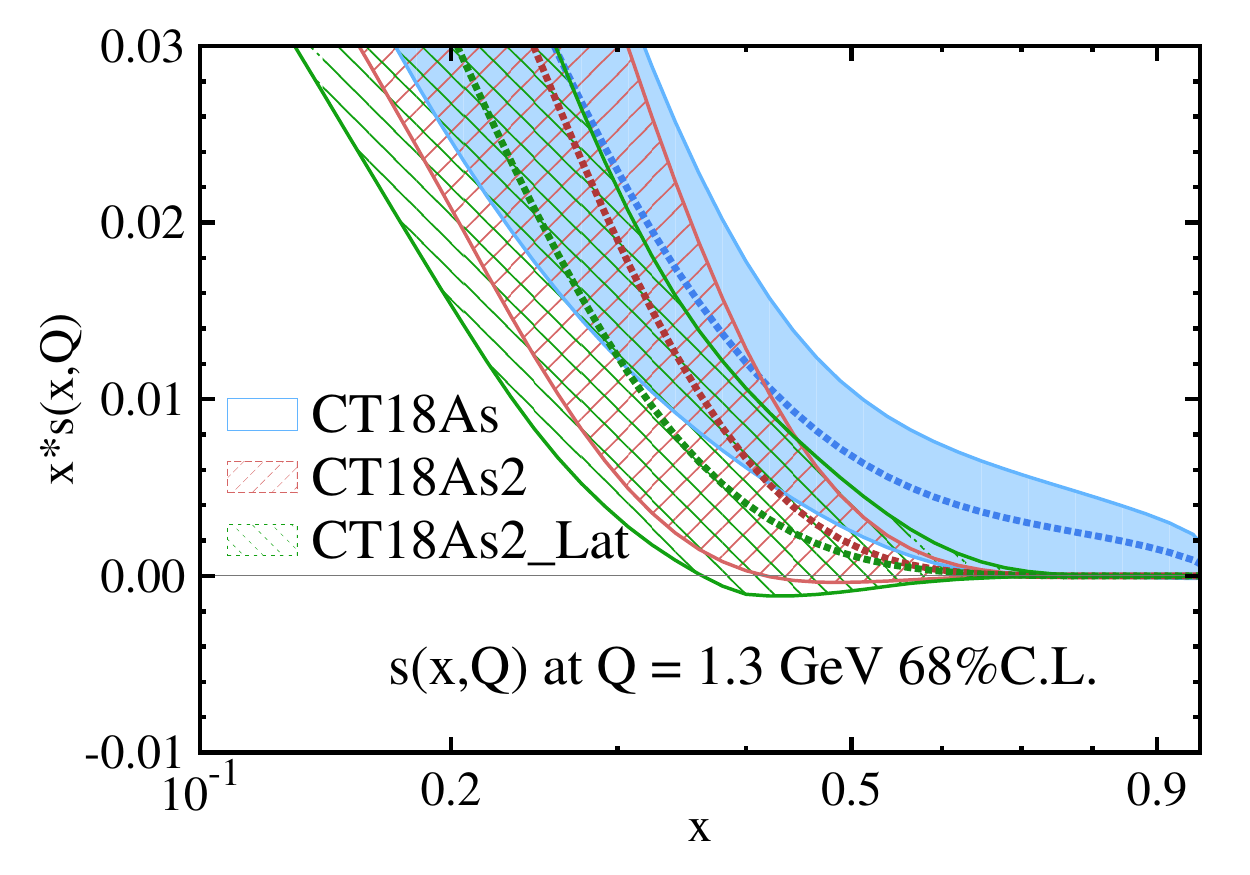}
\includegraphics[width=0.49\textwidth]{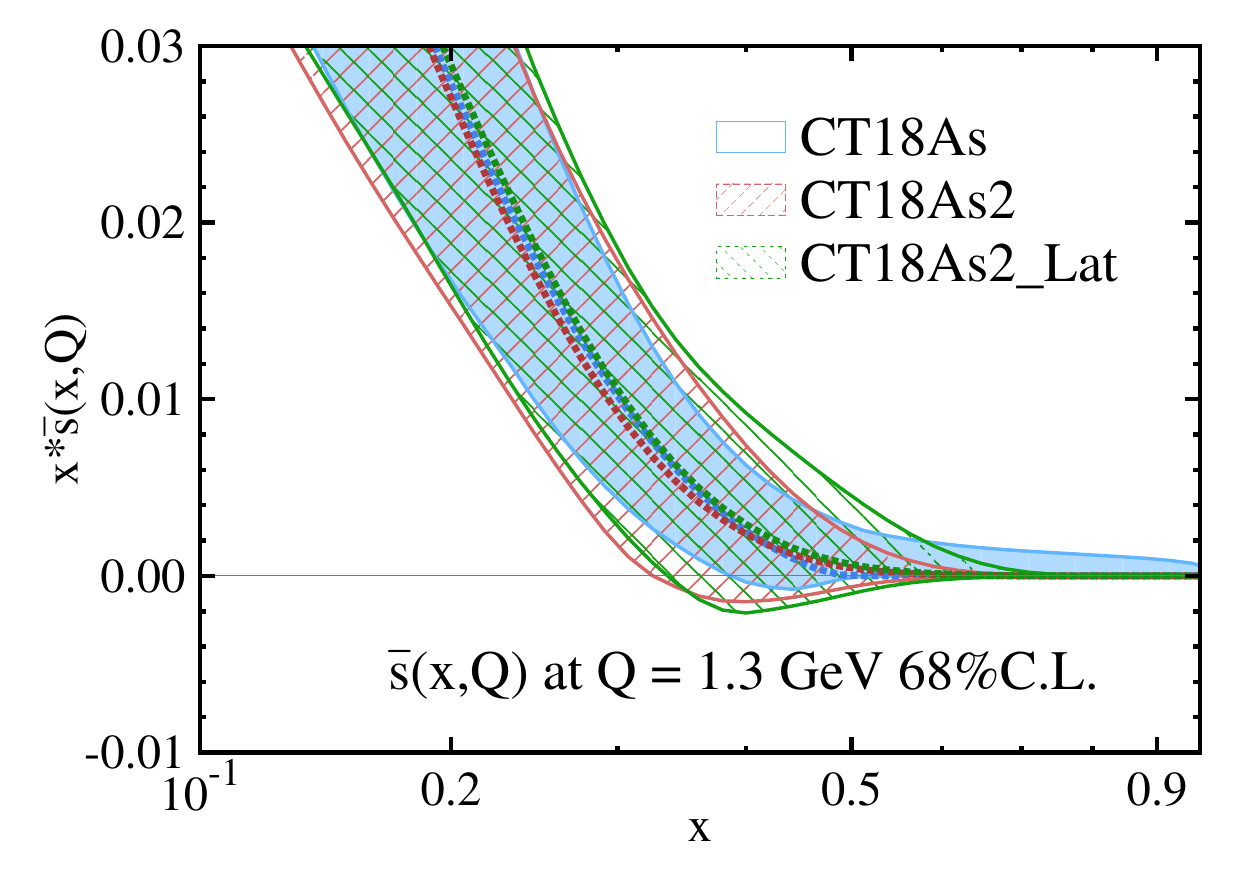}
\caption{
\label{fig:CT18As2_figs_1}
The comparison of $s(x)$ (left), $\bar{s}(x)$ (right) PDFs at the initial $Q_0(=1.3 \ {\rm GeV})$ scale for CT18As, CT18As2, and CT18As2\_Lat.
}
\end{figure}

\begin{figure}[htbp]
\includegraphics[width=0.49\textwidth]{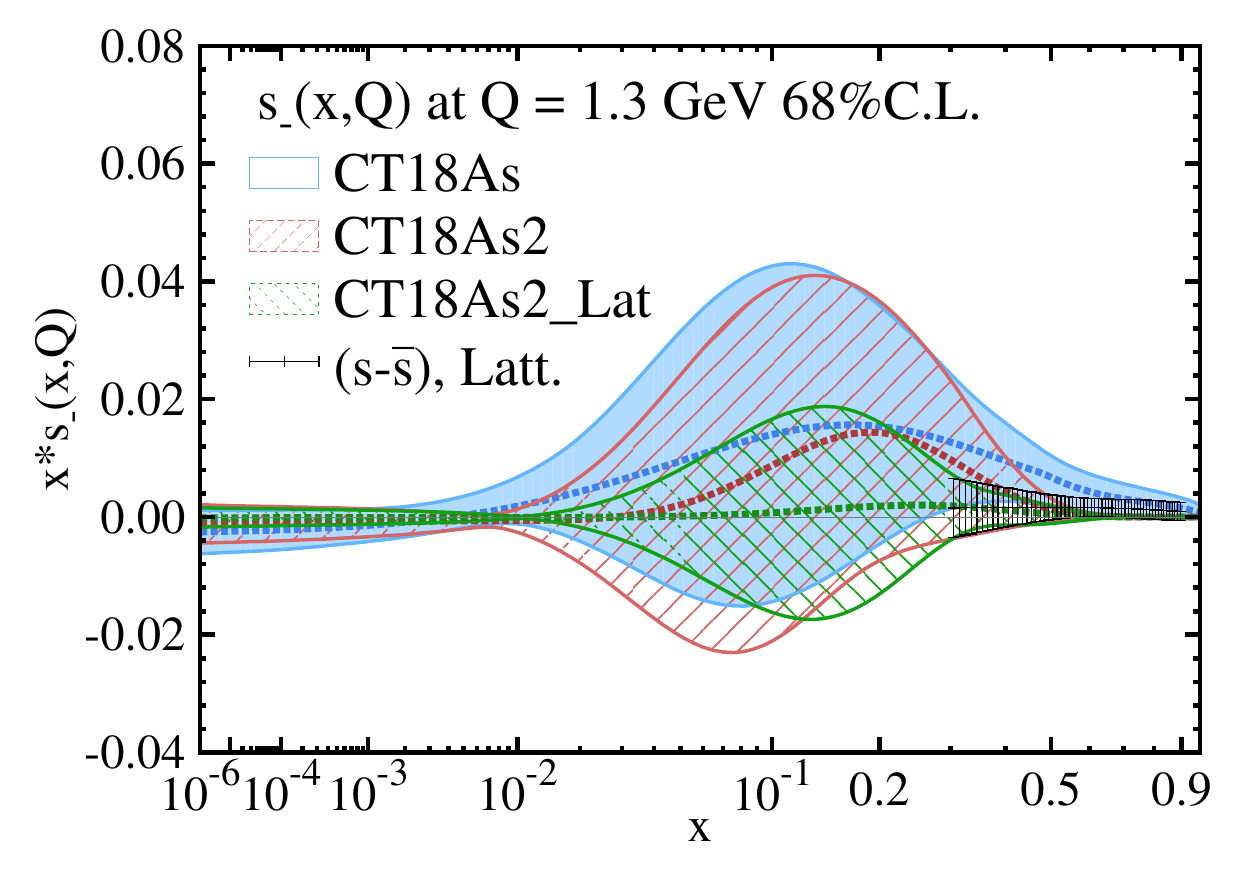}
\includegraphics[width=0.49\textwidth]{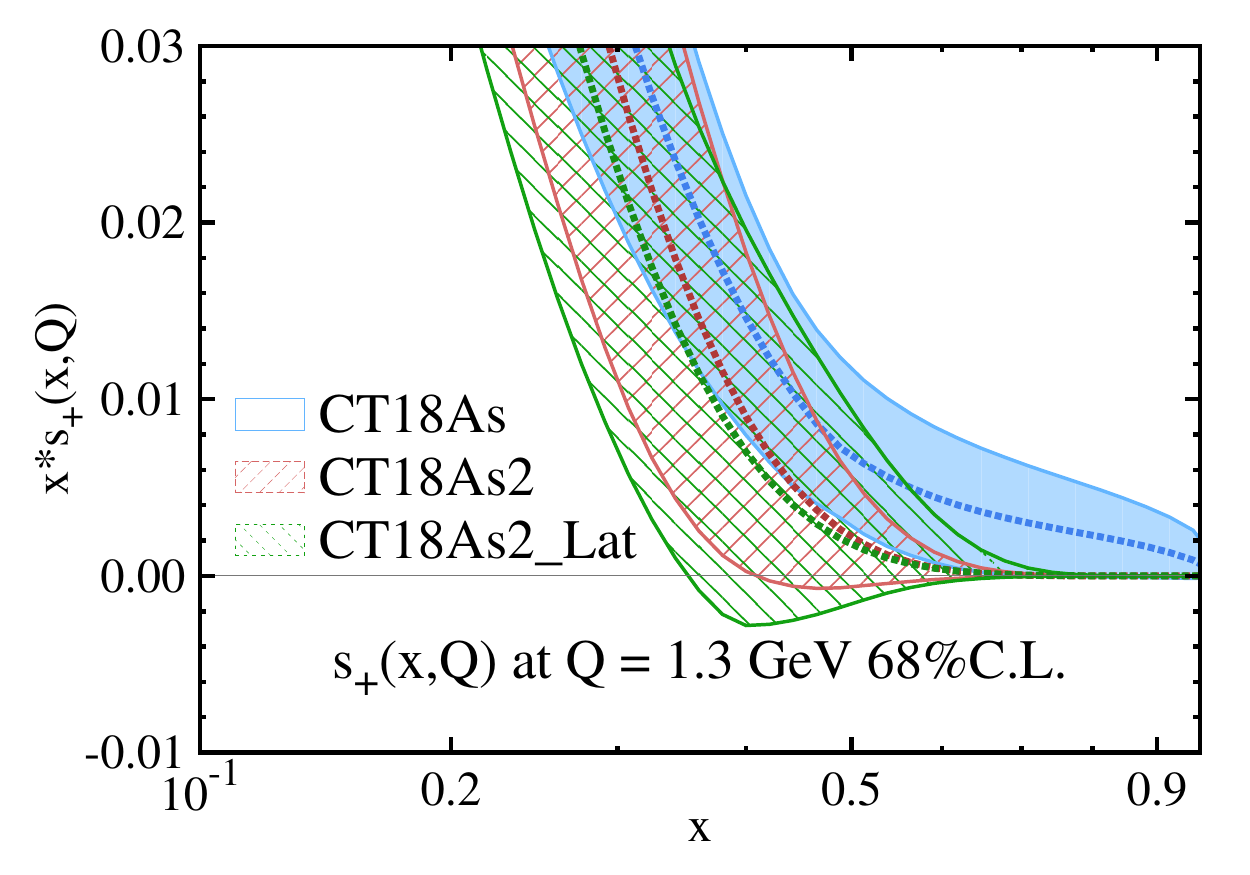}
\includegraphics[width=0.49\textwidth]{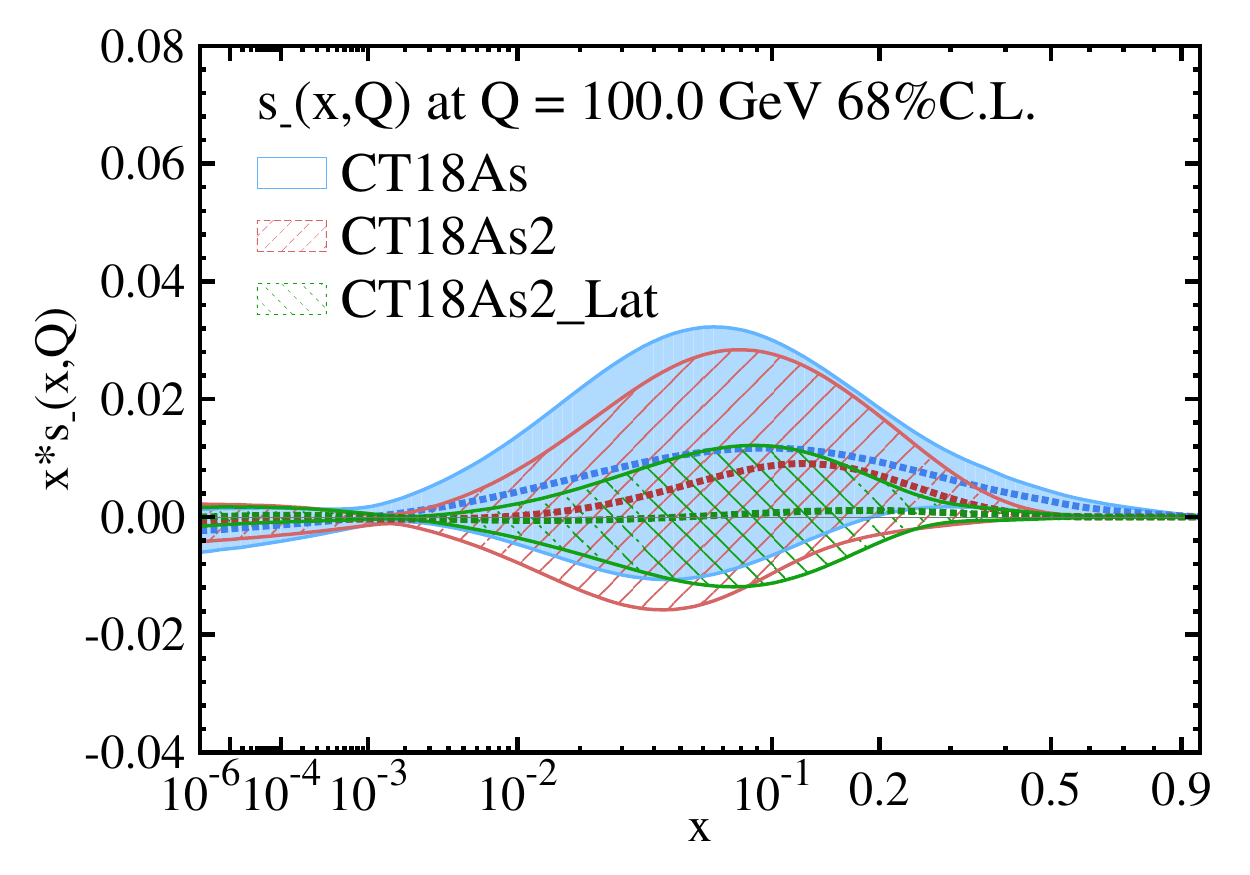}
\includegraphics[width=0.49\textwidth]{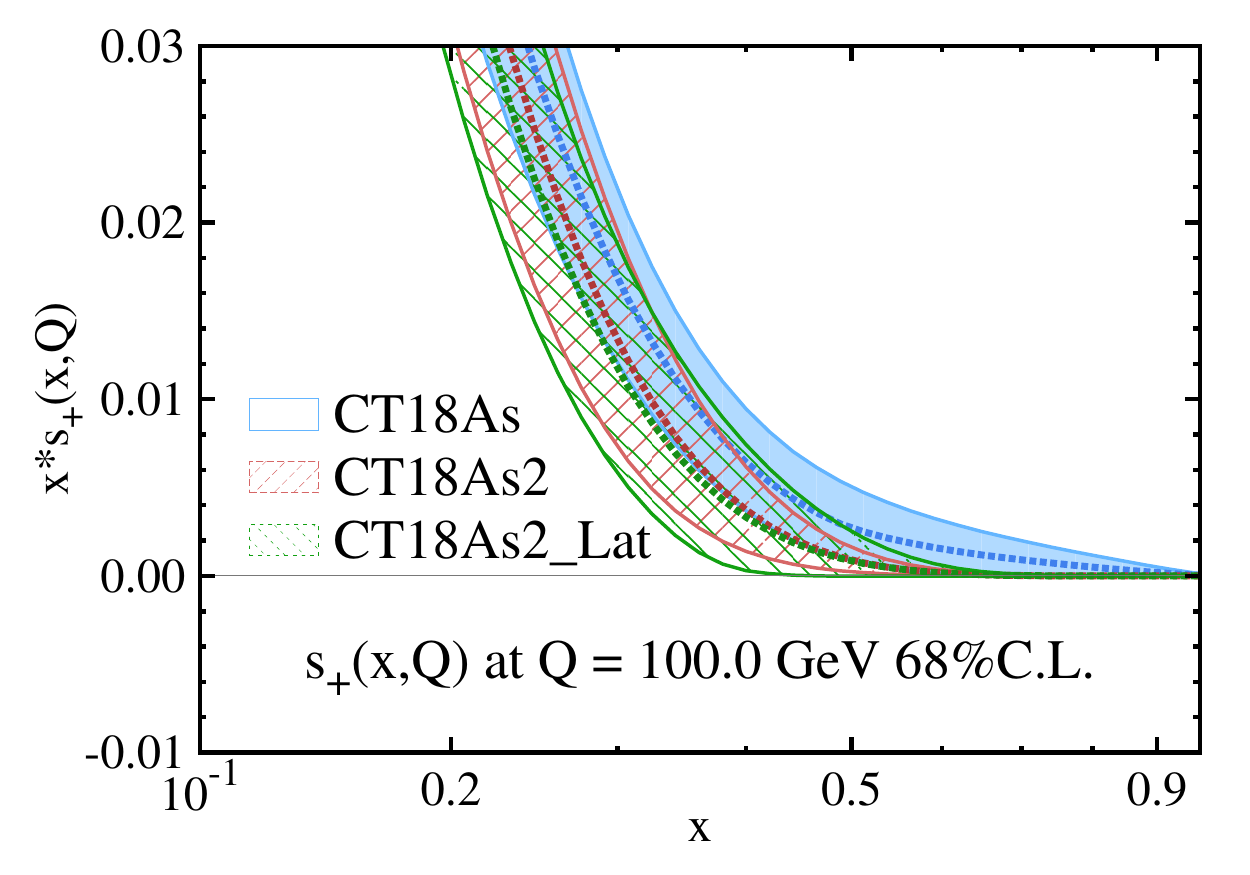}
\includegraphics[width=0.49\textwidth]{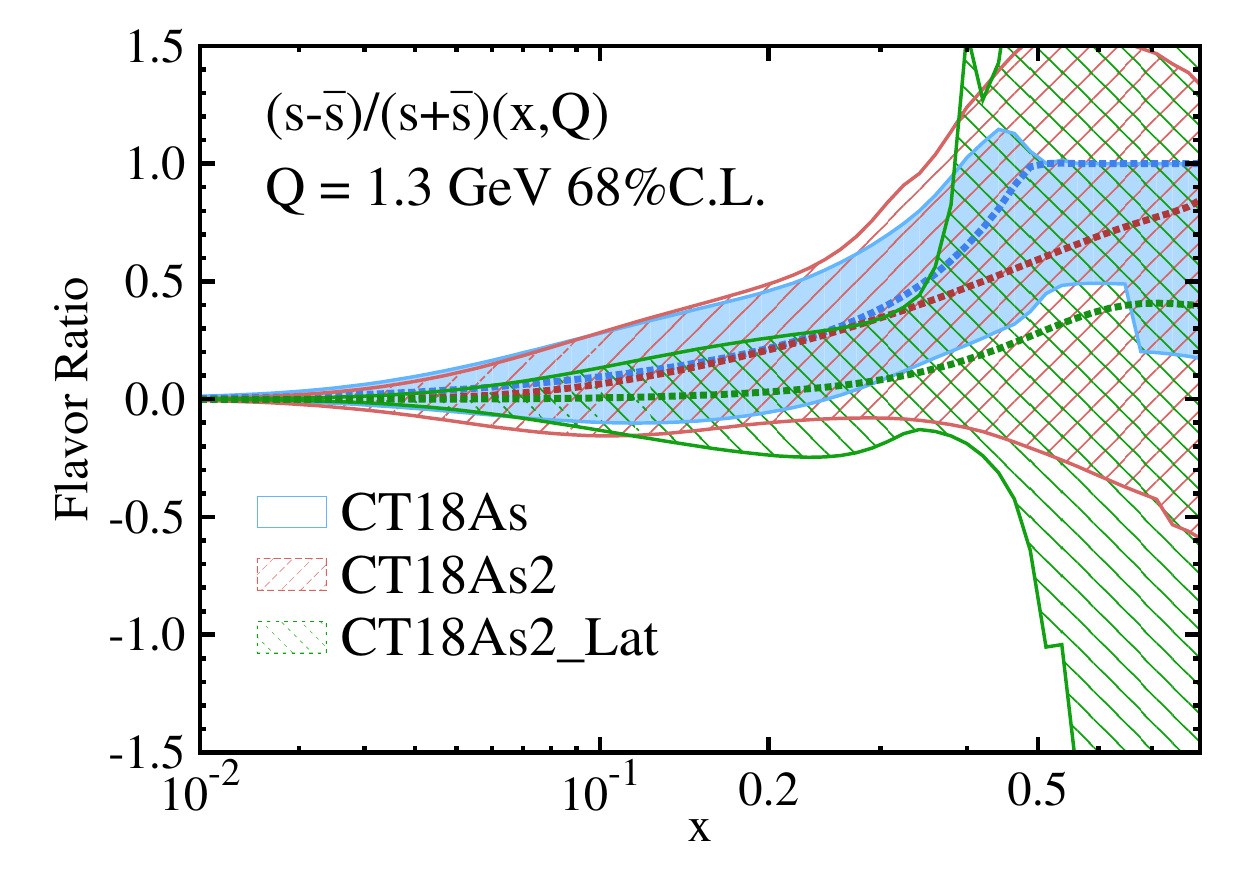}
\includegraphics[width=0.49\textwidth]{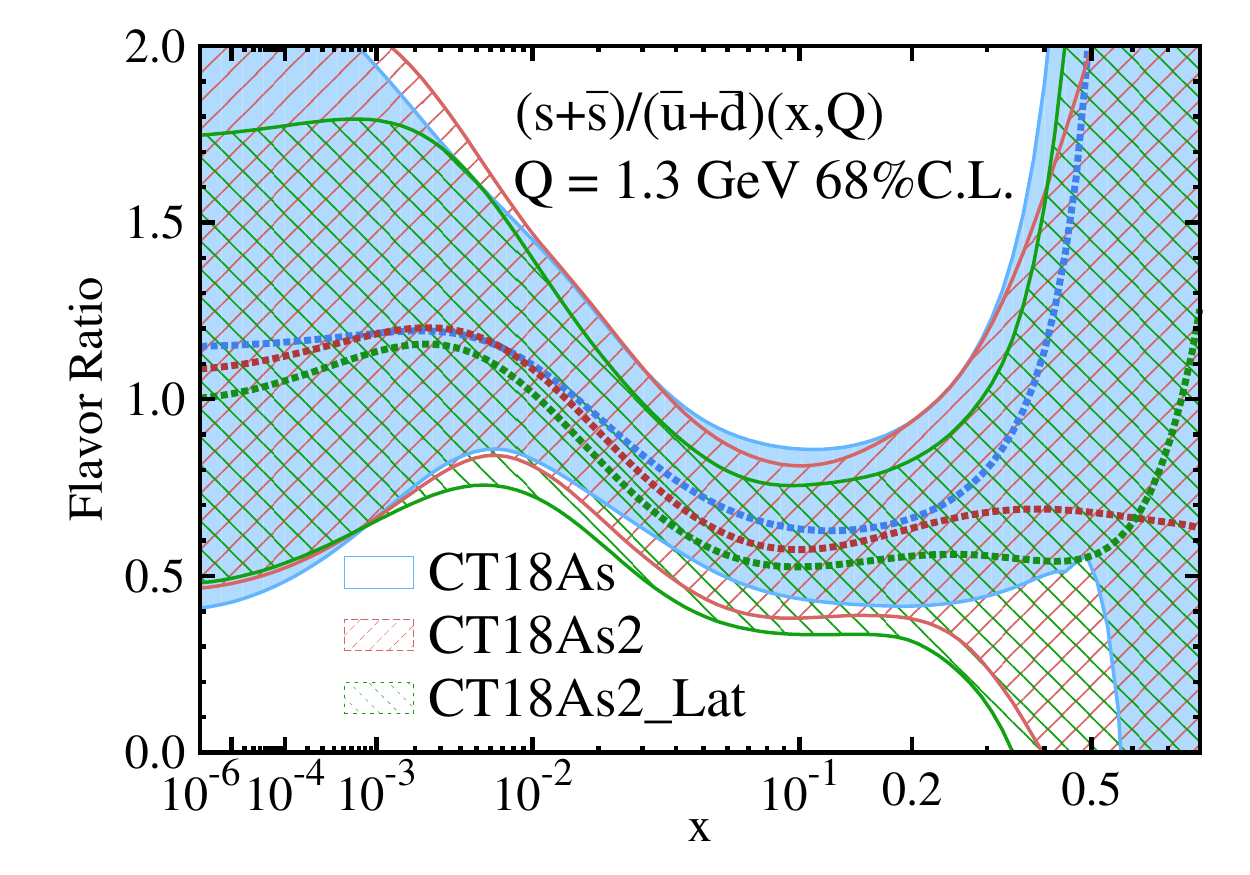}
\caption{
\label{fig:CT18As2_figs_2}
The comparison of $s_-(x)$, and $s_+(x)$ PDFs at the initial $Q_0(=1.3 \ {\rm GeV})$ scale and $Q = 100$~GeV, as well as PDF ratio $(s-\bar{s})/(s+\bar{s})(x)$ and $(s+\bar{s})/(\bar{u}+\bar{d})(x)$ at $Q_0 = 1.3$~GeV, for CT18As, CT18As2, and CT18As2\_Lat.
}
\end{figure}

The alternative parametrization impacts mostly in the strange PDF in the large-$x$ region.
As shown in in Fig.~\ref{fig:CT18As2_figs_1},
the strange PDF in CT18As2 for $x > 0.3$, at $Q=1.3$~GeV, is suppressed as compared to that in CT18As.
We note that the apparent negative CT18As2 ${\bar s}$-PDF for $x$ around 0.4 arises from the numerical precision for calculating the PDF error band.
In Fig.~\ref{fig:CT18As2_figs_2} the strangeness asymmetry $s_-(x)$ in CT18As2 is fairly similar to that in CT18As, except for $x>0.4$ where the  $s_-(x)$ in CT18As2 vanishes, for the strange and antistrange PDFs themselves vanishing fast in this region.
The comparisons of PDF ratios
$(s-\bar{s})/(s+\bar{s})(x)$ and
$(s+\bar{s})/(\bar{u}+\bar{d})(x)$, at $Q_0 = 1.3$~GeV,
are respectively displayed in the bottom row panels of Fig.~\ref{fig:CT18As2_figs_2}.
It is evident that CT18As and CT18As2 have different trends in those two PDF ratios as $x$ approaching to 1.
Nevertheless, CT18As and CT18As2 provide comparable descriptions to the experimental data analyzed in this work. For the CT18As2, the total $\chi^2_{\text{tot}}$ is 4362, higher than CT18As by 18 unites.
For the CT18As2\_Lat, the total $\chi^2_{\text{tot}}$ is 4370, only slightly higher than CT18As\_Lat by 8 units.
The difference in $\chi^2_{\text{tot}}$ is much smaller than the tolerance ($\Delta \chi^2 = 100$ for 90\% CL) used in the CT18 analysis.
Finally, we note that after including the lattice data, the resulting CT18As\_Lat and CT18As2\_Lat fits lead to similar conclusion.

\section{Non-perturbative parametrization form of strangeness}
\label{sec_app:para}

The general CT18 parametrization is reviewed in Appendix C of Ref.~\cite{Hou:2019efy}. The CT18 parametrization strategy is, out of a wide range of parametric forms at the starting scale $Q_0=1.3$ GeV, to find a flexible parametrization, which fits high-energy data without overfitting, and to understand the uncertainties associated to these parameters.
In this study,  we adopt the CT18 non-perturbative parametrization of $u$, $\bar{u}$, $d$, $\bar{d}$, and $g$ PDFs.
To obtain fits with non-vanishing strangeness asymmetry of this study, we choose to parametrize $s(x,Q_0)$ and $\bar{s}(x,Q_0)$ PDFs separately, unlike in CT18 parametrization form where $s(x,Q_0) = \bar{s}(x,Q_0)$ is taken. The specific parametrization of $s(x,Q_0)$ and $\bar{s}(x,Q_0)$ for fits in this study is summarized below.

CT PDFs are parametrized with a set of Bernstein polynomials (also called a B\'ezier curve), such that for the PDF of flavour $i$,
\begin{equation}
f_{(i)}(x,Q_0) = a_0 x^{a_1 - 1}(1-x)^{a_2} P_{(i)}\big{(}y(a_3,x), a_4, a_5, \dots \big{)} = a_0 \tilde{f}_{(i)}(x,Q_0).
\label{eq:PDF_para_general}
\end{equation}
As stated in Eq.~(\ref{eq:sv_nsr}), the strangeness asymmetry satisfies the number sum rule, that the net number of strange quarks subtracted by antistrange quark number is zero,
\begin{equation}
\int_0^1 dx \; \big{(} s(x) - \bar{s}(x) \big{)} = 0.
\label{eq:sqk_sbr_nsr}
\end{equation}
Combining with Eq.~(\ref{eq:PDF_para_general}), the number sum rule for strangeness asymmetry applies as a constraint on the parametrization form,
\begin{equation}
\int_0^1 dx \; \big{(} a_0^{(s)} \tilde{f}_{(s)}(x) - a_0^{(\bar{s})} \tilde{f}_{(\bar{s})}(x) \big{)} = 0 .
\label{eq:svl_para_constraint}
\end{equation}
By given the $a_0^{(\bar{s})}$ fitted value, which is determined as in usual CT18, the overall factor for $s(x,Q_0)$ is found via Eq.~(\ref{eq:svl_para_constraint}),
\begin{equation}
a_0^{(s)} = \frac{\int dx \ a_0^{(\bar{s})} \tilde{f}_{(\bar{s})}(x) }{\int dx \ \tilde{f}_{(s)}(x) }.
\end{equation}
The functional form of strangeness distributions is similar to the case in CT18,
\begin{eqnarray}
P(x) &=&             (1-y)^5 +
       a_4   5 y   (1-y)^4 +
       a_5  10 y^2 (1-y)^3 +
       a_6  10 y^3 (1-y)^2   \nonumber \\
\label{eq:func_1}
     &\quad& + \ a_7   5 y^4 (1-y)   +
       a_8     y^5\ , \\
y(x) &=& 1 - (1-\sqrt{x})^{a_3}.
\label{eq:func_2}
\end{eqnarray}
The best-fit values of $s(x)$ and $\bar{s}(x)$ PDFs parameters for the central PDF of CT18As\_Lat and CT18As2\_Lat fits are summarized in Table~\ref{tab:opt_parm_value}.

\begin{table}[htbp]
\begin{center}
\begin{tabular}{c|rr|rr}
 & \multicolumn{2}{c|}{ CT18As\_Lat } & \multicolumn{2}{c}{CT18As2\_Lat} \\
 cent. fitted value & $s(x)$ & $\bar{s}(x)$ & $s(x)$ & $\bar{s}(x)$ \\ \hline
 $a_1$ & -0.013$^{(c)}$ & -0.013$^{(c)}$ & -0.021$^{(c)}$ & -0.021$^{(c)}$ \\
 $a_2$ & 0.256$\quad$ & 3.727$\quad$ &  7.744$^{(d)}$ &  7.744$^{(d)}$ \\
 $a_3$ & 4.000$^{(a)}$ & 4.000$^{(a)}$ &  1.048$\quad$ &  1.048$\quad$ \\
 $a_4$ & 0.408$^{(b)}$ & 0.408$^{(b)}$ & -0.059$^{(b)}$ & -0.059$^{(b)}$ \\
 $a_5$ & 0.408$^{(b)}$ & 0.408$^{(b)}$ & -0.059$^{(b)}$ & -0.059$^{(b)}$ \\
 $a_6$ & 0.169$^{(c)}$ & 0.169$^{(c)}$ &  1.750$^{(c)}$ &  1.750$^{(c)}$ \\
 $a_7$ & 0.150$\quad$ & 0.228$\quad$ & -1.254$\quad$ & -1.219$\quad$ \\
 $a_8$ & 0.004$\quad$ & 0.068$\quad$ &  2.948$\quad$ &  1.530$\quad$ \\
\end{tabular}
\end{center}
\caption{
\label{tab:opt_parm_value}
Fitted parameter values obtained for the central CT18As\_Lat and CT18As2\_Lat NNLO fits. The functional forms for each parameterzation are specified in Eqs.~(\ref{eq:func_1}) and~(\ref{eq:func_2}).
(a) Value of $a_3$ parameter in Eq.~(\ref{eq:func_2}) is fixed at 4.0.
(b) Values of $a_4$ and $a_5$ parameters for $s(x)$ and $\bar{s}(x)$ PDFs are bound together.
(c) Values of the denoted parameters for $s(x)$ and $\bar{s}(x)$ PDFs are bound together.
(d) Values of the large-$x$ parameter $a_2$ for both $s(x)$ and $\bar{s}(x)$ distributions are bound with that of $\bar{u}(x)$ and $\bar{u}(x)$.
}
\end{table}

-------------
\clearpage
\bibliographystyle{utphys}
\bibliography{CT18As_Lat}

\end{document}